\documentclass[10pt]{article}

\usepackage[T1]{fontenc}
\usepackage{lmodern}
\usepackage{microtype}
\usepackage[american]{babel}

\usepackage[hmargin=2.6cm,vmargin=2.4cm]{geometry}
\usepackage{mathtools}
\usepackage{amssymb}
\usepackage{amsthm}

\usepackage{graphicx}
\usepackage{float}
\usepackage{subcaption}

\usepackage{array}
\usepackage{multirow}
\usepackage{makecell}
\usepackage{siunitx}

\usepackage[multiple]{footmisc}
\usepackage{titlesec}

\usepackage{xcolor}

\newcommand{\cryptosym}{%
    \leavevmode
    \vtop{%
        \offinterlineskip
        \setbox0=\hbox{\texttt{C}}%
        \setbox2=\hbox to\wd0{%
            \hfil\hskip .08em
            \vrule height .3ex width .15ex\hfil
        }%
        \vbox{\copy2\box0}\box2
    }%
}

\usepackage{parskip}
\usepackage{cite}
\usepackage[hidelinks]{hyperref}

\begin{document}

\title{Where Are the Rates? Derivative-Implied Interest Rates in Cryptocurrency Markets\thanks{An earlier version of this paper circulated under the title ``Cryptocurrencies and Interest Rates: Inferring Yield Curves in a Bondless Market'' (arXiv:2509.03964).}}

\author{Philippe \textsc{Bergault}\footnote{Université Paris Dauphine-PSL, Ceremade, Paris, France, bergault@ceremade.dauphine.fr.} \and Sébastien \textsc{Bieber}\footnote{Université Paris Dauphine-PSL, Ceremade, Paris, France, bieber@ceremade.dauphine.fr.} \and Olivier \textsc{Guéant}\footnote{Université Paris Cité, Laboratoire de Probabilités, Statistique et Modélisation, Paris, France, olivier.gueant@u-pariscite.fr.} \and Wenkai \textsc{Zhang}\footnote{LCH SA, Paris, France, wenkai.zhang@lseg.com.}}
\date{}
\maketitle

\begin{abstract}
In traditional finance, liquid markets for bonds, interest rate swaps, and other interest rate derivatives
provide a basis for constructing yield curves. In the cryptocurrency ecosystem, lending and staking yields
are predominantly variable and adjust over short timescales, whereas fixed-rate instruments remain scarce
and, as of early 2026, widely accepted benchmark yield curves were still lacking. We jointly infer USD
and cryptocurrency discount factors from inverse options and futures by combining put-call parity with
the futures basis in a closed-form estimator. We introduce a novel RANSAC-based quote filter tailored
to cryptocurrency market microstructure. We apply the method to hourly BTC and ETH derivatives data
from Deribit between February~2023 and February~2026. The two underlying assets imply broadly similar USD
term structures, with rates averaging approximately 8\% per year and rising during the 2024 cryptocurrency
rallies. Implied BTC and ETH rates are typically close to zero across the maturities considered, despite
positive Aave lending yields and ETH staking yields. Implied USD rates co-move with Aave USDC rates but
exhibit persistent differences. These findings suggest limited integration between derivatives markets
and decentralized lending and staking.\\


\end{abstract}

\section{Introduction}

\subsection{Yield Curves: A Familiar Tool}

Yield curves are a fundamental component of financial market data, describing how interest rates vary across
maturities. Central bank policy rates and interbank rates provide key reference points at the short end, while
bond and interest rate derivatives markets provide information at longer horizons. Issuer-specific curves
reflect the borrowing conditions of governments and corporations, whereas discount curves used in derivatives
valuation depend on the currency of the cash flows and the terms of collateralization. These curves support
the valuation of future cash flows and the calibration of stochastic interest rate models.

Depending on the instruments available, yield curves can be constructed from bond prices, money market instruments, or interest rate derivatives using bootstrapping, spline-based methods, or parametric and penalized regression approaches (see, e.g., \cite{nelson1987parsimonious, svensson1994estimating, fisher1995fitting, hagan2006interpolation}, and \cite{brigo2001interest} for a broader treatment of interest rate models). In many established financial markets, their construction is facilitated by the availability of liquid instruments across a broad range of maturities, reflecting the historical development of markets in debt securities and interest rate derivatives.\footnote{For a historical overview of interest rates and their term structure, see, for example, \cite{gueant2026historical} and the references therein.}

This observation raises a natural question: how can one infer maturity-dependent interest rates in markets where fixed-income instruments are scarce? Cryptocurrency markets illustrate this challenge: observable yields are predominantly variable, adjusting over very short intervals rather than being fixed in advance over a specified horizon.

\subsection{The Emergence of Cryptocurrencies}

From ancient coinage bearing the faces of kings and emperors to modern fiat currencies managed by central
banks, money has historically been intimately tied to a central authority. While money fulfills core economic
functions -- serving as a medium of exchange, a unit of account, and a store of value -- it has also
frequently operated as an instrument of governance, a symbol of regional or national sovereignty, and, at
times, a tool of diplomacy or coercion.

The recent emergence of cryptocurrencies represents a significant departure from this historical pattern.
Arising from the convergence of cryptography, computer science, and libertarian ideas, Bitcoin and many other
cryptocurrencies embody a model of money that is neither issued by a central authority nor backed by a
government.\footnote{The link between cryptocurrency markets and traditional finance has been facilitated by the
introduction of stablecoins. These digital assets are designed to maintain a stable value relative to a fiat
currency, most commonly the US dollar. Various types of stablecoins exist, including centralized ones such
as USDC and USDT, issued by Circle and Tether, respectively, and decentralized alternatives such as DAI and USDS. Stablecoins represent a major segment of the cryptocurrency market, with a total capitalization exceeding \$300~billion as of early 2026 -- see \url{https://www.coingecko.com} and \url{https://www.defillama.com}.} Instead, they rely on open-source protocols that specify issuance and transaction-validation rules, with cryptographic verification and consensus among network participants allowing the ledger to be maintained without a central operator.

One intellectual antecedent is Friedrich Hayek's \emph{Denationalisation of Money}~\cite{hayek1976denationalisation}, which advocates competition among privately issued currencies. Although cryptocurrencies differ from the arrangements envisaged by Hayek, they revive the question of how monetary systems can operate without a central issuer. Bitcoin, in particular, was introduced in the wake of the~2008 financial crisis as an alternative to state-backed fiat currencies and as a response to the perceived structural failures of the traditional banking
system.  In the years that followed, Bitcoin inspired a growing ecosystem of other blockchains and
cryptocurrencies with new technical and economic features. The launch of Ethereum in~2015 marked a major
turning point. By supporting smart contracts -- programs executed on the blockchain -- Ethereum extended the
scope of blockchain applications beyond peer-to-peer payments and enabled a new generation of decentralized
applications~(dApps). This development laid the foundations for decentralized finance (DeFi), whose protocols provide services such as lending, borrowing, and trading with reduced reliance on traditional financial intermediaries. Interest rates in these lending protocols are generally variable, adjusting algorithmically as the supply of funds and demand for borrowing change.

\subsection{Interest Rates in the Cryptocurrency Landscape}

Although markets for fixed-income instruments remain limited in the cryptocurrency ecosystem, yields are directly observable through staking and lending. These yields are predominantly variable and provide reference points for short-term remuneration only. Yet the absence of established benchmark yield curves does not mean that the prices of future payments cannot be inferred: they may be embedded in other traded instruments. We first discuss staking and lending yields before turning to the information contained in cryptocurrency option and futures prices.

\subsubsection{Staking Yields as Short-Term Reference Rates}

Unlike Bitcoin's blockchain, which relies on proof-of-work (PoW), many other blockchains use proof-of-stake (PoS) mechanisms.\footnote{Ethereum completed its transition from PoW to PoS in 2022 through ``The Merge.''} In PoS systems, validators commit native tokens as stake and participate in proposing and validating blocks according to protocol-specific rules. They receive rewards, typically paid in the native token, for contributing to network security and transaction validation. Token holders can participate directly as validators or indirectly through pooled staking services such as Lido and Rocket Pool or the staking services of centralized exchanges such as Binance.\footnote{See \cite{carre2024liquid} and \cite{gogol2024sok} for discussions of staking and the role of liquid staking in DeFi.}

Staking rewards compensate validators for both the capital committed and the services provided. The resulting yield is variable: the remuneration per unit of stake depends on protocol rules and network conditions, including the amount staked and the transaction fees distributed to validators. Under simplifying assumptions that set aside the associated costs and risks, this yield can be viewed as an endogenous analog of a short-term risk-free rate denominated in the blockchain's native token.\footnote{This interpretation abstracts from the costs of operating a validator and from blockchain and protocol failures, validator penalties and slashing, and risks associated with any staking intermediary or smart contract used. It concerns remuneration in units of the native token and does not imply a risk-free return in USD.} This analogy concerns the remuneration of capital over short intervals; it does not imply that staked capital can always be withdrawn immediately or that the yield is fixed over the holding period.

\subsubsection{Short-Term Rates in Decentralized Lending Markets}

Decentralized lending provides another source of observable yields. Protocols such as Aave, Compound, and Morpho Blue allow users to supply tokens to lending pools and earn interest paid by borrowers.\footnote{See \cite{gudgeonDeFiProtocolsLoanable2020} for further details on lending protocols.} Pseudonymity allows borrowers in permissionless protocols to participate without disclosing their legal identity, limiting lenders' ability to rely on conventional legal recourse. Lending therefore typically relies on overcollateralization: borrowers pledge collateral worth more than the amount borrowed, with collateral requirements and liquidation rules enforced through smart contracts. These mechanisms are designed to limit credit losses, although bad debt can arise when collateral prices fall rapidly or liquidations cannot be executed as intended.

These lending markets can be viewed as counterparts to money markets. Their interest rates are generally determined algorithmically, often as functions of pool utilization, with parameters reflecting the characteristics of the assets and collateral involved. See \cite{frangella2022aave} for Aave's interest rate model, \cite{leshner2019compound} for Compound's, and \cite{cohen2023economics, bertucci2025agents, baude2025optimal} for alternative modeling approaches. As of early 2026, variable-rate lending without a predetermined maturity remained the prevailing model. Rates can adjust over very short intervals as pool conditions change.\footnote{Fixed-rate alternatives exist, although variable-rate lending continues to dominate. Term Finance provides fixed-rate lending and borrowing through auctions -- see \url{https://www.term.finance}. Pendle separates a yield-bearing asset into principal and future yield claims. Investors can purchase the principal token at a discount and hold it to maturity, thereby locking in an implied yield in the relevant accounting asset, subject to the product's redemption terms -- see \url{https://docs.pendle.finance/pendle-v2/ProtocolMechanics/YieldTokenization/PT}. A subsequent development is Morpho Midnight, a fixed-rate, fixed-term credit protocol launched in July~2026, while we were finalizing the manuscript -- see \url{https://morpho.org/blog/now-live-morpho-midnight/}.}

The relationship between DeFi interest rates and conventional monetary conditions remains an empirical question. Research on decentralized lending, such as that of Chaudhary, Kozhan, and Viswanath-Natraj~\cite{chaudharyInterestRateParity2023}, emphasizes the role of cryptocurrency trading and leverage demand. Evidence on monetary policy transmission points to some connection with traditional rates, but also to substantial and persistent departures from them. Barthélemy et al.~\cite{banquedefranceInterestRatesDecentralised} highlight the importance of crypto-specific demand, while Barbon, Barthélemy, and Nguyen~\cite{barbonDeFiyingFedMonetary2023a} document monetary policy effects whose strength and direction depend on the interaction between interest rate arbitrage and changes in demand for leverage. Taken together, these findings suggest limited and uneven integration rather than either complete disconnection or close alignment.

\subsubsection{Inferring Fixed Interest Rates from Cryptocurrency Options and Futures}

To move from these variable short-term yields to rates associated with specified maturities, we turn to cryptocurrency options and futures. Building on existing approaches in the derivatives literature, we reverse the usual pricing perspective: rather than taking interest rates as inputs to value derivatives, we use observed derivative prices to infer implicit fixed interest rates. The availability of contracts at multiple maturities provides information about the valuation of future payments in both the cryptocurrency and the reference currency.

Derivative prices can provide information about financing conditions and convenience yields. Spot-futures relationships reflect financing costs net of the income and other benefits associated with holding the underlying asset, while option prices can be used to infer interest rates. In cryptocurrency markets, Arkorful et al.~\cite{arkorfulWhatCanWe2023} compare the convenience yield of Bitcoin with that of gold during the COVID-19 crisis and relate its decline to speculative trading, while Hilliard and Ngo~\cite{hilliardBitcoinJumpsConvenience2022} estimate daily Bitcoin convenience yields from Deribit data using put-call parity and the cost-of-carry relation. They also develop and estimate a jump-diffusion option pricing model with stochastic convenience yield. Schmeling, Schrimpf, and Todorov~\cite{schmelingCryptoCarry2023} document large and variable crypto futures carry, which they relate to demand for leveraged exposure and limits to arbitrage arising from various frictions.

The extraction of interest rates from option prices dates back at least to Brenner and Galai~\cite{brennerImpliedInterestRates1986}, who infer rates from put-call relationships for American stock options traded on the CBOE, accounting for the possibility of early exercise, while Ronn and Ronn~\cite{ronn1989box} study the arbitrage conditions of box spreads, which combine put-call parity relations at two strikes. More recently, van Binsbergen, Diamond, and Grotteria~\cite{vanbinsbergenRiskfreeInterestRates2022} construct term structures of interest rates from put-call parity on European S\&P~500 index options, an approach equivalent to pricing box spreads. They interpret the spread between these option-implied rates and Treasury yields as a measure of the convenience yield associated with holding Treasury securities. Diamond and Van Tassel~\cite{diamondRiskFreeRatesConvenience2021} extend the use of index-option-implied rates to the measurement of safe-asset convenience yields across currencies. Azzone and Baviera~\cite{azzone2021synthetic} exploit the fact that, by put-call parity, the difference between call and put prices is linear in the strike, with a slope equal to minus the discount factor. Applying this idea to European S\&P~500 and EURO STOXX~50 index options, they find that the implied discount curve coincides with the EUR OIS curve in the EURO STOXX~50 market, whereas it embeds a cost of funding of about 34 basis points over the USD OIS curve in the S\&P~500 market. From a methodological perspective, Kamau and Mwaniki~\cite{kamauRobustEstimationOptionimplied2023} jointly estimate interest rates and dividend yields from European S\&P~500 index options using robust regression methods, including Theil-Sen and repeated-median estimators. These studies demonstrate the usefulness of option-based rate extraction, while Kamau and Mwaniki's robust estimation approach highlights the importance of limiting the influence of noisy observations.

The economic interpretation of derivative-implied rates requires attention to market frictions. Naranjo~\cite{naranjoImpliedInterestRates2009} develops an equilibrium model of the futures market in which borrowing and short-selling costs allow demand imbalances to affect implied interest rates. In this framework, the implied rate reflects both the underlying risk-free rate and a demand component. In cryptocurrency markets, Felföldi-Szűcs, Králik, and Váradi~\cite{felfoldi-szucsBackwardForwardPut2023} examine put-call parity using one-second quotes for USDT-settled European options on Ethereum traded on Binance. They distinguish violations associated with the two directions of the arbitrage strategy and find that significant deviations remain after accounting for bid-ask spreads, trading fees, and funding costs. Their findings indicate that the costs considered in their analysis do not fully explain the observed departures from parity. For our purposes, this motivates particular care in selecting quotes and interpreting the rates extracted from them.

Building on this literature, we jointly estimate cryptocurrency and USD discount factors by combining inverse option and futures prices within a robust quote-filtering procedure. Closest to our work are the study of Hilliard and Ngo~\cite{hilliardBitcoinJumpsConvenience2022}, discussed above, and that of Alexander, Chen, Deng, and Wang~\cite{alexander2024arbitrage}, who derive a fiat-currency-free put-call parity relationship for BTC and ETH options and use it to test the joint efficiency of options and perpetual futures markets. In equity index markets, Azzone and Baviera~\cite{azzone2021synthetic} and van Binsbergen et al.~\cite{vanbinsbergenRiskfreeInterestRates2022} recover implied discount factors from the strike dimension of put-call parity. Our approach differs in three respects: we estimate the USD and cryptocurrency discount factors jointly, maturity by maturity, by combining inverse options with the inverse futures basis; we filter quotes with a model-agnostic procedure based on the geometry of bid and ask prices; and we use the resulting curves to discuss cross-market coherence. Because the inverse futures basis pins down, up to estimation error, the ratio of the two discount factors (Eq.~\eqref{eq:zc_crypto_from_inverse_futures}), the spread between implied USD and cryptocurrency rates is essentially the annualized log-basis of the futures contract, that is, the crypto carry studied by Schmeling, Schrimpf, and Todorov~\cite{schmelingCryptoCarry2023}, who show that it is poorly explained by differentials between observable interest rates. Option prices allow us to go one step further and split this spread into a USD rate and a cryptocurrency rate. We apply our approach to BTC and ETH derivatives traded on Deribit. The resulting estimates allow us to ask whether the two underlying markets imply a common USD term structure and how the inferred USD and cryptocurrency rates relate to lending and staking yields on decentralized protocols.

Our main findings are as follows. First, BTC and ETH inverse options imply broadly similar USD term structures: the median difference between the two estimates is about 1\% at the 60-, 90-, and 180-day tenors. Second, these implied USD rates average about 8\% per year over the sample, with peaks above 15\% during the 2024 rallies and episodes of term-structure inversion. Third, implied BTC and ETH rates remain close to zero across the maturities considered, despite positive lending yields on Aave and ETH staking rewards. Fourth, implied USD rates co-move with Aave USDC rates without matching their level. We interpret these results as evidence of limited integration between centralized derivatives venues and decentralized lending and staking, and we discuss the respective roles of margin and collateral frictions and of Deribit's quoting conventions.

The paper proceeds as follows. Section~\ref{sec:market} describes the cryptocurrency derivatives landscape, motivates our
focus on Deribit, and presents descriptive market statistics. Section~\ref{sec:method} develops the replication arguments and
estimation procedure. Section~\ref{sec:data} presents the empirical results and discusses what the comparisons with lending
and staking yields suggest about integration across the cryptocurrency ecosystem. Section~\ref{sec:conclusion}~concludes.

\section{Market Fragmentation and Platform Focus}\label{sec:market}

Cryptocurrency markets encompass traditional financial exchanges, crypto-native centralized exchanges,
and decentralized protocols. This section briefly describes these venues before focusing on derivatives
and the choice of Deribit for our empirical analysis.

\subsection{A Fragmented Market}

\subsubsection{Trading Venues and Regulatory Frameworks}

Three types of venues provide access to cryptocurrency markets. Traditional financial exchanges,
such as the Chicago Mercantile Exchange (CME) and Eurex, offer cryptocurrency exposure through listed
financial instruments. CME cryptocurrency futures, for example, are cash-settled in USD based on reference prices, without physical delivery of the underlying tokens. Crypto-native centralized exchanges, such as Binance, Deribit, and Coinbase, typically match orders and maintain customer
account balances off-chain. Decentralized finance (DeFi) protocols instead use smart contracts to provide
services such as trading and lending, with varying degrees of decentralization. Some rely on oracles,
which supply smart contracts with external data, such as asset prices. Derivatives are available across
all three categories: CME lists futures and options on futures, crypto-native exchanges offer a range
of futures and options, and some DeFi protocols provide access to options, such as Derive (formerly Lyra).\footnote{For more details on Derive, see \url{https://blog.derive.xyz/a-technical-overview-of-lyra-v2}.}

Regulatory requirements depend on the jurisdiction, legal entity, and products offered. In the United
States, CME cryptocurrency futures and options on futures fall under the oversight of the Commodity
Futures Trading Commission (CFTC). In the European Union, the Markets in
Crypto-Assets Regulation (MiCA) covers crypto-assets and related services within its scope, while
crypto-assets qualifying as financial instruments and cryptocurrency derivatives fall under the relevant
financial-markets framework, notably MiFID~II, rather than MiCA.\footnote{See Article~2(4)(a) of MiCA:
\url{https://www.esma.europa.eu/publications-and-data/interactive-single-rulebook/mica/article-2-scope}.}

\subsubsection{The Specific Case of Derivatives}

Contract specifications vary across cryptocurrency derivatives, notably in
their payoff structure, quotation conventions, margin requirements, and
settlement currency. A key distinction is between linear contracts\footnote{In this terminology, ``linear'' refers to the payoff and settlement convention,
not to a delta-one exposure: a linear option still has a nonlinear payoff as a function of the underlying
price.}, whose
payoffs are expressed and settled in a reference currency, such as USD or a stablecoin, and inverse
contracts, whose payoffs are settled in the underlying cryptocurrency. For the inverse contracts studied
here, strikes and futures prices are expressed in USD, while settlement takes place in BTC or ETH.
Margin arrangements depend on the venue and account type.\footnote{For an introduction to cryptocurrency
options, see \cite{alexanderCryptoQuantoInverse2021}.}

\begin{figure}[!h]
    \centering
    \includegraphics[width=0.7\linewidth]{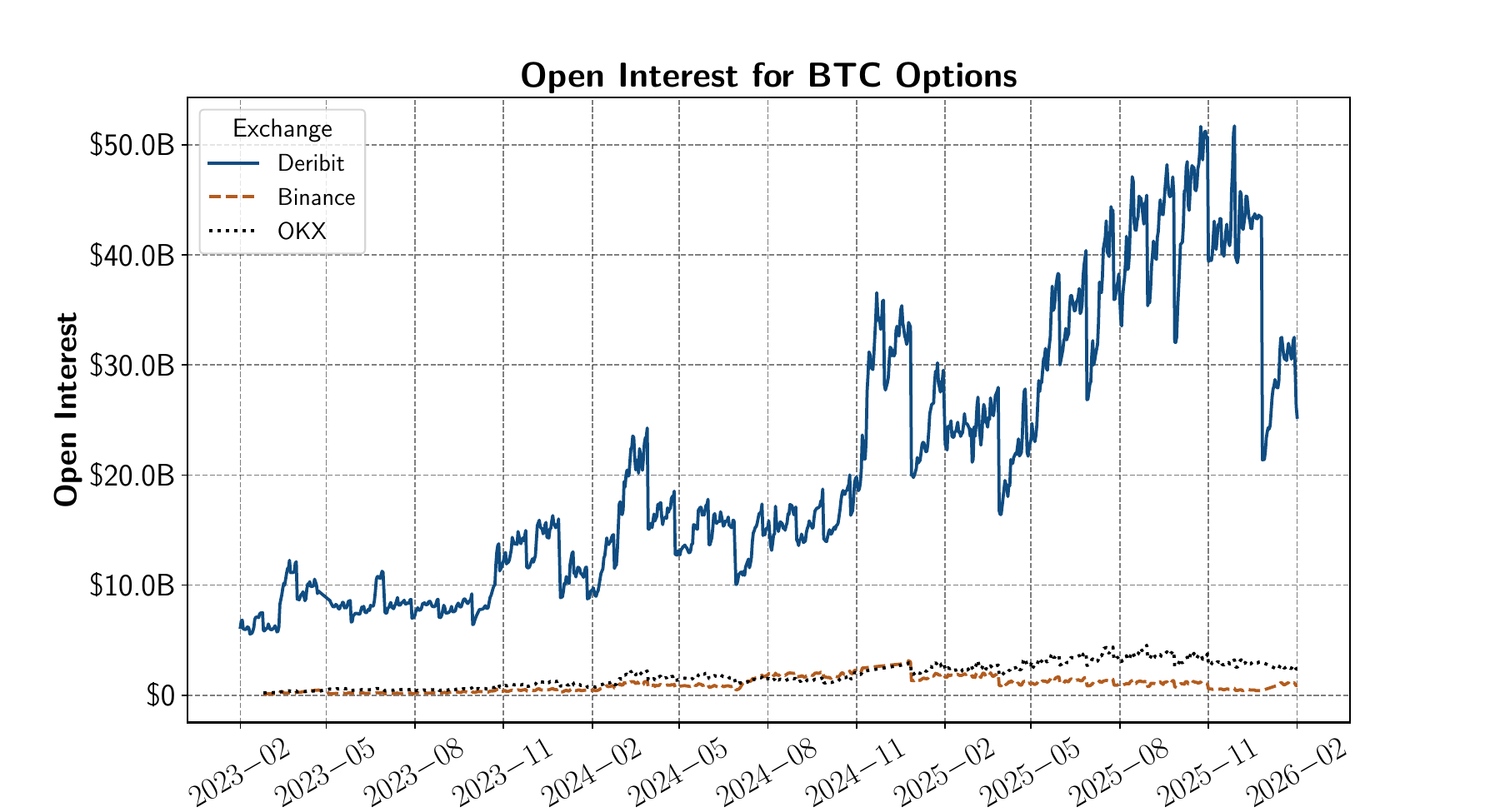}

    \par\vspace{0.3cm}

    \includegraphics[width=0.7\linewidth]{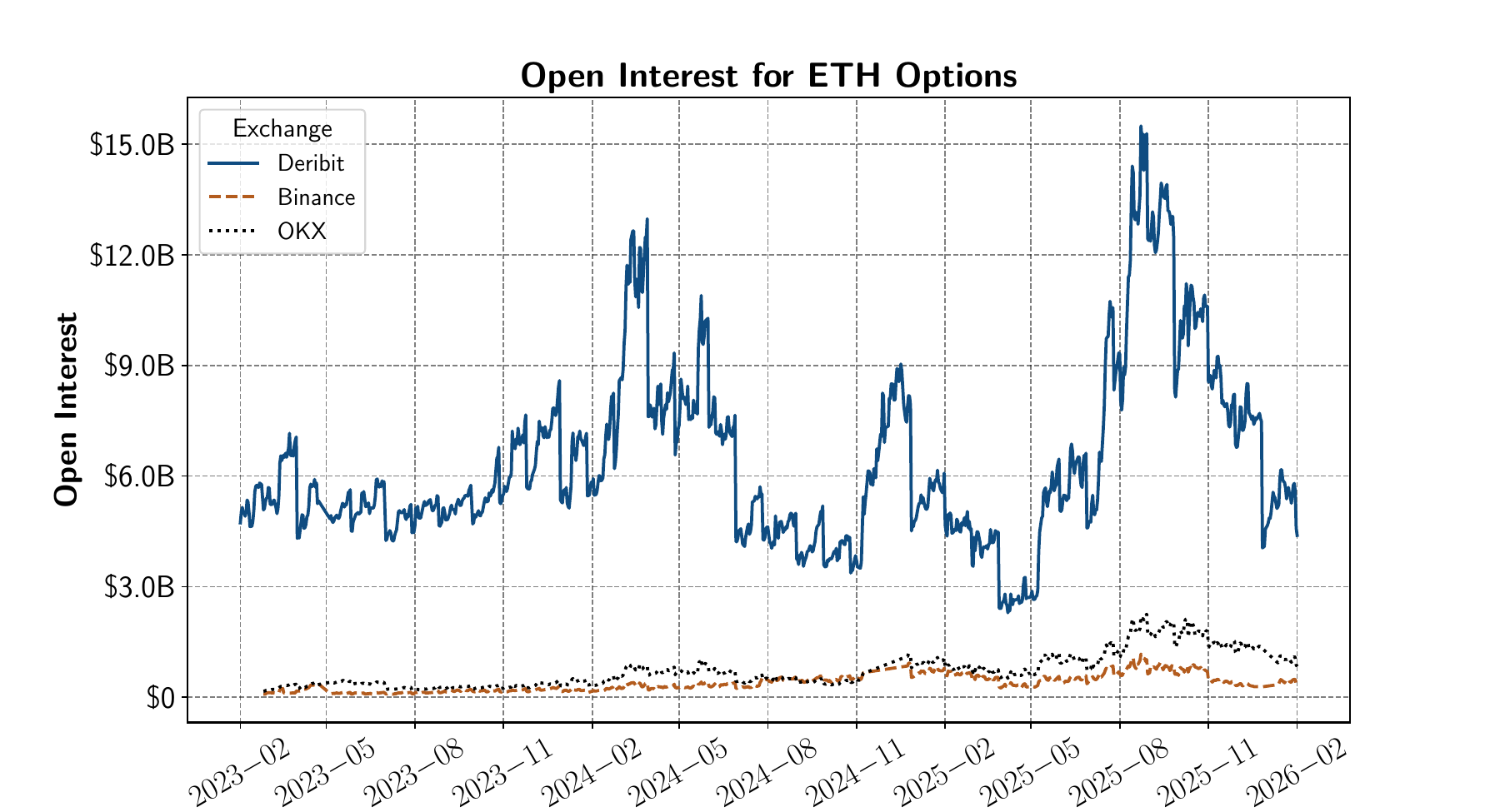}

    \caption{Daily aggregated open interest for options on BTC (top) and ETH (bottom) from February~1, 2023 to February~1, 2026. Source: \url{https://www.coinglass.com}.}
    \label{fig:OI_Option}
\end{figure}

Among crypto-native centralized exchanges, Deribit dominates the options market.
Figure~\ref{fig:OI_Option} illustrates this dominance by comparing open interest in BTC and ETH options
across three major venues.\footnote{As of January~1, 2026, the market capitalizations of BTC and ETH were approximately
\$1.77~trillion and \$362~billion, respectively. These figures provide a scale for comparison with the
notional open interest shown in Figure~\ref{fig:OI_Option}, although open interest measures outstanding
derivative exposure rather than capital invested. Source:
\url{https://coinmarketcap.com/historical/20260101/}.}

Deribit's dominant position and the availability of inverse options and futures across multiple maturities motivate our choice of venue.\footnote{Deribit was acquired by Coinbase in a transaction completed in August 2025. Because the synthetic zero-coupon bonds replicated in Section~\ref{sec:method} can be viewed as claims on the trading platform, such a change in ownership could affect the platform-specific credit risk embedded in the implied rates. This could explain the behavior of implied USD rates around this date.} Focusing on a single exchange also avoids combining quotes obtained under
different contract specifications and trading arrangements.

\subsection{Derivatives on Deribit}

We study two types of contracts traded on Deribit: fixed-maturity inverse futures\footnote{Our analysis excludes perpetual futures, which have no expiration date and
therefore do not provide the fixed horizons required by our approach.}
and European inverse options on BTC and ETH. All these contracts are settled
in the underlying cryptocurrency: BTC for Bitcoin contracts and ETH for Ether
contracts. For options, strike prices are expressed in USD, whereas option prices are
quoted in BTC for Bitcoin options and ETH for Ether options. Futures prices
are quoted in USD, but gains and losses are settled in BTC or ETH, depending
on the underlying asset. Deribit also offers USDC-settled linear derivative products, including BTC and ETH options
introduced on August~19, 2025.\footnote{See
\url{https://insights.deribit.com/education/usdc-settled-btc-eth-options-launch/}.}
Since these options are available only during the final months of our sample, we restrict our analysis
to inverse contracts to maintain consistent coverage over the study period.

Available option maturities include day, day+1, day+2, and day+3, as well as weekly, monthly, and quarterly
expirations. Contracts expire at 08:00~UTC on their maturity date. Weekly contracts expire on Fridays,
while monthly and quarterly contracts expire on the last Friday of the corresponding month or quarter.

Table~\ref{tab:deribit_quotes_moneyness_maturity} reports maturities, strike ranges, and moneyness ranges
for BTC inverse options listed on March~1, 2024. Moneyness is the strike divided by the underlying
price.\footnote{During that day, Bitcoin traded from slightly below \$61,000 to slightly
above \$63,000. For consistency, all moneyness percentages in the table are calculated using a common
reference price of \$62,000 and rounded to the nearest integer.}
Options at different strikes for each maturity provide the cross-sectional data used in our estimation procedure.

\begin{table}[h!]
    \centering
    \renewcommand{\arraystretch}{1.2}
    \begin{tabular}{
            |c|c|
            >{\centering\arraybackslash}p{2.3cm} >{\centering\arraybackslash}p{2.3cm}|
            >{\centering\arraybackslash}p{2.3cm} >{\centering\arraybackslash}p{2.3cm}|
        }
        \hline
        \multirow{2}{*}{\textbf{Expiry date}} &
        \multirow{2}{*}{\makecell{\textbf{Days} \\ \textbf{to expiry}}} &
        \multicolumn{2}{c|}{\textbf{Strike (USD)}} &
        \multicolumn{2}{c|}{\textbf{Moneyness range}} \\
        & & \textbf{Min} & \textbf{Max} & \textbf{Min} & \textbf{Max} \\
        \hline
        March 1, 2024 & 0 & 55,000 & 68,000 & 90\% & 111\% \\
        March 2, 2024 & 1 & 52,000 & 75,000 & 84\% & 121\% \\
        March 3, 2024 & 2 & 54,000 & 75,000 & 87\% & 121\% \\
        March 4, 2024 & 3 & 54,000 & 75,000 & 87\% & 121\% \\
        March 8, 2024 & 7 & 42,000 & 76,000 & 68\% & 123\% \\
        March 15, 2024 & 14 & 42,000 & 76,000 & 68\% & 123\% \\
        March 22, 2024 & 21 & 48,000 & 80,000 & 77\% & 129\% \\
        March 29, 2024 & 28 & 19,000 & 120,000 & 31\% & 194\% \\
        April 26, 2024 & 56 & 15,000 & 95,000 & 24\% & 153\% \\
        May 31, 2024 & 91 & 20,000 & 100,000 & 32\% & 161\% \\
        June 28, 2024 & 119 & 10,000 & 120,000 & 16\% & 194\% \\
        September 27, 2024 & 210 & 10,000 & 160,000 & 16\% & 258\% \\
        December 27, 2024 & 301 & 10,000 & 200,000 & 16\% & 323\% \\
        \hline
    \end{tabular}
    \caption{Maturities, strike ranges, and indicative moneyness ranges for BTC inverse options listed on Deribit as of March~1, 2024. Source: \url{https://coinmetrics.io}.}
    \label{tab:deribit_quotes_moneyness_maturity}
\end{table}

To illustrate quote sizes, trading activity, and bid-ask spreads, we use unfiltered data for BTC inverse
options from July~1 to July~7, 2024. This observation window allows us to examine options expiring
on September~27, 2024, December~27, 2024, and June~27, 2025, spanning horizons of approximately 3, 6,
and 12~months.

\begin{figure}[htbp]
    \centering

    \includegraphics[width=1\textwidth]{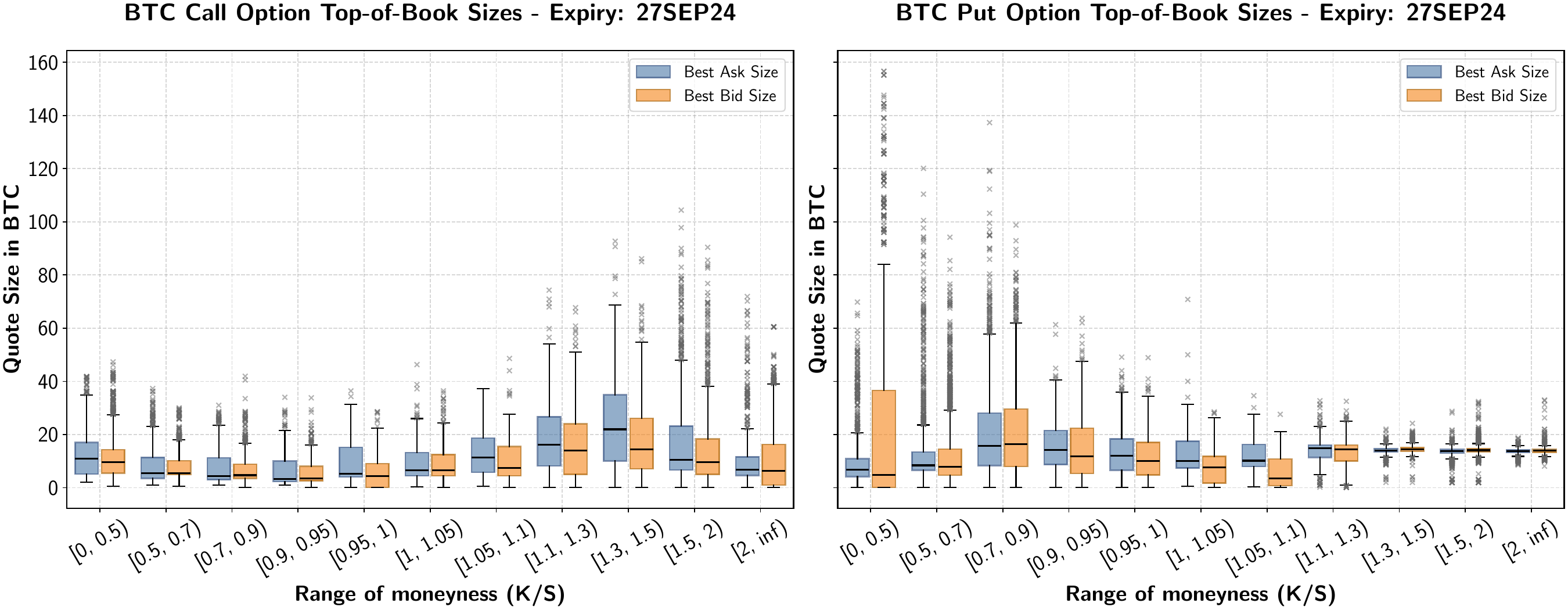}

    \vspace{0.5em} 

    \includegraphics[width=1\textwidth]{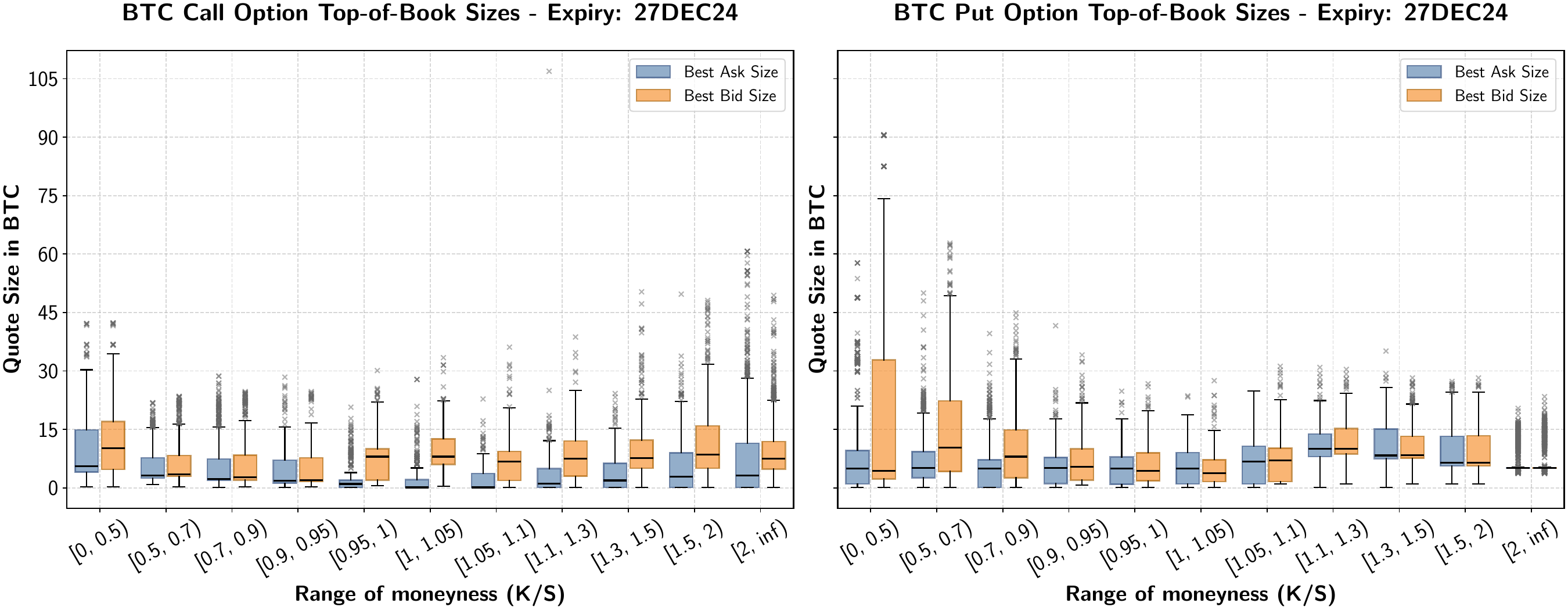}

    \vspace{0.5em} 

    \includegraphics[width=1\textwidth]{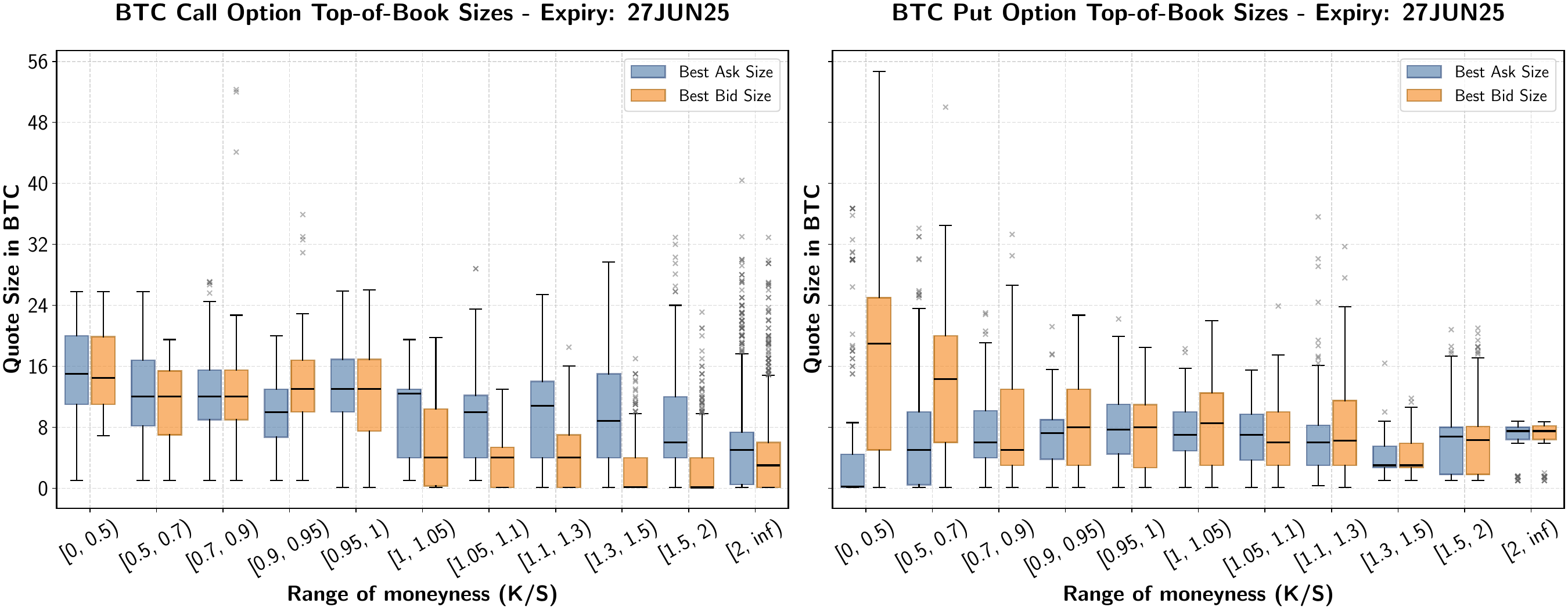}

    \caption{Boxplot of the top-of-book sizes of BTC inverse option contracts listed on Deribit for different maturities. Option expiries: September~27, 2024 (top), December~27, 2024 (middle), and June~27, 2025 (bottom). Unfiltered hourly data from July~1, 2024 to July~7, 2024. Source:~\mbox{\url{https://coinmetrics.io}}.}
    \label{fig:BTC_box_plot_best_quotes_size}
\end{figure}

Figure~\ref{fig:BTC_box_plot_best_quotes_size} shows the distribution of top-of-book sizes, i.e., quantities
available at the best bid and ask prices, grouped by moneyness. For the nearest quarterly expiration,
out-of-the-money (OTM) options exhibit larger quoted sizes, with some observations corresponding to
particularly large quantities at the best quotes. Top-of-book sizes
tend to decrease as time to expiration increases, indicating less displayed liquidity at the best quotes
for longer-dated options in this sample.

\begin{figure}[htbp]
    \centering

    \includegraphics[width=\textwidth]{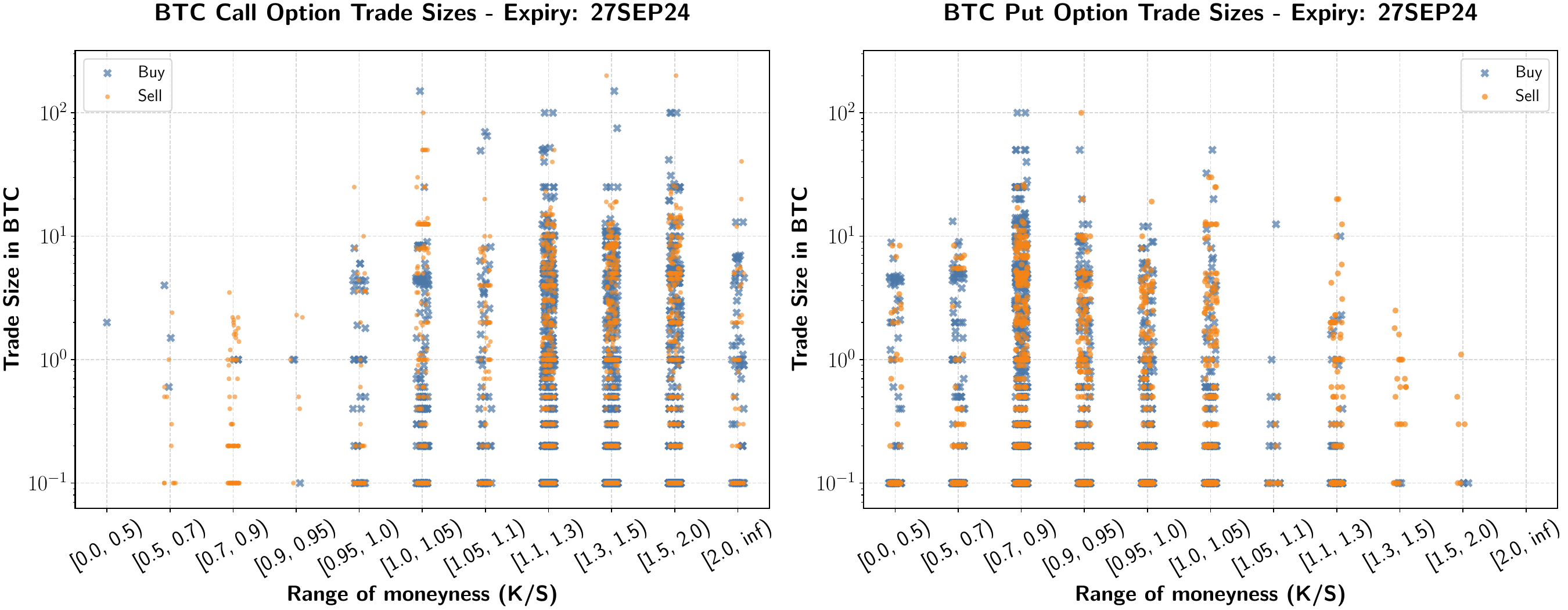}

    \vspace{0.5em} 

    \includegraphics[width=\textwidth]{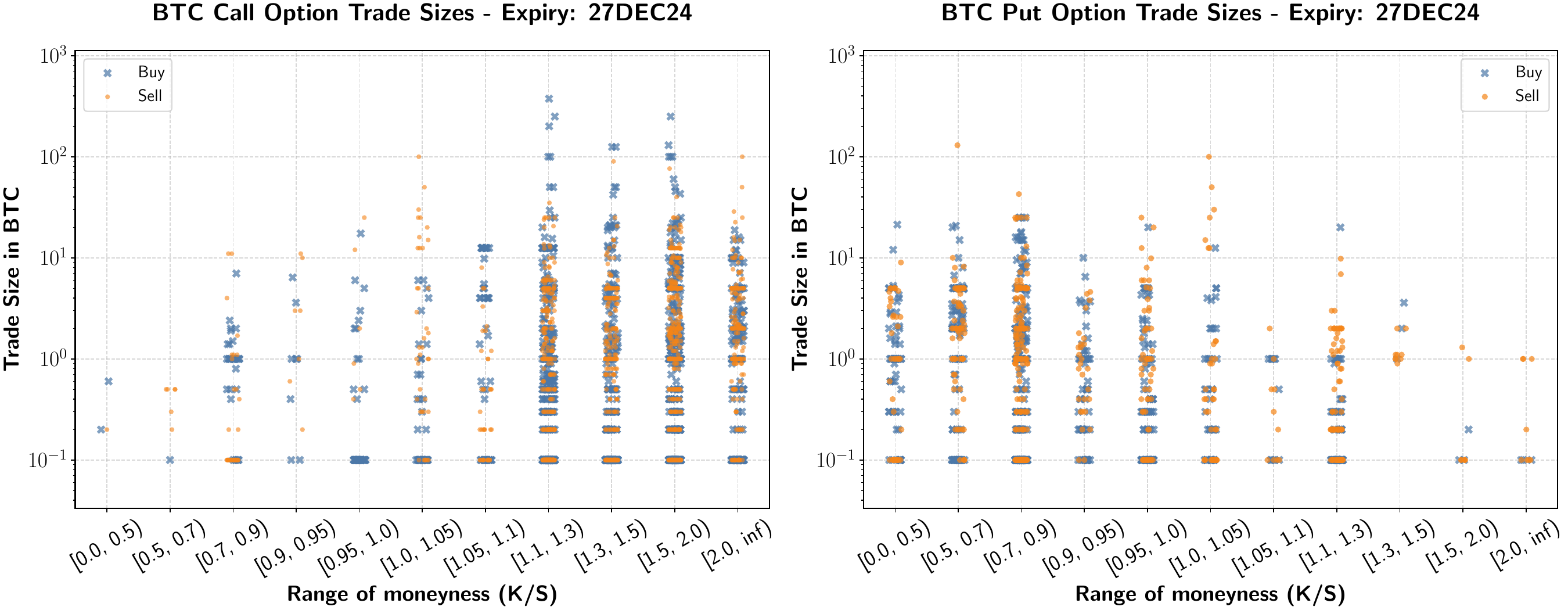}

    \vspace{0.5em} 

    \includegraphics[width=\textwidth]{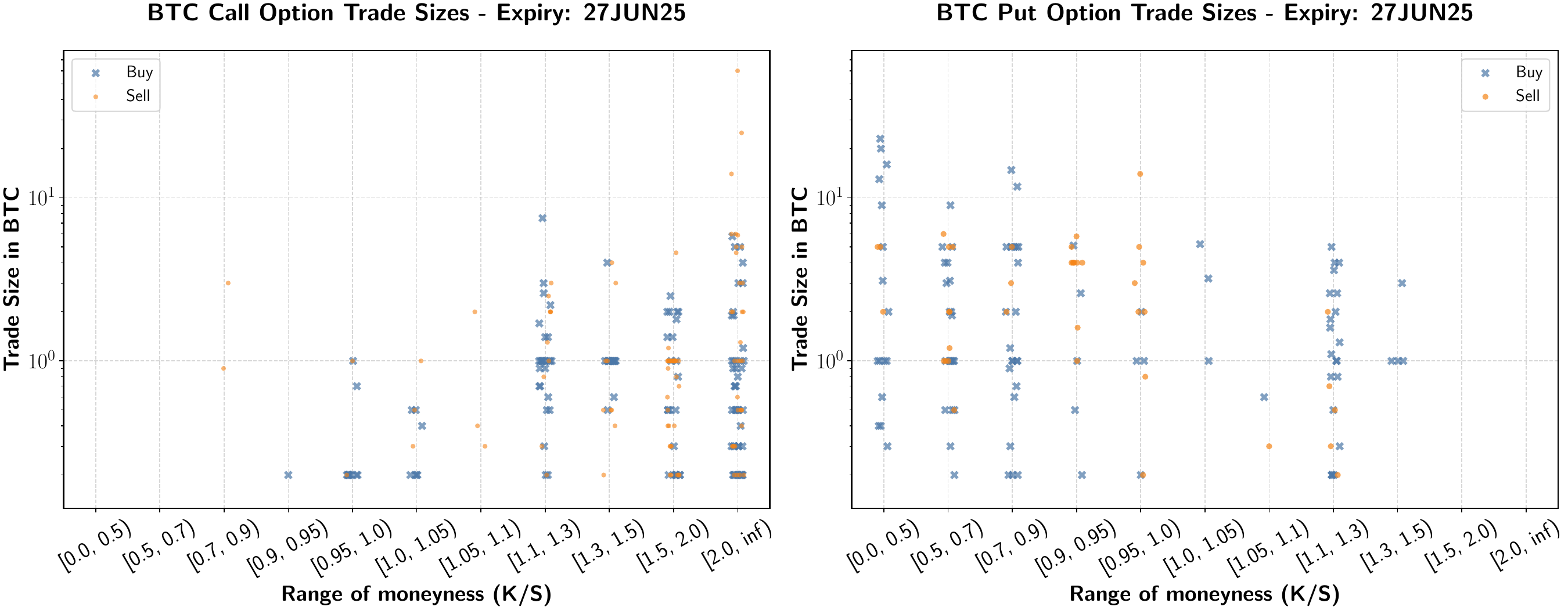}

    \caption{Distribution (logarithmic scale) of trade sizes, expressed in BTC, for BTC inverse option contracts listed on Deribit for different maturities. Option expiries: September~27, 2024 (top), December~27, 2024 (middle), and June~27, 2025 (bottom). Unfiltered data from July~1, 2024 to July~7, 2024. Source:~\mbox{\url{https://coinmetrics.io}}.}

    \label{fig:BTC_trade_amount}
\end{figure}

Figure~\ref{fig:BTC_trade_amount} presents executed trade sizes, expressed in BTC, for the same options
over the same week.\footnote{The minimum trade size for these options corresponds to a notional amount
of \(0.1\) BTC.} Trades are concentrated in OTM options for both calls and puts. Relatively few
transactions occur at the longest horizon, where top-of-book sizes are also smaller.
These observations indicate lower trading activity and less displayed liquidity
for the longest-dated options during this week.

\begin{figure}[htbp]
    \centering

    \includegraphics[width=0.7\textwidth]{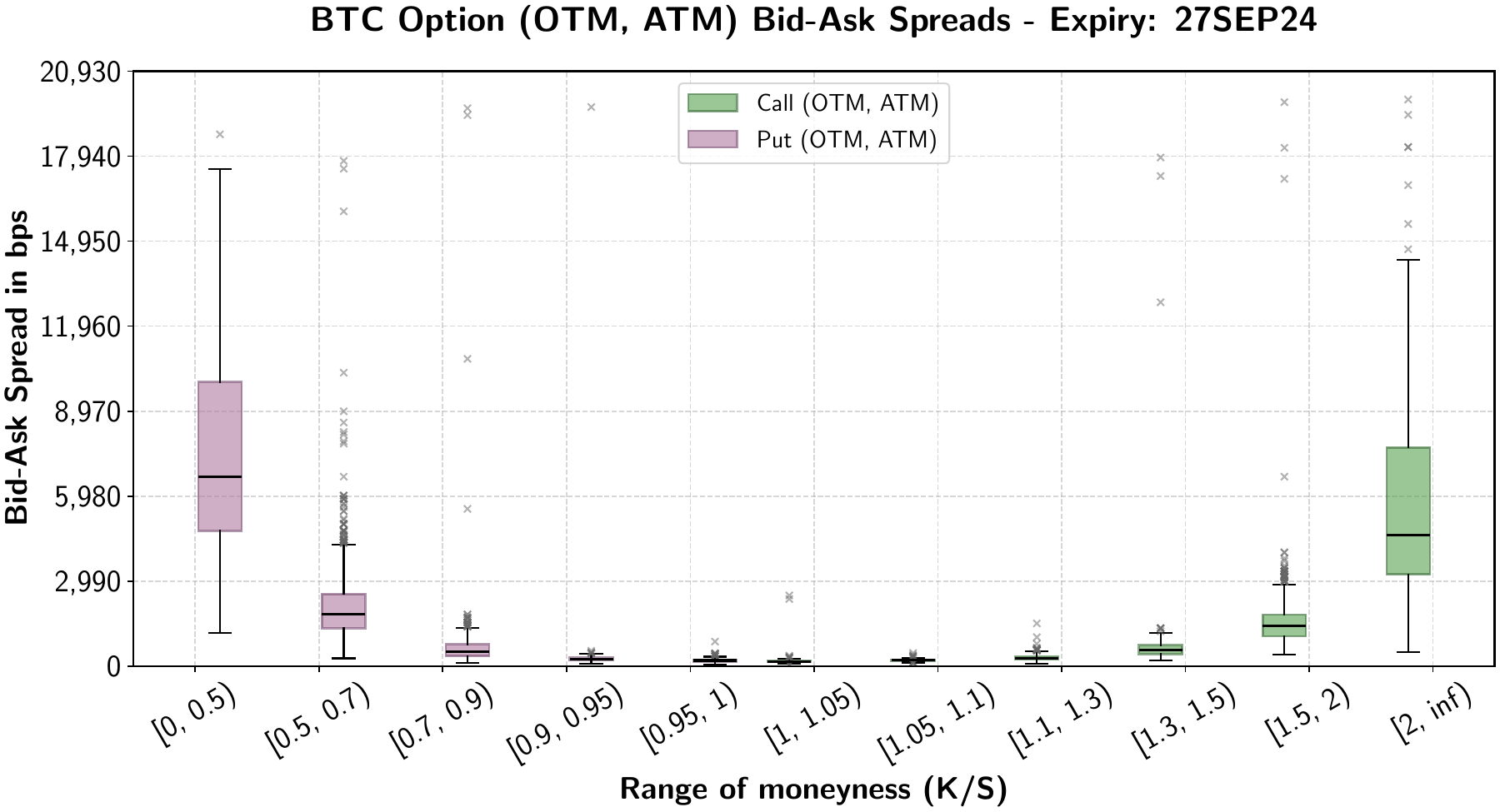}

    \vspace{0.5em} 

    \includegraphics[width=0.7\textwidth]{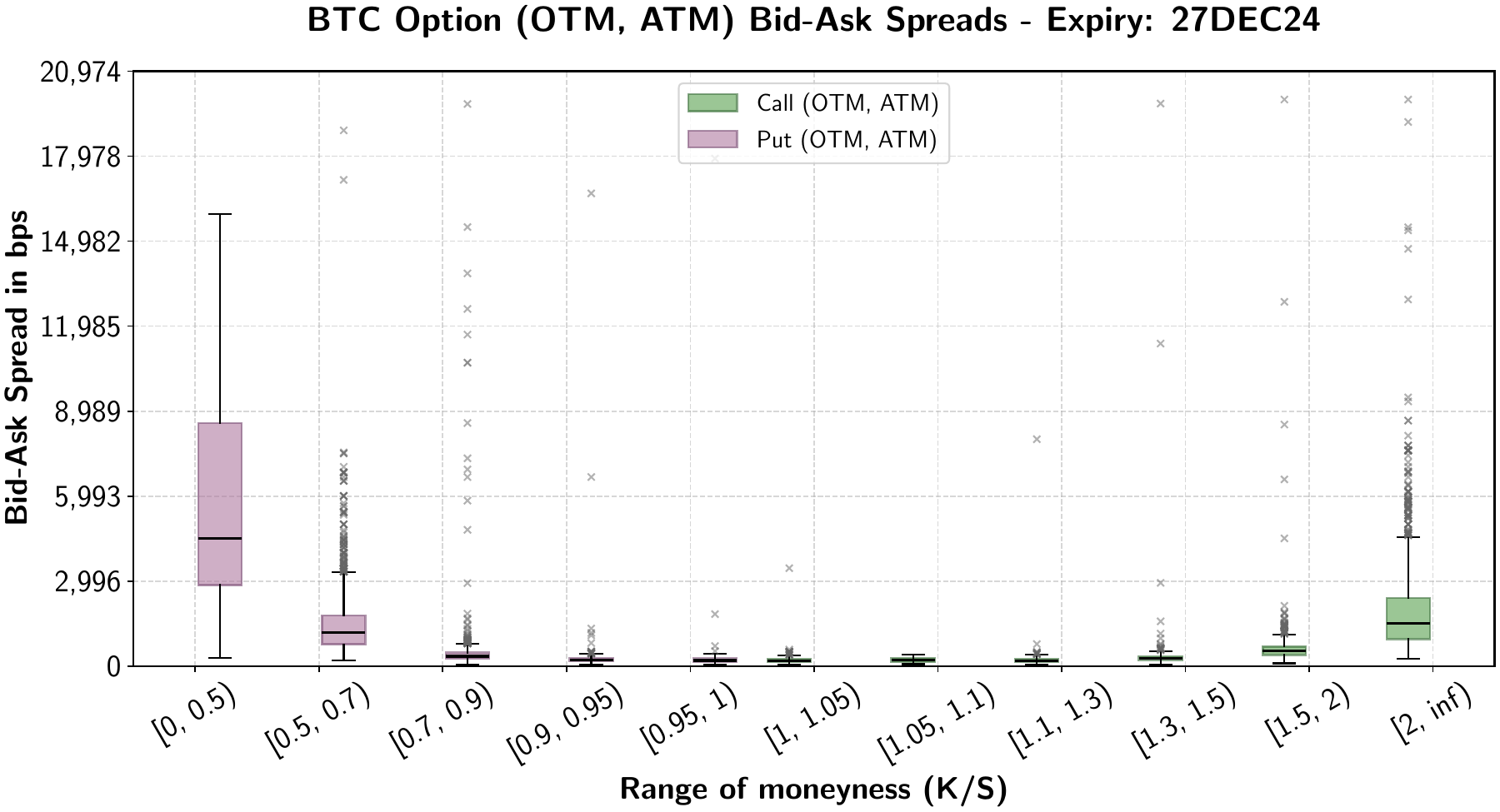}

    \vspace{0.5em} 

    \includegraphics[width=0.7\textwidth]{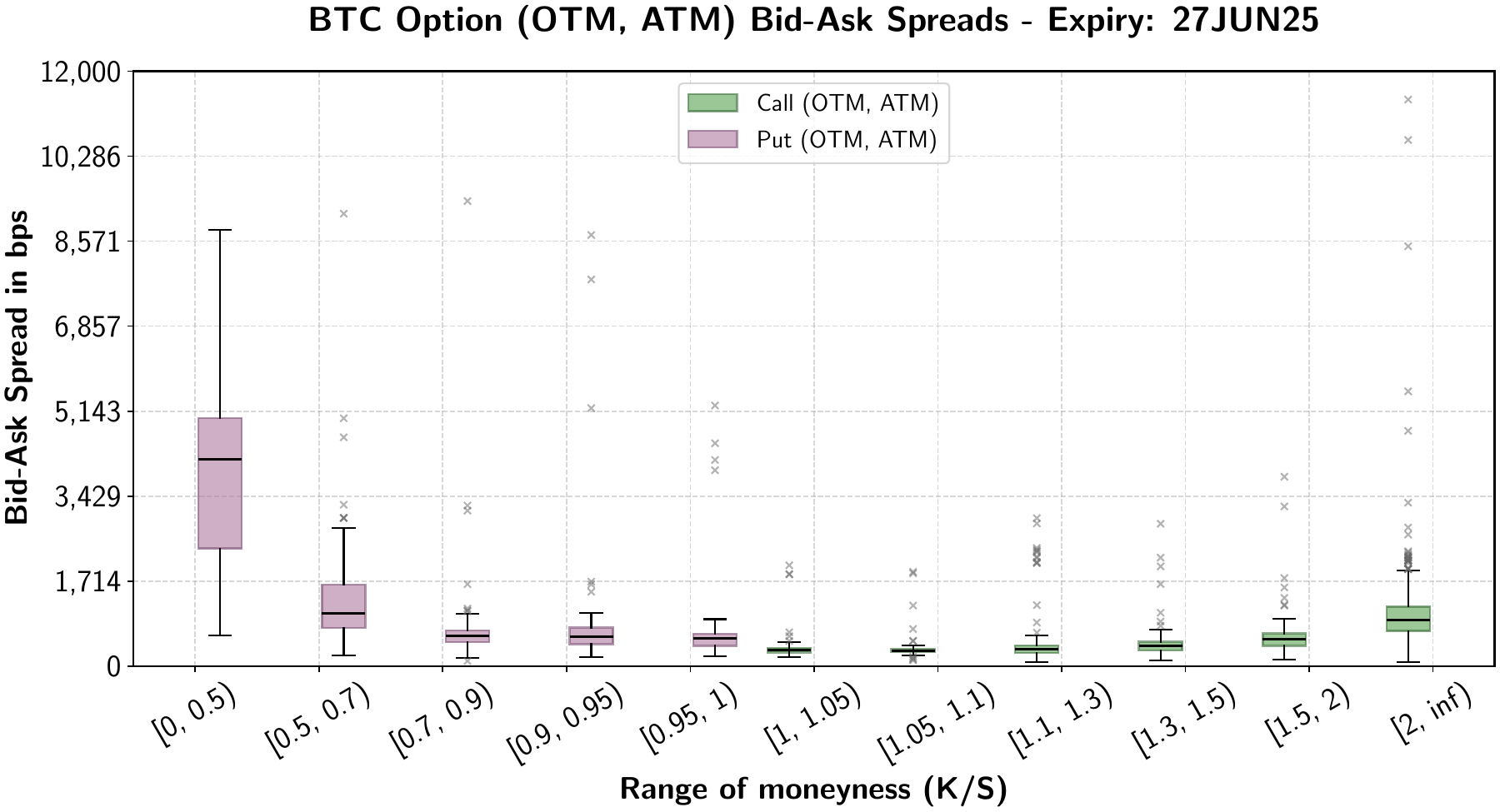}

    \caption{Boxplot of bid-ask spreads (in bps of the mid-price) for BTC inverse option contracts listed on Deribit for different maturities. Option expiries: September~27, 2024 (top), December~27, 2024 (middle), and June~27, 2025 (bottom). Only out-of-the-money and at-the-money options for both call and put contracts are included. Unfiltered hourly data from July~1, 2024 to July~7, 2024. Source:~\mbox{\url{https://coinmetrics.io}}.}
    \label{fig:btc_bid_ask_spread_quotes}
\end{figure}

Figure~\ref{fig:btc_bid_ask_spread_quotes} shows relative bid-ask spreads for at-the-money (ATM) and
OTM BTC options, expressed in basis points of the mid-price. These spreads are smallest near the
money and increase for far-OTM options. Part of this pattern may reflect the low prices of far-OTM
options, which can produce large relative spreads even when absolute spreads are small. Together,
these figures illustrate how quoted sizes, executed trade sizes, and relative spreads vary across
strikes and maturities. This heterogeneity motivates careful treatment of quote quality in the
filtering and estimation procedure introduced next.

\newcommand{\threemthmatsize}{./BTC_box_plot_best_quotes_size_27SEP24_2024-07-01_2024-07-07.pdf}
\newcommand{\sixmthmatsize}{./BTC_box_plot_best_quotes_size_27DEC24_2024-07-01_2024-07-07.pdf}
\newcommand{\oneyrmatsize}{./BTC_box_plot_best_quotes_size_27JUN25_2024-07-01_2024-07-07.pdf}

\newcommand{\threemthmattrade}{./BTC_traded_amount_27SEP24_2024-07-01_2024-07-07.pdf}
\newcommand{\sixmthmattrade}{./BTC_traded_amount_27DEC24_2024-07-01_2024-07-07.pdf}
\newcommand{\oneyrmattrade}{./BTC_traded_amount_27JUN25_2024-07-01_2024-07-07.pdf}

\newcommand{\threemthmatbidask}{./BTC_box_plot_bid_ask_spd_pct_27SEP24_2024-07-01_2024-07-07.pdf}
\newcommand{\sixmthmatbidask}{./BTC_box_plot_bid_ask_spd_pct_27DEC24_2024-07-01_2024-07-07.pdf}
\newcommand{\oneyrmatbidask}{./BTC_box_plot_bid_ask_spd_pct_27JUN25_2024-07-01_2024-07-07.pdf}

\section{Methodology}\label{sec:method}

\subsection{Back to First Principles}

Replication arguments play a central role in derivative pricing. In a frictionless
market without arbitrage, a derivative and a portfolio that replicates its cash
flows must have the same value. Replication can be dynamic, requiring positions
to be adjusted over time, or static, with positions established at inception
and held until maturity.

The classical Black-Scholes approach illustrates dynamic replication: an option
is replicated by continuously adjusting positions in the underlying asset and
a cash account. The interest rate governing this account enters the
self-financing condition and is therefore an input to the pricing model.
Static replication arguments, such as those underlying put-call parity,
instead establish relationships between the prices of traded instruments
by matching their payoffs at maturity, without specifying the dynamics of
the underlying assets. These relationships can be used either to express derivative prices in terms of discount factors or, conversely, to infer discount factors from observed derivative prices.

We adopt the latter perspective. In cryptocurrency markets, where liquid
fixed-income instruments remain scarce, we combine inverse European options
and fixed-maturity inverse futures to replicate fixed payments in either
the cryptocurrency or the reference currency. For these replication
arguments, we treat futures as contracts settled only at maturity, abstracting
from intermediate margin cash flows and their financing. We also assume
frictionless currency conversion and identify the terminal conversion price
with the common settlement price of the derivatives.
The initial costs of these portfolios define synthetic zero-coupon prices,
from which we can infer fixed interest rates at different horizons.

\subsection{From the Replication of Synthetic Zero-Coupon Bonds to Interest Rate Estimates}

\subsubsection{Revisiting the Put-Call Parity}

Put-call parity follows from static replication and relates European call and put prices to a forward
price and a discount factor. We apply the same reasoning to inverse contracts under the idealized assumptions stated above.

Consider a cryptocurrency denoted by \cryptosym{}, and let \(S_t\) denote its
price at time \(t\), expressed in the reference currency. Let \(F_t(T)\) denote
the futures price at time \(t\) for expiry date \(T\), in the reference currency.
We normalize all positions to correspond to one unit of \cryptosym{}:
the call and put represent the right to buy and sell that unit, respectively,
at a strike \(K\) expressed in the reference currency, while the futures
contract corresponds to exchanging that unit for \(F_t(T)\) at maturity,
abstracting from intermediate margin cash flows. The inverse contracts
settle the net values in \cryptosym{}. Under our replication assumptions, \(S_T\) is both the
price used to determine these payoffs and the exchange rate for currency
conversion at maturity.

At time \(t<T\), we construct the following portfolio:
\begin{itemize}
    \item buying an inverse call option on one unit of \cryptosym{} with
    expiry date \(T\) and strike \(K\);
    \item selling an inverse put option on one unit of \cryptosym{} with
    the same expiry date \(T\) and strike \(K\);
    \item entering into a short inverse futures position corresponding
    to one unit of \cryptosym{}, with expiry date \(T\).
\end{itemize}

At maturity, this portfolio delivers:
\[
    \underbrace{\left(1 - \frac{K}{S_T}\right)^+}_{\shortstack{\scriptsize Payoff of the\\\scriptsize inverse call}}
    \; - \; \underbrace{\left(\frac{K}{S_T} - 1\right)^+}_{\shortstack{\scriptsize Payoff of the\\\scriptsize inverse put}}
    \; - \;
    \underbrace{\left(1 - \frac{F_t(T)}{S_T}\right)}_{\shortstack{\scriptsize Payoff of the\\\scriptsize inverse futures contract}}
    \; = \quad
    \frac{F_t(T) - K}{S_T} \quad \text{\cryptosym{}}.
\]

If \(F_t(T) \ge K\), converting at time \(T\) this amount into the reference currency at price \(S_T\) gives the
fixed payment \(F_t(T)-K\). The portfolio therefore replicates \(F_t(T)-K\) units of a synthetic zero-coupon
claim paying one unit of the reference currency at \(T\).\footnote{These are synthetic claims constructed
from contracts traded on a particular platform. Their implied prices may reflect platform-specific
counterparty and custody risks, as well as collateral requirements and other trading frictions;
they need not coincide with conventional risk-free discount factors.} If \(F_t(T)<K\), the portfolio instead generates a liability equivalent to a short position in
\(K-F_t(T)\) such claims.\footnote{\label{locked_capital}Implementing these portfolios requires margin,
regardless of the sign of the net terminal payoff. The replication relations abstract from the cost
of funding this margin and from possible liquidation before maturity.}

Let \(C_t(K,T)\) and \(P_t(K,T)\) denote the prices, in \cryptosym{}, of the
inverse call and put options with strike \(K\) and expiry date \(T\).
Since the futures position has zero initial value under our assumptions,
the portfolio's initial cost is \(C_t(K,T)-P_t(K,T)\) units of \cryptosym{}.
Denoting by \(\text{ZC}_t^{\text{ref}}(T)\) the price, in the reference currency,
of a zero-coupon bond paying one unit of that currency at \(T\), the law of
one price gives
\begin{equation}
    C_t(K,T)-P_t(K,T)
    = \frac{F_t(T)-K}{S_t}\,\text{ZC}_t^{\text{ref}}(T),
    \label{eq:zc_ref_from_cpp}
\end{equation}
where division by \(S_t\) converts the zero-coupon portfolio's value into
units of \cryptosym{}.

\subsubsection{Inverse Futures and Currency Conversion of Synthetic Zero-Coupon Bonds}

We can also replicate a zero-coupon bond paying one unit of \cryptosym{}
at expiry date \(T\) by combining synthetic zero-coupon bonds in the reference
currency with an inverse futures position. Using the same normalization
as above, consider the following portfolio:
\begin{itemize}
    \item buying \(F_t(T)\) zero-coupon bonds, each paying one unit of the
    reference currency at \(T\);
    \item entering into a long inverse futures position corresponding
    to one unit of \cryptosym{}, with expiry date \(T\) and futures price
    \(F_t(T)\).
\end{itemize}

At maturity, converting the proceeds of the zero-coupon bonds into
\cryptosym{} gives the following total payoff:
\[
    \underbrace{\frac{F_t(T)}{S_T}}_{\shortstack{\scriptsize Payoff of the zero-coupon bonds\\
    \scriptsize converted into \cryptosym{}}}
    +
    \underbrace{\left(1-\frac{F_t(T)}{S_T}\right)}_{\shortstack{\scriptsize Payoff of the long\\
    \scriptsize inverse futures position}}
    =\quad  1 \quad \cryptosym{}.
\]

Denote by \(\text{ZC}_t^{\text{\cryptosym{}}}(T)\) the price, in \cryptosym{},
of a zero-coupon bond paying one unit of \cryptosym{} at time \(T\).
Since the futures position has zero initial value under our assumptions,
the law of one price gives
\begin{equation}
    \text{ZC}_t^{\text{\cryptosym{}}}(T)
    = \frac{F_t(T)}{S_t}\,\text{ZC}_t^{\text{ref}}(T),
    \label{eq:zc_crypto_from_inverse_futures}
\end{equation}
where division by \(S_t\) converts the initial cost of the reference-currency
bonds into units of \cryptosym{}. Combining this relation with
Eq.~\eqref{eq:zc_ref_from_cpp} gives
\[
    C_t(K,T)-P_t(K,T)
    = \text{ZC}_t^{\text{\cryptosym{}}}(T)
    - \frac{K}{S_t}\,\text{ZC}_t^{\text{ref}}(T),
\]
which provides the cross-strike relation used in our estimation.

\subsubsection{General Principles of Estimation}

For an expiry date \(T\) and a given time \(t<T\), Eq.~\eqref{eq:zc_ref_from_cpp} implies that the ratio
\[
    \frac{C_t(K,T)-P_t(K,T)}{(F_t(T)-K)/S_t}
\]
should be constant across strikes \(K\ne F_t(T)\). In observed quotes, this ratio varies,
and small pricing errors can be greatly amplified when the strike is close to the futures price.
Averaging such ratios is therefore potentially unstable. Regression avoids division by a small
strike-specific denominator, but remains sensitive to noisy quotes. We combine quote filtering with
joint estimation of the two discount factors, using options across strikes and the futures price
as complementary sources of information.

In our empirical implementation, \(S_t\) is the Deribit index price, which aggregates prices from multiple
spot exchanges using safeguards against outlying observations. This avoids relying on the order book
of a single spot venue. It does not eliminate residual index noise or synchronization errors between
the index and derivative quotes. Option and futures prices used in the estimation are mid-prices,
with calls and puts paired by strike and expiry date.

More precisely, we estimate the two synthetic zero-coupon prices,
\(\text{ZC}_t^{\text{\cryptosym{}}}(T)\) and \(\text{ZC}_t^{\text{ref}}(T)\), by balancing two types of
pricing discrepancies:
\begin{enumerate}
    \item the distance between \(C_t(K,T) - P_t(K,T)\) and \(\text{ZC}_t^{\text{\cryptosym{}}}(T) - \frac{K}{S_t} \, \text{ZC}_t^{\text{ref}}(T)\) across all available strikes~\(K\), and
    \item the distance between \(\frac{F_t(T)}{S_t}\,\text{ZC}_t^{\text{ref}}(T)\)
    and \(\text{ZC}_t^{\text{\cryptosym{}}}(T)\).
\end{enumerate}

To obtain closed-form estimators, we minimize a weighted sum of these squared
discrepancies (see Eq.~\eqref{equation:regression} below).

The futures relation enters as a penalty rather than an exact constraint,
allowing the fit to accommodate discrepancies between option and futures quotes.

For each cryptocurrency and expiry date \(T\), positive estimated discount factors yield continuously
compounded interest rates over the horizon \(T-t\), expressed in years. We denote the rates in the
reference currency and the cryptocurrency by \(r_{t,T-t}^{\text{ref}}\) and
\(r_{t,T-t}^{\text{\cryptosym{}}}\), respectively.\footnote{An extension could combine observations across
underlyings to estimate a common reference-currency discount factor, provided that this restriction
is appropriate for the markets considered.}

\subsection{Estimation Procedure}
\label{est_proc}

\subsubsection{Objective Function and Closed-Form Estimators}

For an expiry date \(T\) and a given time \(t<T\), let $(K^i)_{1\leq i\leq n_{t,T}}$ denote the strikes available in the option dataset. Following the relations established above, we define the objective function
\begin{equation}
\label{equation:regression}
f_{t,T}(\alpha,\beta)
:=\sum_{i=1}^{n_{t,T}}\omega_{t,T}^i\bigl(y_{t,T}^i-\alpha+\beta m_t^i\bigr)^2
+\lambda_{t,T}\left(\beta\frac{F_{t,T}}{S_t}-\alpha\right)^2.
\end{equation}
We estimate the synthetic discount factors in the cryptocurrency and the reference currency by minimizing $f_{t,T}$ over $(\alpha,\beta)\in\mathbb{R}_+^2$.

Here, $y_{t,T}^i:=C_t(K^i,T)-P_t(K^i,T)$ is the call--put mid-price difference, $m_t^i:=K^i/S_t$ is the corresponding moneyness, and $F_{t,T}:=F_t(T)$. The variables $\alpha$ and $\beta$ represent the synthetic zero-coupon prices in the cryptocurrency and the reference currency, respectively. The option weights satisfy $\omega_{t,T}^i\geq 0$, while $\lambda_{t,T}\geq 0$ controls the contribution of the futures quote.

When $\lambda_{t,T}=0$, both discount factors are inferred from the cross-strike relation between call and put prices. The futures term penalizes discrepancies between $\alpha/\beta$ and $F_{t,T}/S_t$, thus helping to determine the relative levels of the two discount factors and hence the difference between the corresponding implied interest rates. We use a penalty rather than an exact constraint to allow for discrepancies between option and futures quotes.

The objective function is quadratic in $(\alpha,\beta)$, with gradient
\[
\nabla f_{t,T}(\alpha,\beta)=2\begin{pmatrix}
\displaystyle
\alpha\left(\lambda_{t,T}+\sum_{i=1}^{n_{t,T}}\omega_{t,T}^i\right)-\beta\left(\lambda_{t,T}\frac{F_{t,T}}{S_t}+\sum_{i=1}^{n_{t,T}}\omega_{t,T}^i m_t^i\right)-\sum_{i=1}^{n_{t,T}}\omega_{t,T}^i y_{t,T}^i\\[1.2em]
\displaystyle
-\alpha\left(\lambda_{t,T}\frac{F_{t,T}}{S_t}+\sum_{i=1}^{n_{t,T}}\omega_{t,T}^i m_t^i\right)+\beta\left(\lambda_{t,T}\frac{F_{t,T}^2}{S_t^2}+\sum_{i=1}^{n_{t,T}}\omega_{t,T}^i(m_t^i)^2\right)+\sum_{i=1}^{n_{t,T}}\omega_{t,T}^i m_t^i y_{t,T}^i
\end{pmatrix}.
\]

Setting the gradient to zero yields a linear system of two equations. Provided its determinant $d$ is strictly positive, this system admits a unique solution:
\[
\begin{aligned}
\alpha_{t,T}^{\star}
&=\frac{1}{d}\sum_{i=1}^{n_{t,T}}\omega_{t,T}^i y_{t,T}^i
\left[
\lambda_{t,T}\frac{F_{t,T}^2}{S_t^2}
+\sum_{j=1}^{n_{t,T}}\omega_{t,T}^j(m_t^j)^2
-m_t^i\left(
\lambda_{t,T}\frac{F_{t,T}}{S_t}
+\sum_{j=1}^{n_{t,T}}\omega_{t,T}^j m_t^j
\right)
\right],\\[0.8em]
\beta_{t,T}^{\star}
&=\frac{1}{d}\sum_{i=1}^{n_{t,T}}\omega_{t,T}^i y_{t,T}^i
\left[
\lambda_{t,T}\frac{F_{t,T}}{S_t}
+\sum_{j=1}^{n_{t,T}}\omega_{t,T}^j m_t^j
-m_t^i\left(
\lambda_{t,T}
+\sum_{j=1}^{n_{t,T}}\omega_{t,T}^j
\right)
\right],
\end{aligned}
\]
where
\[
d=
\left(\lambda_{t,T}+\sum_{i=1}^{n_{t,T}}\omega_{t,T}^i\right)
\left(\lambda_{t,T}\frac{F_{t,T}^2}{S_t^2}+\sum_{i=1}^{n_{t,T}}\omega_{t,T}^i(m_t^i)^2\right)
-\left(\lambda_{t,T}\frac{F_{t,T}}{S_t}+\sum_{i=1}^{n_{t,T}}\omega_{t,T}^i m_t^i\right)^2.
\]

In particular, $d>0$ whenever at least two distinct strikes have strictly positive weights. In our empirical application, the  choice of the weights $\omega_{t,T}^i$, described below, consistently yielded positive unconstrained solutions $\alpha_{t,T}^{\star}$ and $\beta_{t,T}^{\star}$. Consequently, the nonnegativity constraints were never binding.

The estimators of the synthetic zero-coupon prices are therefore
\[
\widehat{\mathrm{ZC}}_t^{\text{\cryptosym{}}}(T):=\alpha_{t,T}^{\star},
\qquad
\widehat{\mathrm{ZC}}_t^{\mathrm{ref}}(T):=\beta_{t,T}^{\star}.
\]

The corresponding continuously compounded interest rate estimates are
\[
\widehat{r}_{t,T-t}^{\text{\cryptosym{}}}:=-\frac{1}{T-t}\log\!\left(\widehat{\mathrm{ZC}}_t^{\text{\cryptosym{}}}(T)\right),
\qquad
\widehat{r}_{t,T-t}^{\mathrm{ref}}:=-\frac{1}{T-t}\log\!\left(\widehat{\mathrm{ZC}}_t^{\mathrm{ref}}(T)\right),
\]
where $T-t$ is expressed in years.

\subsubsection{Quote Weighting and Filtering}

It remains to specify the weights in Eq.~\eqref{equation:regression}. We pool hourly observations within each day to obtain daily estimates, assuming that the discount factors vary sufficiently slowly to be treated as approximately constant over that interval. We experimented with continuous option weights, notably weights inversely proportional to the combined call and put bid-ask spread. In the empirical analysis, however, we favor binary weights that exclude observations identified as unreliable by Random Sample Consensus (RANSAC) -- see \cite{fischlerRandomSampleConsensus1981}.

Importantly, RANSAC is not fitted to the pricing relation in Eq.~\eqref{equation:regression}. Instead, it is applied to two bid-ask relationships constructed from opposite call-put positions. For a given strike and expiry, define the quoted costs of buying the call and selling the put, and of taking the opposite position, respectively, by
\[
x_{t,T}^i:=C_t^{\mathrm{ask}}(K^i,T)-P_t^{\mathrm{bid}}(K^i,T),
\qquad
z_{t,T}^i:=P_t^{\mathrm{ask}}(K^i,T)-C_t^{\mathrm{bid}}(K^i,T).
\]
For each pair $(t,T)$, RANSAC is applied separately in the two regression orientations
\begin{equation}
\label{eq:spread_relation}
\begin{aligned}
z_{t,T}^i&=\zeta_{t,T}^{(1)}+\xi_{t,T}^{(1)}x_{t,T}^i+\eta_{t,T}^{i,(1)},\\
x_{t,T}^i&=\zeta_{t,T}^{(2)}+\xi_{t,T}^{(2)}z_{t,T}^i+\eta_{t,T}^{i,(2)}.
\end{aligned}
\end{equation}

The rationale is that the sum of these two quoted costs equals the combined call and put bid-ask spread:
\[
x_{t,T}^i+z_{t,T}^i
=\bigl(C_t^{\mathrm{ask}}(K^i,T)-C_t^{\mathrm{bid}}(K^i,T)\bigr)
+\bigl(P_t^{\mathrm{ask}}(K^i,T)-P_t^{\mathrm{bid}}(K^i,T)\bigr).
\]
It is also the quoted cost of entering and immediately reversing either call-put position, excluding trading fees. If this combined spread is approximately constant across a group of strikes, the observations in Eq.~\eqref{eq:spread_relation} lie near a line with slope $-1$ and intercept equal to the common spread. We exploit this regularity to detect outliers, while allowing RANSAC to estimate both the slope and the intercept rather than imposing a slope of $-1$. This is a quote-quality filter based on the cross-strike pattern of combined bid-ask spreads, not a direct test of put-call parity at mid-prices.

For each orientation, RANSAC repeatedly fits candidate lines to randomly selected subsets and identifies the observations whose squared residuals do not exceed a prescribed threshold. It selects the candidate with the largest consensus set and refits the regression on its inliers. Let $\widehat{\eta}_{t,T}^{i,(j)}$ denote the residual associated with the selected candidate in orientation $j\in\{1,2\}$, before the final refit. The corresponding consensus sets are
\[
\mathcal{I}_{t,T}^{(j)}:=\left\{i\in\{1,\ldots,n_{t,T}\}:\bigl(\widehat{\eta}_{t,T}^{i,(j)}\bigr)^2\leq M_{t,T}\right\},\qquad j\in\{1,2\},
\]
where $M_{t,T}\geq 0$ is the squared-residual threshold. Although the two equations in Eq.~\eqref{eq:spread_relation} describe the same bid-ask information, their fitted lines and consensus sets need not coincide: ordinary least squares residuals depend on which variable is treated as the dependent variable. We therefore retain only observations classified as inliers in both orientations. In other words, we set
\[
\omega_{t,T}^i:=\mathbf{1}_{\{i\in\mathcal{I}_{t,T}\}} \quad \text{where} \quad \mathcal{I}_{t,T}:=\mathcal{I}_{t,T}^{(1)}\cap\mathcal{I}_{t,T}^{(2)}.
\]
This selection retains equal weights for accepted observations and assigns zero weight to the others. It requires neither a parametric option-pricing model nor preliminary estimates of the discount factors. Its usefulness nevertheless depends on the empirical regularity of the combined spreads and on retaining enough observations with sufficient moneyness variation for identification.

Finally, we specify the futures weight $\lambda_{t,T}$. Futures are generally more liquid than options, suggesting that futures quotes can provide a useful anchor when the option-based fit is noisy. We adopt a two-step, data-dependent rule. First, we solve Eq.~\eqref{equation:regression} using the RANSAC-selected observations with $\lambda_{t,T}=0$, obtaining preliminary estimates $\widehat{\alpha}_{t,T}^{(0)}$ and $\widehat{\beta}_{t,T}^{(0)}$. We then compute their option residuals,
\[
\varepsilon_{t,T}^i:=y_{t,T}^i-\widehat{\alpha}_{t,T}^{(0)}+\widehat{\beta}_{t,T}^{(0)}m_t^i,\qquad i\in\mathcal{I}_{t,T},
\]
and set
\[
\lambda_{t,T}:=\frac{S_t^2}{\bigl(\psi_{t,T}^F\bigr)^2}\sum_{i\in\mathcal{I}_{t,T}}\bigl(\varepsilon_{t,T}^i\bigr)^2,
\qquad
\psi_{t,T}^F:=F_{t,T}^{\mathrm{ask}}-F_{t,T}^{\mathrm{bid}}.
\]
This heuristic increases the influence of the futures relation when the option-only residual sum of squares is large or the relative futures bid-ask spread is small. Conversely, a wider futures spread reduces its influence. The final discount factors and implied rates are computed by solving Eq.~\eqref{equation:regression} with this value of $\lambda_{t,T}$.

\section{Empirical Analysis}
\label{sec:data}
As previously discussed, Deribit dominates the cryptocurrency options market, which justifies the use of their
market data in the following analysis to derive interest rates using the approach described above. While
competing crypto-native exchanges and fully decentralized applications offering derivatives have emerged,
their options markets currently lack the deep, consistent liquidity required to extract stable interest rate
term structures. In such illiquid environments, sparse order books and excessively wide bid-ask spreads would
lead to instability of the estimators obtained.

\subsection{Market Data}

\begin{figure}[H]
    \centering
    \includegraphics[width=0.83\textwidth]{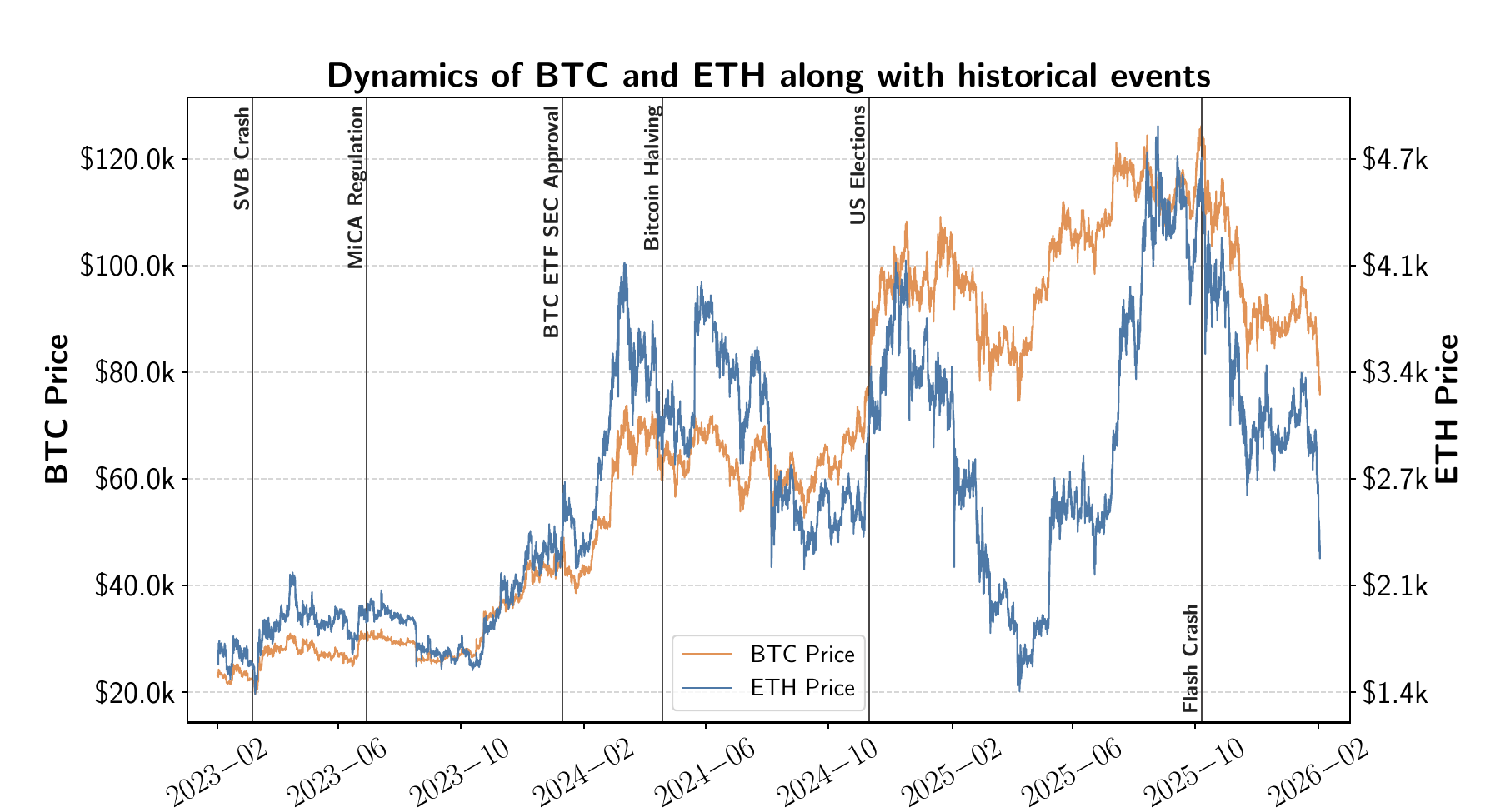}
    \caption{BTC and ETH index prices from February~1, 2023 to February~1, 2026, with selected historical events. Source: CoinMetrics (\url{https://coinmetrics.io}).}
    \label{fig:BTCTimeSeries}
\end{figure}

The dataset consists of Deribit's options and futures top-of-book quotes, together with BTC and ETH index prices, sourced from the data provider CoinMetrics. Observations are recorded at hourly intervals from February~1, 2023 to February~1, 2026. This period covers several significant market events, including the collapse of Silicon Valley Bank in March~2023, the adoption of the European Markets in Crypto-Assets Regulation (MiCA) in June~2023, the Bitcoin halving in April~2024, the US presidential election and subsequent cryptocurrency rally in late 2024, and the flash crash of October~2025 (see Figure~\ref{fig:BTCTimeSeries}). For options and futures, mid-prices are computed as the averages of the best bid and ask prices.

We apply two preliminary filters to the options data. First, we exclude options with maturity less than one month, since short maturities tend to produce particularly noisy interest rate estimates. Second, we discard quotes with excessively wide bid-ask spreads relative to their mid-prices. The remaining observations are then subjected to the RANSAC filtering procedure, whose empirical results are examined below.

\subsection{The RANSAC Step}

We now examine the results of the RANSAC filtering procedure to illustrate its role in identifying and removing unreliable option quotes. We focus on BTC options; analogous results are obtained for ETH.

\begin{figure}[htbp]
    \centering
    \begin{subfigure}[b]{0.32\textwidth}
        \centering
        \includegraphics[width=\textwidth]{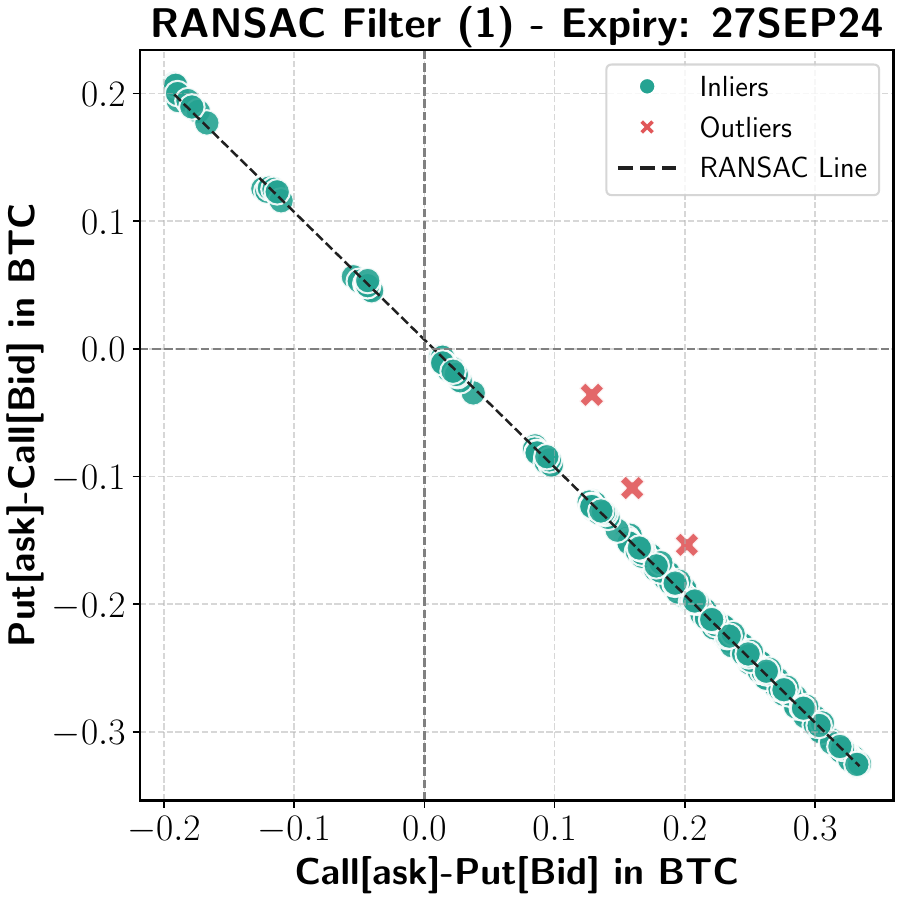}
    \end{subfigure}
    \begin{subfigure}[b]{0.32\textwidth}
        \centering
        \includegraphics[width=\textwidth]{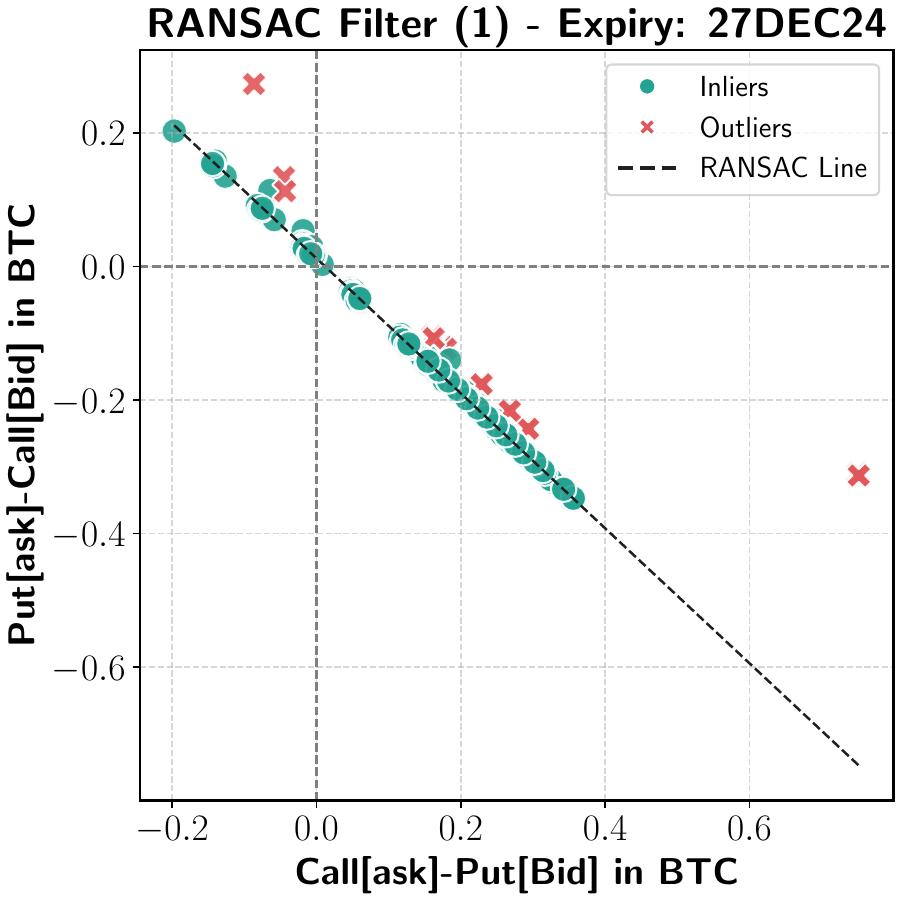}
    \end{subfigure}
    \begin{subfigure}[b]{0.32\textwidth}
        \centering
        \includegraphics[width=\textwidth]{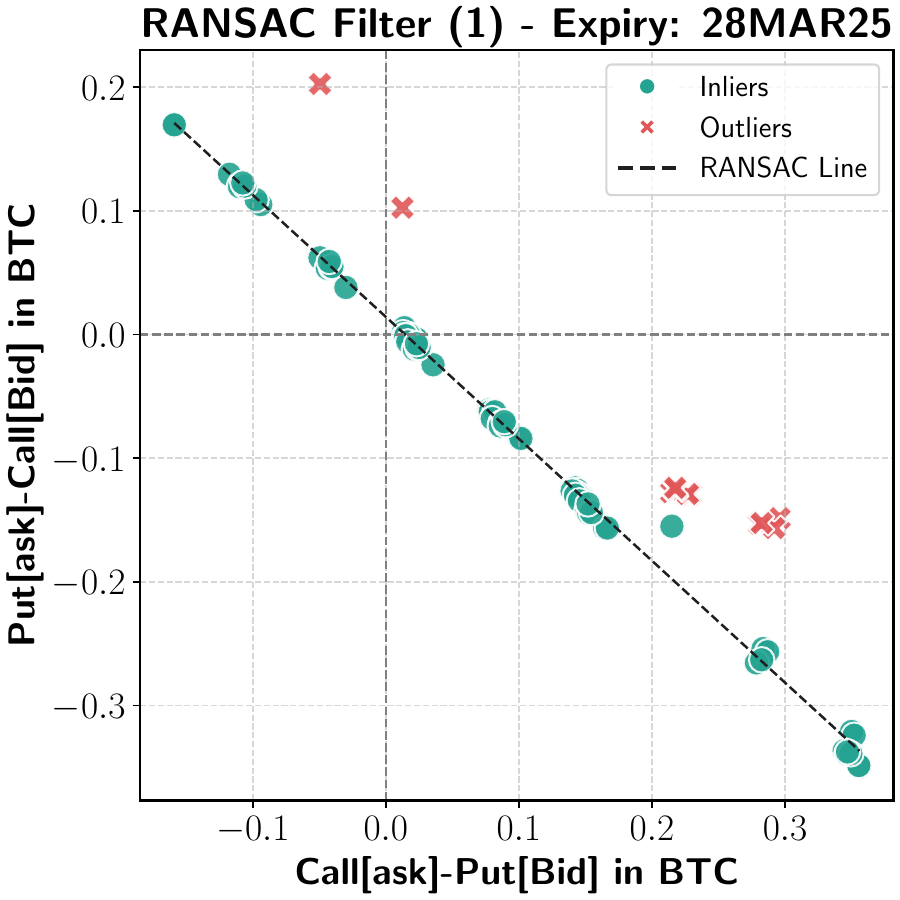}
    \end{subfigure}

    \begin{subfigure}[b]{0.32\textwidth}
        \centering
        \includegraphics[width=\textwidth]{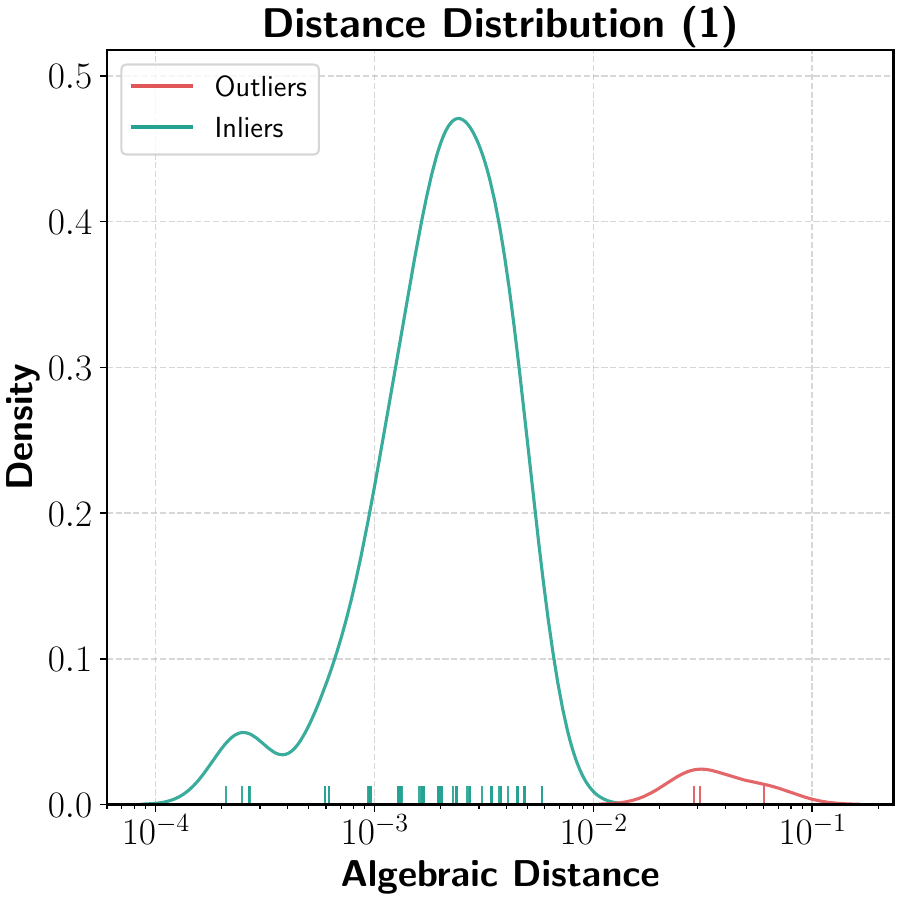}
    \end{subfigure}
    \begin{subfigure}[b]{0.32\textwidth}
        \centering
        \includegraphics[width=\textwidth]{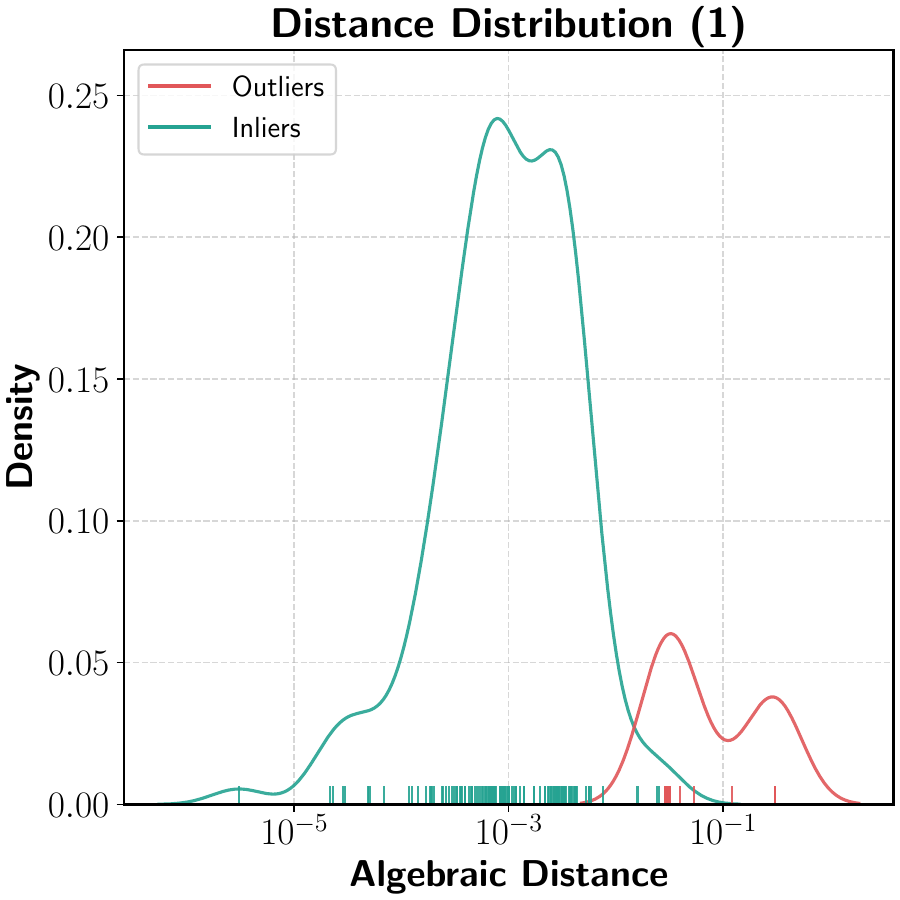}
    \end{subfigure}
    \begin{subfigure}[b]{0.32\textwidth}
        \centering
        \includegraphics[width=\textwidth]{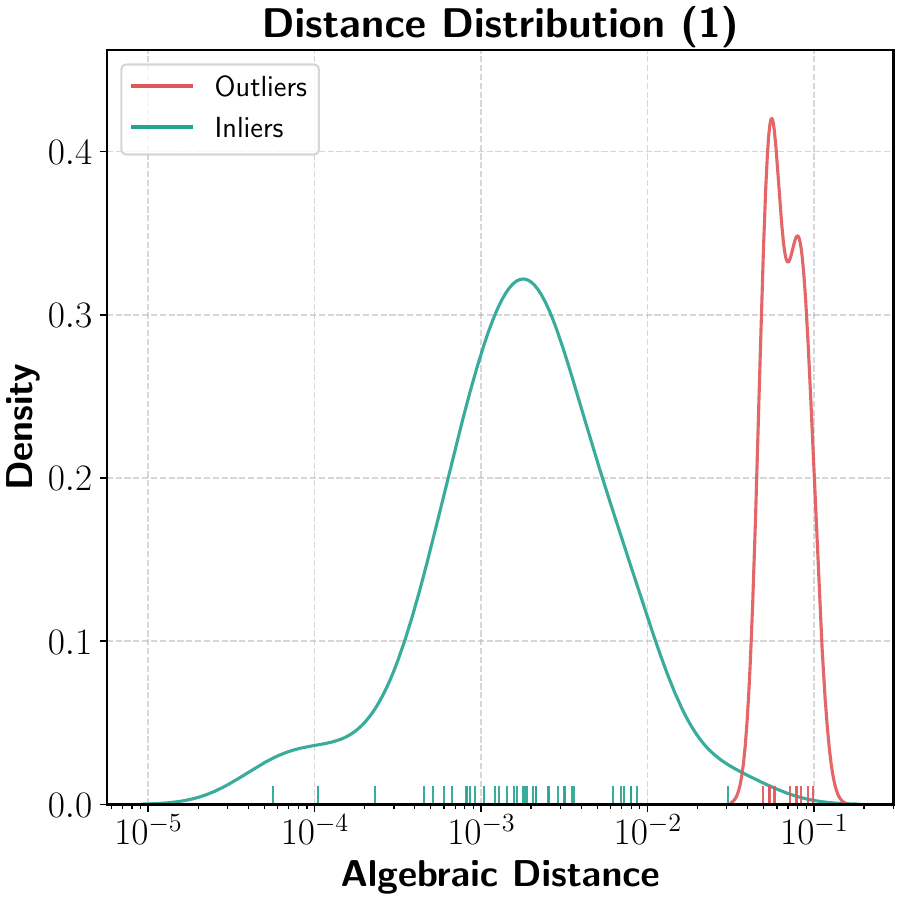}
    \end{subfigure}

    \vspace{1.2em}

    \begin{subfigure}[b]{0.32\textwidth}
        \centering
        \includegraphics[width=\textwidth]{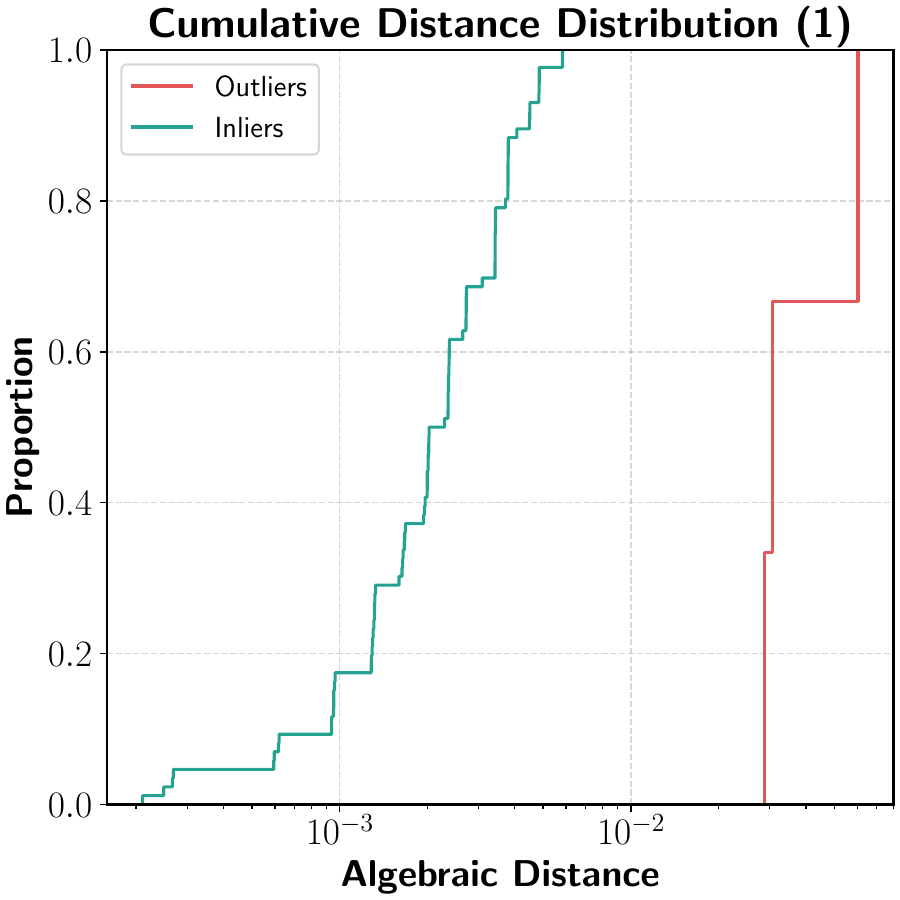}
    \end{subfigure}
    \begin{subfigure}[b]{0.32\textwidth}
        \centering
        \includegraphics[width=\textwidth]{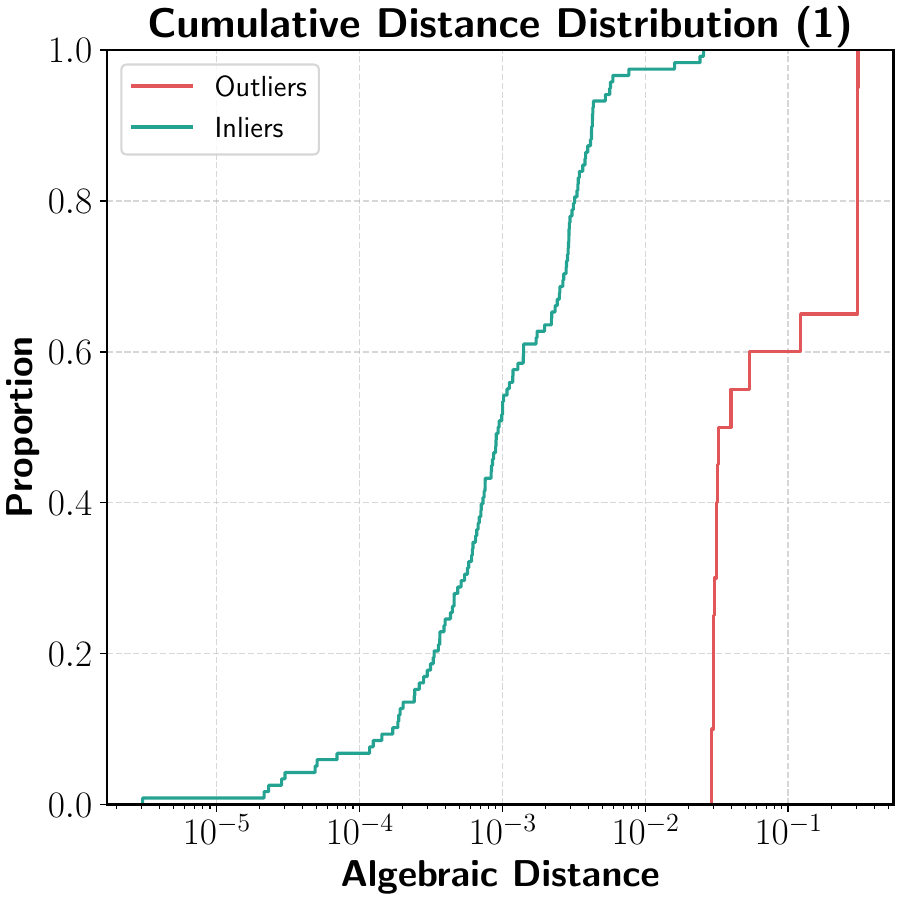}
    \end{subfigure}
    \begin{subfigure}[b]{0.32\textwidth}
        \centering
        \includegraphics[width=\textwidth]{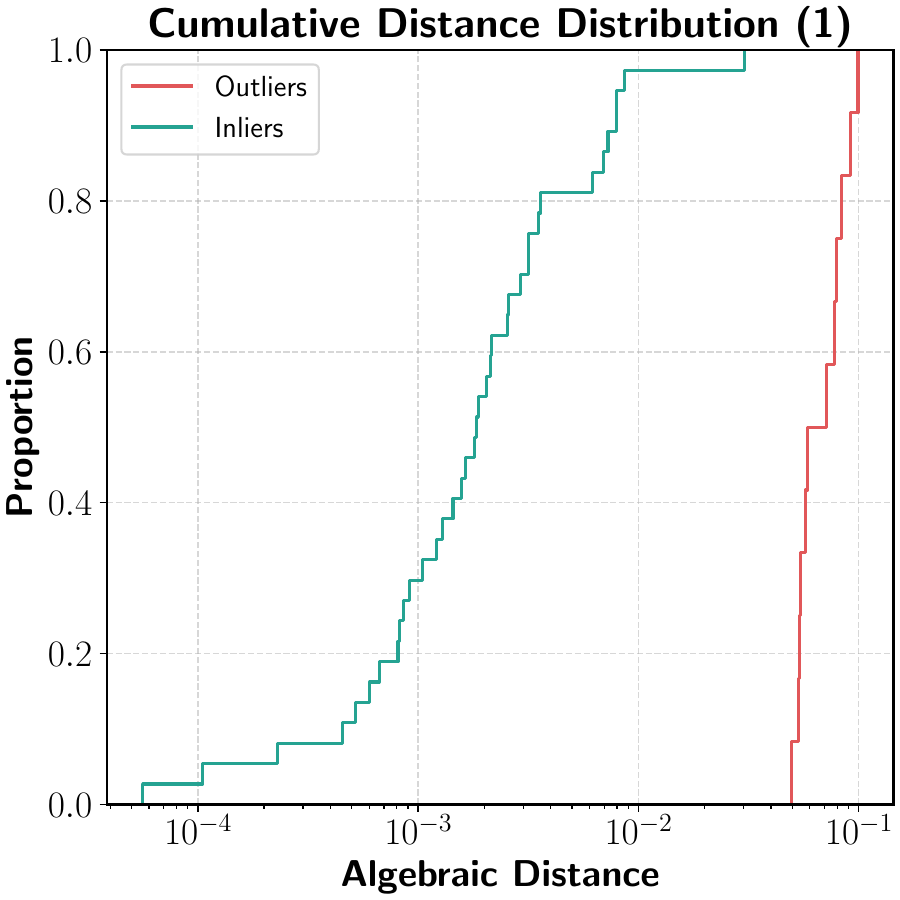}
    \end{subfigure}
    \caption{RANSAC outlier detection for BTC inverse options on May~30, 2024, using orientation~(1) of Eq.~\eqref{eq:spread_relation}. The columns correspond to expiry dates September~27, 2024 (left), December~27, 2024 (middle), and March~28, 2025 (right). The top row shows the fitted RANSAC lines and the classification into inliers and outliers; the middle and bottom rows show the distributions and cumulative distributions of algebraic distances to the fitted lines.}
    \label{fig:outlier_RANSAC_KED_CDF}
\end{figure}

We first consider May~30, 2024, and three BTC option expiry dates: September~27, 2024, December~27, 2024, and March~28, 2025. The top panels of Figure~\ref{fig:outlier_RANSAC_KED_CDF} show the fitted RANSAC lines and the classification of observations into inliers and outliers for orientation~(1) of Eq.~\eqref{eq:spread_relation}. For each observation, we also compute the algebraic (i.e. signed) distance to its orthogonal projection onto the fitted line, distinguishing observations lying above and below it. The middle and bottom panels display the corresponding distributions and cumulative distributions, respectively. The analogous figures for orientation~(2) are provided in the Appendix.

The procedure is applied using a squared-residual threshold of $0.004$ for the RANSAC classification. As explained in Section~\ref{est_proc}, an observation is retained for subsequent interest rate estimation only if it is classified as an inlier in both orientations of Eq.~\eqref{eq:spread_relation}. Table~\ref{tab:ransac_descriptive_stats} reports descriptive statistics of the RANSAC slope and intercept estimates across the full sample period. For both orientations, the fitted slopes are concentrated around $-1$, consistent with relatively stable combined call and put bid-ask spreads within the selected groups of quotes.

\begin{table}[htbp]
    \centering
    \begingroup
    \small
    \setlength{\tabcolsep}{3pt}
    \renewcommand{\arraystretch}{1.3}
    \begin{tabular}{|l|c|c|c|c|c|c|c|}
        \hline
        & \textbf{Mean}
        & \textbf{Std}
        & \textbf{Min}
        & $\boldsymbol{q_{25\%}}$
        & $\boldsymbol{q_{50\%}}$
        & $\boldsymbol{q_{75\%}}$
        & \textbf{Max} \\
        \hline
        Slope
        & $-0.99, -0.99$
        & $0.07, 0.08$
        & $-1.93, -2.65$
        & $-1.01, -1.00$
        & $-1.01, -0.99$
        & $-0.99, -0.98$
        & $0.06, 0.77$ \\
        Intercept
        & $0.01, 0.01$
        & $0.01, 0.01$
        & $-0.02, 0.00$
        & $0.01, 0.01$
        & $0.01, 0.01$
        & $0.02, 0.02$
        & $0.19, 0.17$ \\
        \hline
    \end{tabular}
    \endgroup
    \caption{Descriptive statistics of the RANSAC slope and intercept estimates for BTC inverse options from February~1, 2023 to February~1, 2026. The quantities $q_{25\%}$, $q_{50\%}$, and $q_{75\%}$ denote the first quartile, median, and third quartile, respectively. For each statistic, the two values, separated by a comma, correspond to orientations~(1) and~(2) of Eq.~\eqref{eq:spread_relation}, respectively.}
    \label{tab:ransac_descriptive_stats}
\end{table}

\newcommand{\threemthoutlierRSCline}{./BTC_outlier_detection_eq1_RSCline_27SEP24_2024-05-30.pdf}
\newcommand{\threemthoutlierKDE}{./BTC_outlier_detection_eq1_kde_27SEP24_2024-05-30.pdf}
\newcommand{\threemthoutlierCDF}{./BTC_outlier_detection_eq1_cdf_27SEP24_2024-05-30.pdf}

\newcommand{\sixmthoutlierRSCline}{./BTC_outlier_detection_eq1_RSCline_27DEC24_2024-05-30.pdf}
\newcommand{\sixmthoutlierKDE}{./BTC_outlier_detection_eq1_kde_27DEC24_2024-05-30.pdf}
\newcommand{\sixmthoutlierCDF}{./BTC_outlier_detection_eq1_cdf_27DEC24_2024-05-30.pdf}

\newcommand{\oneyroutlierRSCline}{./BTC_outlier_detection_eq1_RSCline_28MAR25_2024-05-30.pdf}
\newcommand{\oneyroutlierKDE}{./BTC_outlier_detection_eq1_kde_28MAR25_2024-05-30.pdf}
\newcommand{\oneyroutlierCDF}{./BTC_outlier_detection_eq1_cdf_28MAR25_2024-05-30.pdf}

Together, Figure~\ref{fig:outlier_RANSAC_KED_CDF} and Table~\ref{tab:ransac_descriptive_stats} illustrate how RANSAC identifies observations that deviate from the prevailing pattern of bid-ask spreads. Combined with the preliminary exclusion of excessively wide spreads, this procedure helps retain quotes that are more likely to reflect current market conditions. Indeed, materially stale executable quotes expose market makers to adverse execution by informed or faster traders, creating incentives to update their quotes. The filtered data therefore provide a more reliable basis for estimating interest rates, although this diagnostic does not, by itself, establish put-call parity.

\subsection{Results}

We now present the interest rates inferred from BTC and ETH derivatives traded on Deribit. At each observation date~$t$ and for each available expiry~$T$, BTC derivatives provide estimates of the USD and BTC interest rates, $\widehat{r}_{t,T-t}^{\mathrm{USD}}$ and $\widehat{r}_{t,T-t}^{\mathrm{BTC}}$. Likewise, ETH derivatives yield estimates $\widehat{r}_{t,T-t}^{\mathrm{USD}}$ and $\widehat{r}_{t,T-t}^{\mathrm{ETH}}$. We first examine how these rates evolve over time and then compare them with lending and staking yields observed in decentralized finance.

\subsubsection{Temporal Evolution of Implied Yield Curves}

The derivative contracts have fixed expiry dates, so their remaining maturities decrease as time passes. To compare interest rates over time at a constant investment horizon, we therefore construct series for three benchmark tenors: 60, 90, and 180~days. At each observation date, we linearly interpolate the estimated rates between the two available expiries whose remaining maturities bracket the benchmark tenor. We do not extrapolate when such a pair is unavailable. The resulting series describe the evolution of implied rates at fixed horizons rather than at fixed expiry dates.

\begin{figure}[htbp]
    \centering
    \begin{subfigure}[b]{0.48\textwidth}
        \centering
        \includegraphics[width=\textwidth]{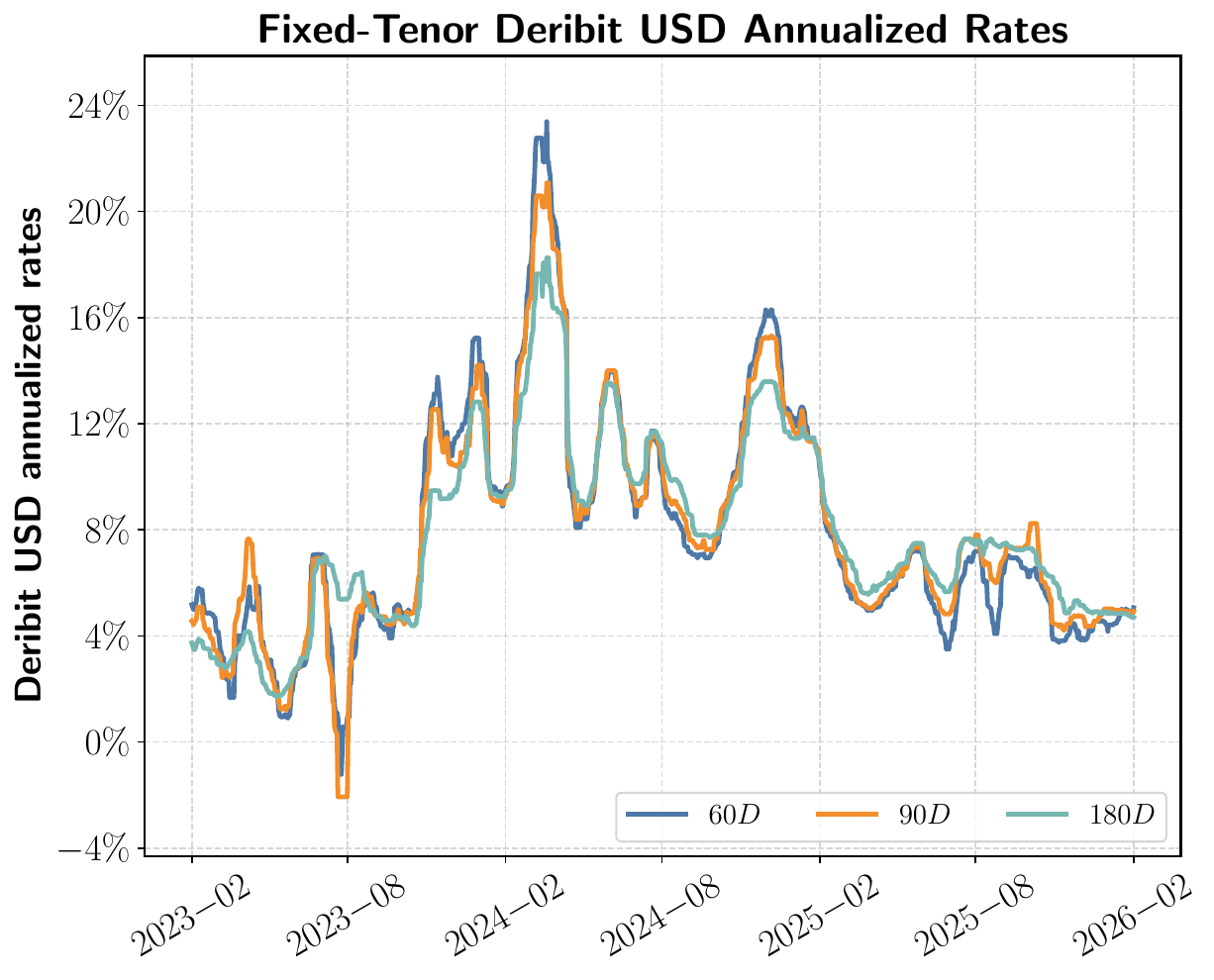}
    \end{subfigure}%
    \hspace{0.02\textwidth}%
    \begin{subfigure}[b]{0.48\textwidth}
        \centering
        \includegraphics[width=\textwidth]{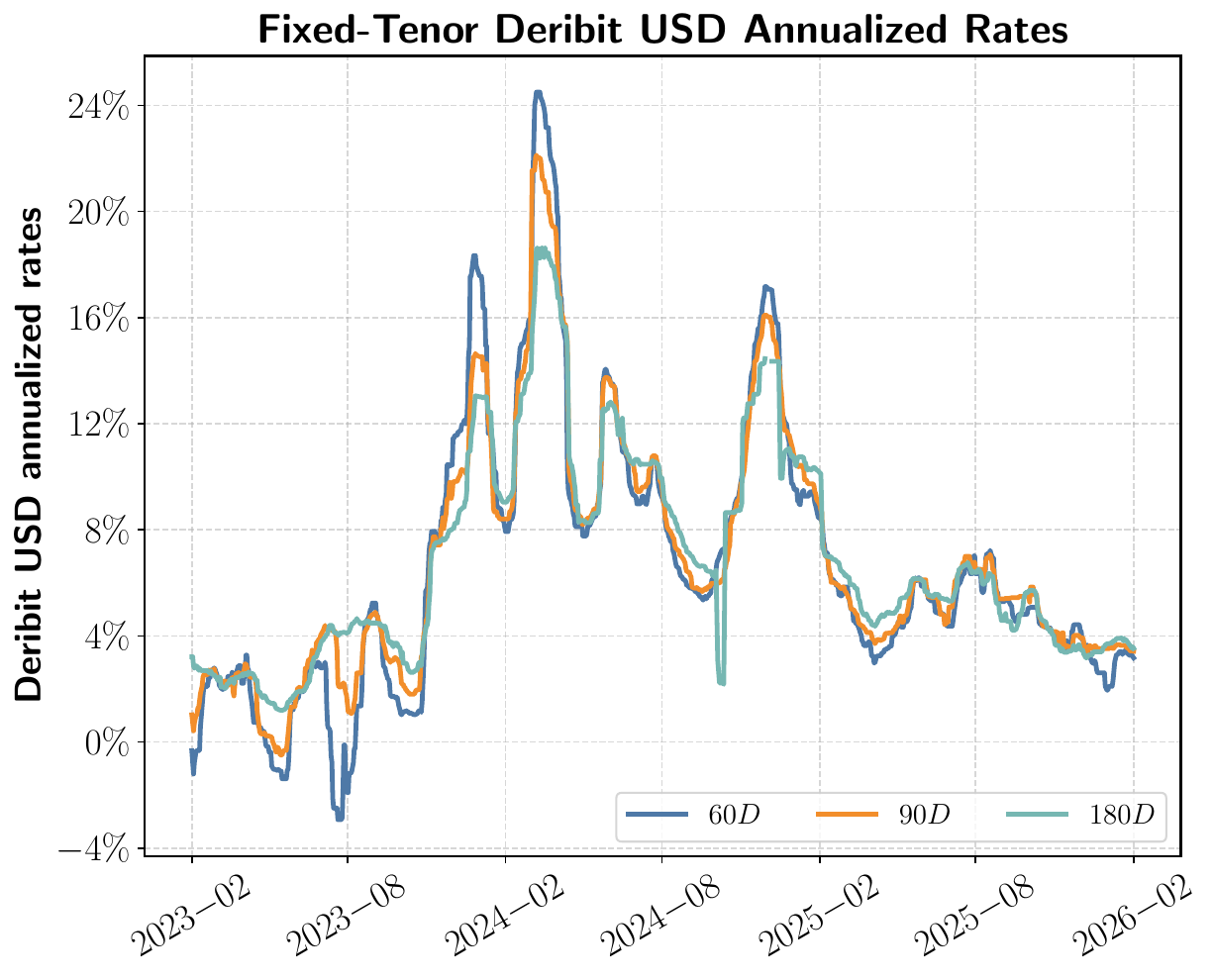}
    \end{subfigure}
    \caption{Time series of USD interest rates inferred from BTC inverse derivatives (left) and ETH inverse derivatives (right) on Deribit. The curves show centered 21-day rolling medians for constant tenors of 60~days (blue), 90~days (orange), and 180~days (green), from February~1, 2023 to February~1, 2026.}
    \label{fig:rtenorcurves}
\end{figure}

Figure~\ref{fig:rtenorcurves} displays the USD rates inferred from BTC inverse derivatives (left panel) and ETH inverse derivatives (right panel), while Table~\ref{tab:statsUSDratetenor} reports descriptive statistics for the BTC-based estimates. For readability, the plotted series are smoothed using centered 21-day rolling medians. All statistics in this section are calculated from the underlying unsmoothed estimates; the corresponding raw time series are provided in the Appendix. The rolling medians are used solely for graphical presentation and are not available in real time at the end of their respective windows.

\newcommand{\BTCrtenorsmooth}{./BTC_smoothed_r_tenor_curves.pdf}
\newcommand{\ETHrtenorsmooth}{./ETH_smoothed_r_tenor_curves.pdf}

\begin{table}[htbp]
    \centering
    \begingroup
    \small
    \setlength{\tabcolsep}{3pt}
    \renewcommand{\arraystretch}{1.3}
    \begin{tabular}{|l|c|c|c|c|c|c|c|}
        \hline
        & \textbf{Mean} & \textbf{Std} & \textbf{Min} & $\boldsymbol{q_{25\%}}$ & $\boldsymbol{q_{50\%}}$ & $\boldsymbol{q_{75\%}}$ & \textbf{Max} \\
        \hline
        60 days & 0.0803 & 0.0528 & -0.2550 & 0.0479 & 0.0708 & 0.1063 & 0.3110 \\
        90 days & 0.0808 & 0.0460 & -0.1171 & 0.0508 & 0.0738 & 0.1041 & 0.2554 \\
        180 days & 0.0806 & 0.0377 & -0.0568 & 0.0537 & 0.0751 & 0.1030 & 0.2126 \\
        \hline
    \end{tabular}
    \endgroup
    \caption{Descriptive statistics of annualized USD interest rates inferred from BTC inverse derivatives on Deribit, for constant tenors of 60, 90, and 180~days, from February~1, 2023 to February~1, 2026. The statistics are based on unsmoothed estimates. The quantities $q_{25\%}$, $q_{50\%}$, and $q_{75\%}$ denote the first quartile, median, and third quartile, respectively.}
    \label{tab:statsUSDratetenor}
\end{table}

The USD interest rates inferred independently from BTC and ETH derivatives exhibit broadly similar levels and time patterns. Their movements also coincide with distinct cryptocurrency market regimes. After relatively low levels in early 2023, in the aftermath of the FTX bankruptcy, the inferred rates rose markedly during two episodes: in early 2024, around the approval of US spot Bitcoin exchange-traded funds, and in late 2024, around the US presidential election and the subsequent market rally. At times, the implied rates exceeded 15\%. These temporal associations are consistent with changes in demand for leveraged cryptocurrency exposure, although the figures do not establish a causal relationship.

The movements are predominantly changes in the overall level of the implied curves. During some market rallies, however, the 60-day rate exceeds the 180-day rate, producing temporary inversions of the term structure. Such episodes are consistent with particularly strong demand for shorter-horizon exposure and funding, but their interpretation should also account for market-specific risk premia and trading frictions.

To assess the agreement between the two independent USD estimates, Table~\ref{tab:deltaUSDderibitrates} summarizes their signed difference at each benchmark tenor. The median difference is approximately 1.1 percentage points for all three tenors. The dispersion of this difference declines with maturity. Similar estimates from the two derivatives markets are encouraging, but a nonzero difference may reflect liquidity conditions, market segmentation, or estimation error. This comparison is related to the cross-index option-implied rate analysis of \cite{vanbinsbergenRiskfreeInterestRates2022} for the S\&P~500 and Dow Jones indices.

\begin{table}[htbp]
    \centering
    \begingroup
    \small
    \setlength{\tabcolsep}{3pt}
    \renewcommand{\arraystretch}{1.3}
    \begin{tabular}{|l|c|c|c|c|c|c|c|}
        \hline
        & \textbf{Mean} & \textbf{Std} & \textbf{Min} & $\boldsymbol{q_{25\%}}$ & $\boldsymbol{q_{50\%}}$ & $\boldsymbol{q_{75\%}}$ & \textbf{Max} \\
        \hline
        60 days & 0.0116 & 0.0286 & -0.1681 & -0.0011 & 0.0110 & 0.0222 & 0.3914 \\
        90 days & 0.0107 & 0.0191 & -0.1263 & 0.0017 & 0.0110 & 0.0194 & 0.0883 \\
        180 days & 0.0101 & 0.0138 & -0.0891 & 0.0042 & 0.0109 & 0.0166 & 0.0549 \\
        \hline
    \end{tabular}
    \endgroup
    \caption{Descriptive statistics of the signed difference between USD interest rates inferred from BTC and ETH inverse derivatives on Deribit, for constant tenors of 60, 90, and 180~days, from February~1, 2023 to February~1, 2026. Rates and differences are expressed in annualized decimal units. The quantities $q_{25\%}$, $q_{50\%}$, and $q_{75\%}$ denote the first quartile, median, and third quartile, respectively.}
    \label{tab:deltaUSDderibitrates}
\end{table}

\begin{figure}[htbp]
    \centering
    \begin{subfigure}[b]{0.48\textwidth}
        \centering
        \includegraphics[width=\textwidth]{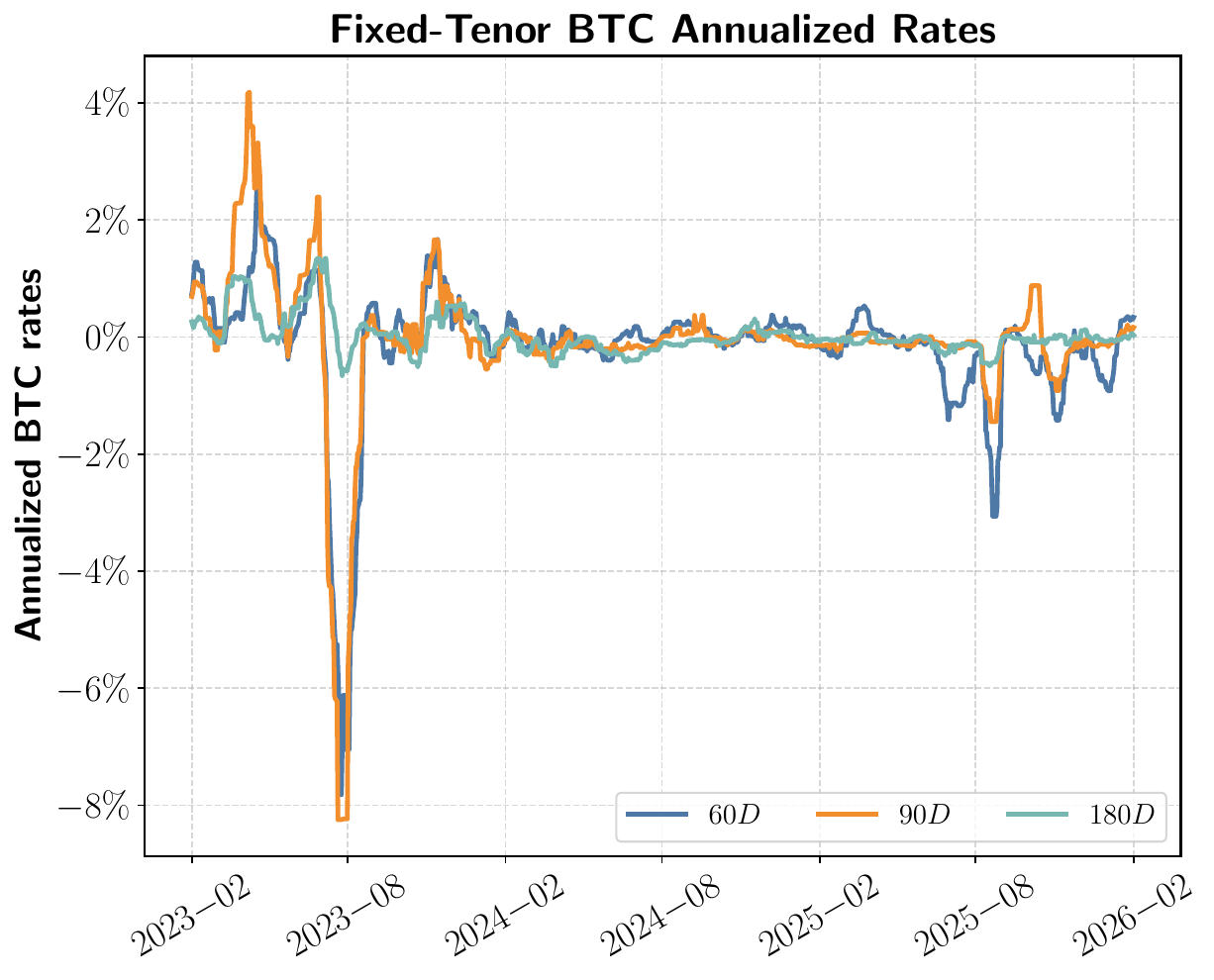}
    \end{subfigure}%
    \hspace{0.02\textwidth}%
    \begin{subfigure}[b]{0.48\textwidth}
        \centering
        \includegraphics[width=\textwidth]{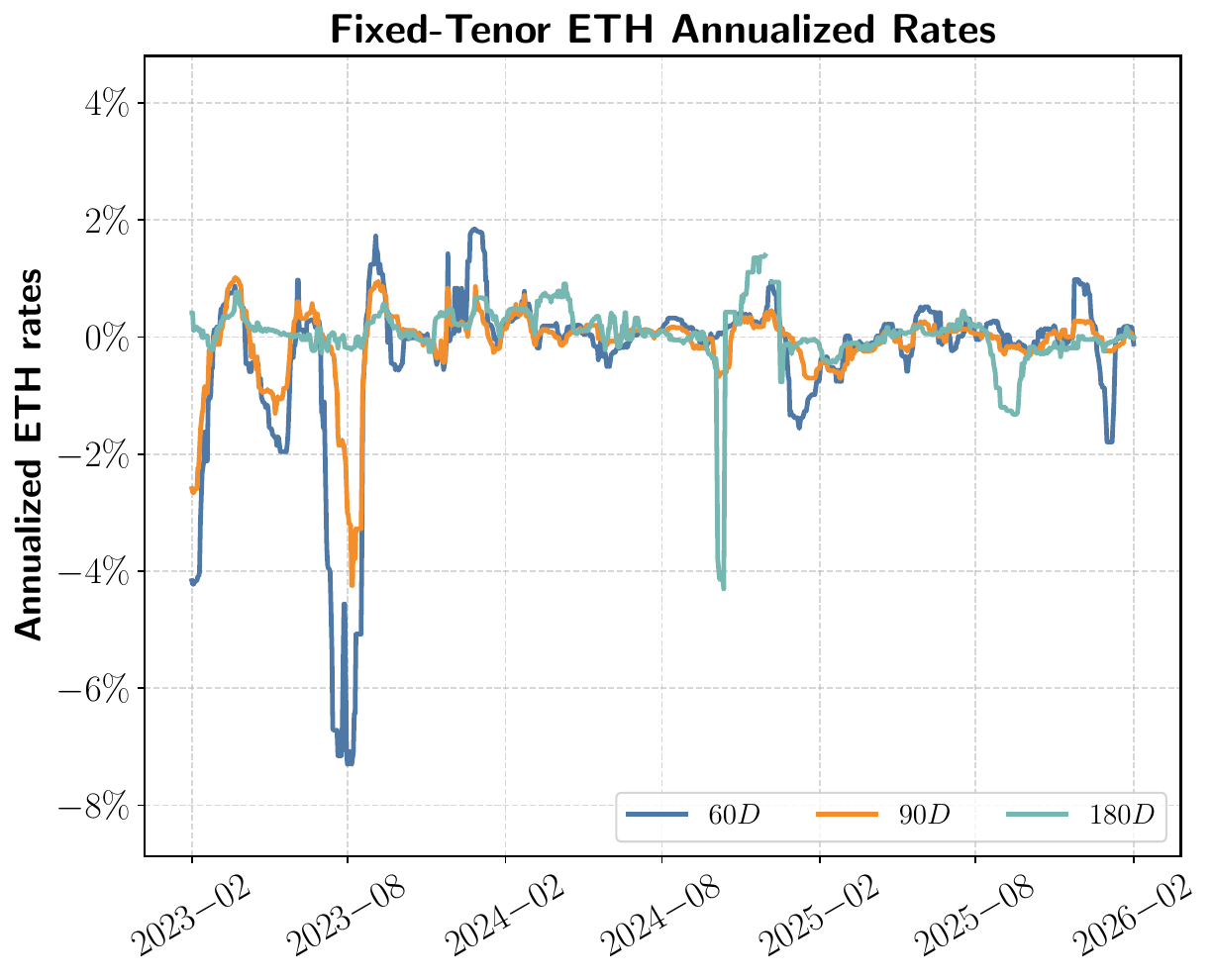}
    \end{subfigure}
    \caption{Time series of BTC interest rates (left) and ETH interest rates (right) inferred from Deribit inverse derivatives. The curves show centered 21-day rolling medians for constant tenors of 60~days (blue), 90~days (orange), and 180~days (green), from February~1, 2023 to February~1, 2026.}
    \label{fig:qtenorcurves}
\end{figure}

We next consider the interest rates denominated in the underlying cryptocurrencies. Figure~\ref{fig:qtenorcurves} shows the implied BTC rates (left panel) and ETH rates (right panel) at the same constant tenors, with the corresponding descriptive statistics reported in Tables~\ref{tab:statsBTCratetenor} and \ref{tab:statsETHratetenor}.

\newcommand{\BTCqtenorsmooth}{./BTC_smoothed_q_tenor_curves.pdf}
\newcommand{\ETHqtenorsmooth}{./ETH_smoothed_q_tenor_curves.pdf}

\begin{table}[htbp]
    \centering
    \begingroup
    \small
    \setlength{\tabcolsep}{3pt}
    \renewcommand{\arraystretch}{1.3}
    \begin{tabular}{|l|c|c|c|c|c|c|c|}
        \hline
        & \textbf{Mean} & \textbf{Std} & \textbf{Min} & $\boldsymbol{q_{25\%}}$ & $\boldsymbol{q_{50\%}}$ & $\boldsymbol{q_{75\%}}$ & \textbf{Max} \\
        \hline
        60 days & -0.0020 & 0.0242 & -0.2954 & -0.0059 & -0.0003 & 0.0046 & 0.1910 \\
        90 days & -0.0012 & 0.0191 & -0.1737 & -0.0039 & -0.0005 & 0.0036 & 0.1428 \\
        180 days & -0.0000 & 0.0110 & -0.1163 & -0.0027 & -0.0004 & 0.0021 & 0.1044 \\
        \hline
    \end{tabular}
    \endgroup
    \caption{Descriptive statistics of annualized BTC interest rates inferred from BTC inverse derivatives on Deribit, for constant tenors of 60, 90, and 180~days, from February~1, 2023 to February~1, 2026. The statistics are based on unsmoothed estimates. The quantities $q_{25\%}$, $q_{50\%}$, and $q_{75\%}$ denote the first quartile, median, and third quartile, respectively.}
    \label{tab:statsBTCratetenor}
\end{table}

The BTC rates remain close to zero for most of the sample, across all three tenors (Figure~\ref{fig:qtenorcurves}, left panel). Although the unsmoothed estimates occasionally exhibit sizeable positive or negative deviations, Table~\ref{tab:statsBTCratetenor} shows that their median values are effectively zero. These findings suggest that the derivatives market prices little systematic yield on BTC over the horizons considered. They do not rule out short-lived funding or liquidity effects, nor do they imply that BTC cannot earn a return when deployed in lending arrangements.

The ETH rates display a similar pattern. Their proximity to zero might initially appear surprising, since Ethereum's proof-of-stake mechanism allows holders who stake ETH to earn rewards, typically of the order of 2 to 4\% annually during the sample period. One might therefore expect some of that yield to be reflected in derivative-implied ETH interest rates. The distinction between immediately available ETH and staked ETH helps explain why this need not be the case.

\begin{table}[htbp]
    \centering
    \begingroup
    \small
    \setlength{\tabcolsep}{3pt}
    \renewcommand{\arraystretch}{1.3}
    \begin{tabular}{|l|c|c|c|c|c|c|c|}
        \hline
        & \textbf{Mean} & \textbf{Std} & \textbf{Min} & $\boldsymbol{q_{25\%}}$ & $\boldsymbol{q_{50\%}}$ & $\boldsymbol{q_{75\%}}$ & \textbf{Max} \\
        \hline
        60 days & -0.0027 & 0.0211 & -0.1698 & -0.0075 & 0.0001 & 0.0049 & 0.0890 \\
        90 days & -0.0012 & 0.0112 & -0.0670 & -0.0048 & -0.0001 & 0.0036 & 0.0440 \\
        180 days & 0.0003 & 0.0084 & -0.0445 & -0.0030 & 0.0003 & 0.0035 & 0.0633 \\
        \hline
    \end{tabular}
    \endgroup
    \caption{Descriptive statistics of annualized ETH interest rates inferred from ETH inverse derivatives on Deribit, for constant tenors of 60, 90, and 180~days, from February~1, 2023 to February~1, 2026. The statistics are based on unsmoothed estimates. The quantities $q_{25\%}$, $q_{50\%}$, and $q_{75\%}$ denote the first quartile, median, and third quartile, respectively.}
    \label{tab:statsETHratetenor}
\end{table}

Inverse ETH derivatives on Deribit are settled in ETH, and open positions create margin and liquidity needs that may change rapidly. Staking ETH, by contrast, entails withdrawal and redemption constraints; liquid staking tokens such as stETH alleviate some of these constraints but introduce basis risk and may be subject to collateral haircuts. Although Deribit supports cross-collateral arrangements, staking claims and immediately available ETH are therefore not perfect substitutes for managing derivative exposures.\footnote{See \url{https://support.deribit.com/hc/en-us/articles/25944777203869-Cross-collateral-specifications}. The collateral arrangements documented there do not necessarily describe all account configurations or all dates in our historical sample.} Consequently, the return obtainable by staking ETH need not be fully reflected in the implied rate associated with ETH on the exchange.

This interpretation also clarifies the contrast with the implied USD rates. USD-denominated funding and dollar-linked assets are connected to a broader set of money markets and lending opportunities, while the substitution between liquid ETH margin balances and yield-bearing staking positions entails additional operational and market risks. The resulting differences in liquidity and collateral use provide a possible explanation for the near-zero ETH rates alongside more variable USD rates. The evidence is consistent with such segmentation, rather than a direct measurement of its magnitude.

Deribit's quoting conventions may reinforce this outcome. Its inverse-option pricing interface uses the futures price as the reference underlying when displaying implied volatilities and converting volatility-based orders into option prices.\footnote{See \url{https://support.deribit.com/hc/en-us/articles/31424939096093-Inverse-Options}.} This convention naturally accommodates prices consistent with little or no separate ETH yield, unless market participants adjust their quotes. It is not, however, an independent arbitrage argument: a sustained nonzero ETH implied rate could arise if funding conditions, relative asset demand, or the practical ability to mobilize staking returns were to change. The same reasoning applies, without the staking component, to BTC.

\subsubsection{Comparison with Decentralized Lending and Staking Rates}

To assess the economic relevance of the inferred rates within the cryptocurrency ecosystem, we compare them with observable rates on decentralized finance protocols. We first consider annualized USDC supply and borrowing rates on Aave v3 over February~1, 2023 to February~1, 2026. These are variable, short-term lending rates, whereas our Deribit estimates correspond to fixed horizons. Moreover, USDC positions on a decentralized lending protocol and USD-denominated synthetic claims on a centralized derivatives venue carry different credit, liquidity, and settlement risks. The comparison is therefore indicative rather than a like-for-like valuation of identical claims.

\begin{figure}[H]
    \centering
    \includegraphics[width=0.87\textwidth]{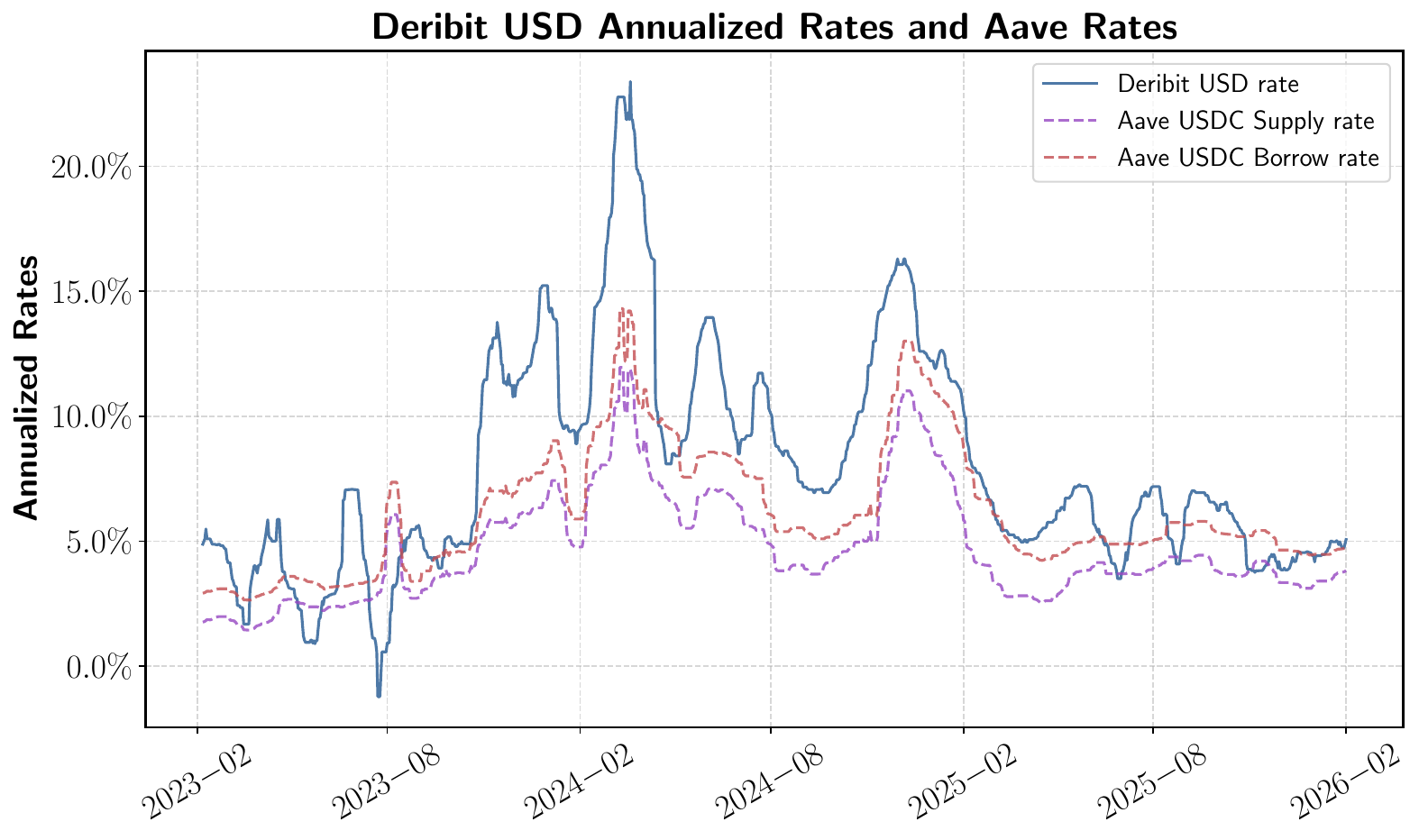}
    \caption{Comparison of the 60-day USD interest rate inferred from BTC inverse derivatives on Deribit with Aave v3 short-term variable USDC supply and borrowing rates, from February~1, 2023 to February~1, 2026. The displayed series are centered 21-day rolling medians. Source of Aave v3 data: \url{https://www.defillama.com}.}
    \label{fig:aave_vs_deribit_rates}
\end{figure}

Figure~\ref{fig:aave_vs_deribit_rates} compares Aave's short-term variable USDC rates with the 60-day Deribit USD rate inferred from BTC inverse derivatives. The series exhibit some common movements, especially during periods of marked increases or decreases, without coinciding in level. This co-movement suggests that the derivative-implied rates are related to broader funding conditions in cryptocurrency markets. The differences in levels are also consistent with incomplete integration and the distinct risks of the two instruments, although the figure alone cannot separate these effects.

\begin{figure}[H]
    \centering
    \includegraphics[width=0.8\textwidth]{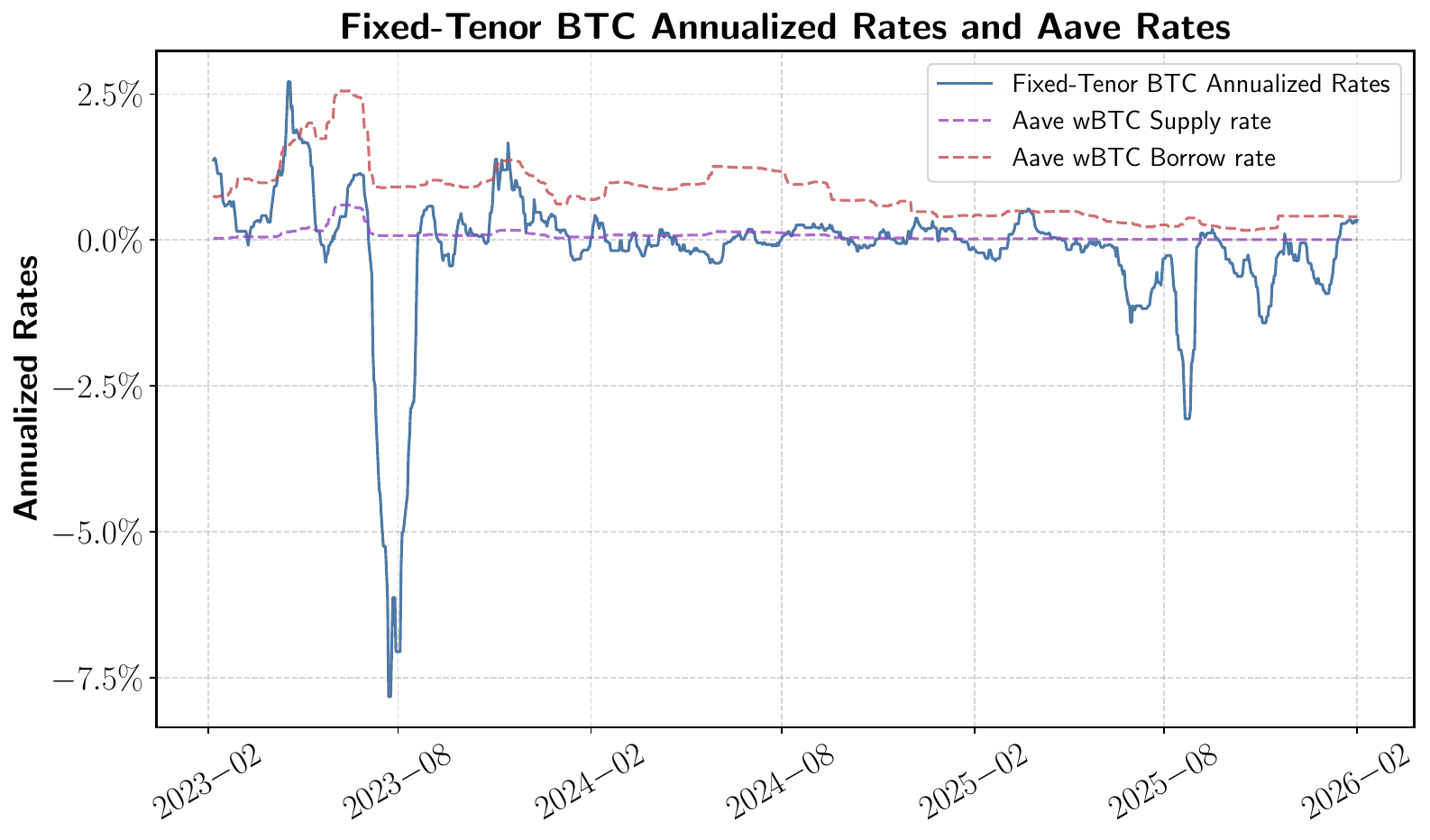}
    \caption{Comparison of the 60-day BTC interest rate inferred from BTC inverse derivatives on Deribit with Aave v3 variable wBTC rates, from February~1, 2023 to February~1, 2026. The displayed series are centered 21-day rolling medians. Source of Aave v3 data: \url{https://www.defillama.com}.}
    \label{fig:wBTC_aave_vs_deribit_rates}
\end{figure}

\begin{figure}[H]
    \centering
    \includegraphics[width=0.8\textwidth]{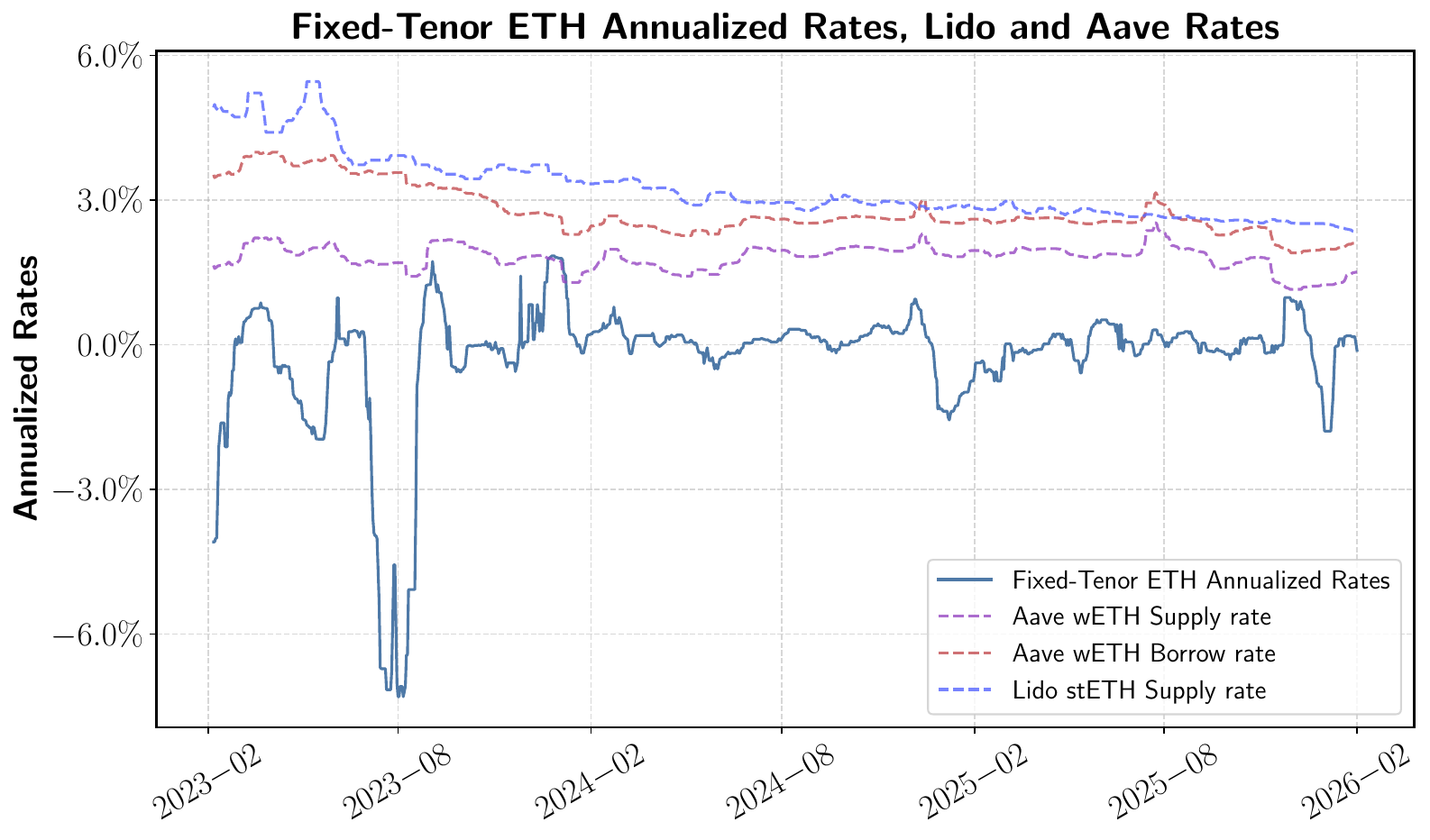}
    \caption{Comparison of the 60-day ETH interest rate inferred from ETH inverse derivatives on Deribit with Aave v3 variable wETH rates and Lido stETH staking yields, from February~1, 2023 to February~1, 2026. The displayed series are centered 21-day rolling medians. Sources of Aave v3 and Lido data: \url{https://www.defillama.com}.}
    \label{fig:wETH_aave_vs_deribit_rates}
\end{figure}

Figures~\ref{fig:wBTC_aave_vs_deribit_rates} and \ref{fig:wETH_aave_vs_deribit_rates} provide analogous comparisons for BTC and ETH. They contrast the 60-day cryptocurrency rates inferred from Deribit derivatives with Aave v3 rates on wrapped BTC (wBTC) and wrapped ETH (wETH), and, for ETH, the staking yield associated with Lido's stETH.\footnote{Wrapped Bitcoin (wBTC) is a tokenized representation of BTC used on Ethereum-compatible networks. Wrapped Ether (wETH) is an ERC-20 representation of ETH, exchangeable for native ETH at a one-to-one ratio through the wrapping contract. Neither instrument is identical, in legal, operational, or settlement terms, to native BTC or to an ETH balance held on a derivatives exchange.}

The BTC and ETH derivative-implied rates remain near zero for much of the sample, whereas the corresponding decentralized lending and staking yields are generally positive. This contrast is consistent with differences in collateral usability, liquidity, counterparty exposure, and access to yield-generating activities. In particular, liquid balances required to support derivatives positions cannot always be deployed simultaneously in lending or staking arrangements without incurring additional risks or costs. Nevertheless, the rates being compared differ in their asset representations, maturity profiles, and contractual risks. Their divergence should not be read as a directly executable arbitrage or as proof that any single institutional friction accounts for the observed gap.

\section{Conclusion}\label{sec:conclusion}

This paper develops a framework for extracting implied interest rate term structures from cryptocurrency derivatives. In a market where conventional fixed-income instruments remain scarce and fixed-maturity benchmark rates are not readily observable, we show how inverse options and futures can be used to infer synthetic zero-coupon discount factors denominated in USD and in the underlying cryptocurrencies. The resulting yield curves provide a means of examining interest rate formation and the degree of integration between different segments of the cryptocurrency ecosystem.

Our methodology combines two complementary steps. We first use a RANSAC-based procedure to identify unreliable option quotes through bid-ask relationships derived from put-call parity. We then jointly exploit the filtered option prices and futures prices to estimate the relevant discount factors. This approach is designed to mitigate the impact of noisy and inconsistent quotes without requiring a parametric specification of the yield curves.

Applying the methodology to Deribit BTC and ETH derivatives over the period from February~2023 to February~2026 reveals several distinctive features. USD implied rates inferred independently from BTC and ETH derivatives exhibit broadly similar dynamics, with pronounced increases during the cryptocurrency market rallies of 2024 and occasional inversions of the term structure. These rates also display similarities in their evolution with the variable USDC lending and borrowing rates observed on Aave, despite differences in maturity, credit exposure, and market structure. In contrast, the implied BTC and ETH rates remain close to zero across the maturities considered. The ETH result is particularly noteworthy given the positive yields available through staking. This difference is consistent with the limited substitutability of liquid exchange collateral and yield-bearing staked assets, as well as with prevailing derivatives quoting conventions.

The framework could be extended to other exchanges as their options markets develop sufficient liquidity. Comparing implied rates across trading venues would help identify differences in funding conditions, collateral arrangements, and platform-specific risks, although interpreting such spreads would require accounting for differences in contract design and market liquidity. Another promising direction is the comparison of derivative-implied rates with directly observable fixed-maturity lending rates. The launch of Morpho Midnight, which offers fixed-rate, fixed-term lending, illustrates the emergence of such instruments within decentralized finance. As these markets develop, comparisons between synthetic and directly traded term structures could provide a more precise assessment of interest rate consistency and market integration in the cryptocurrency ecosystem.

\section*{Acknowledgments and Declarations}

\subsection*{Acknowledgments}

We thank Quentin Archer for his energy, sustained encouragement,
and thoughtful comments, Corentine Poilvet-Clédière for her confidence
in this project, and Philippe de Peretti for insightful discussions
on econometric and statistical methods.

\subsection*{Funding}

This research received financial support from the Research Initiative
``Blockchain et Intelligence Artificielle pour les infrastructures
de marché,'' under the aegis of the Institut Europlace de Finance,
in partnership with LCH SA. The views expressed in this paper are
those of the authors and do not necessarily reflect those of LCH SA.

Olivier Guéant also makes this research part of ANR-24-CE38-7885 (Blockchain and Decentralized Finance - BLOCKFI)

\subsection*{Potential Conflicts of Interest}

Wenkai Zhang is an employee of LCH SA, part of the London Stock
Exchange Group. Sébastien Bieber is a PhD student at Université
Paris Dauphine-PSL whose doctoral research is funded by the Research
Initiative ``Blockchain et Intelligence Artificielle pour les
infrastructures de marché.''

\subsection*{Data Availability}

The data supporting the findings of this study are available
from the authors upon reasonable request, subject to any applicable
third-party licensing restrictions.

\subsection*{Use of Generative AI}

The scientific ideas and contributions in this paper originated exclusively
with the authors. Generative AI tools assisted with language editing and
presentation, with the modest ambition of bringing the English of the four
French authors up to the standard of native speakers. We apologize
for any Gallicisms that may have survived this process. The authors reviewed and revised all AI-assisted text and retain full responsibility for the methodology, calculations, empirical analysis, interpretations, and final content of the paper.

\newpage
\appendix

\section*{Appendix:~Additional Figures}\label{rawplots}


\newcommand{\threemthoutlierRSClineTwo}{./BTC_outlier_detection_eq2_RSCline_27SEP24_2024-05-30.pdf}
\newcommand{\threemthoutlierKDETwo}{./BTC_outlier_detection_eq2_kde_27SEP24_2024-05-30.pdf}
\newcommand{\threemthoutlierCDFTwo}{./BTC_outlier_detection_eq2_cdf_27SEP24_2024-05-30.pdf}

\newcommand{\sixmthoutlierRSClineTwo}{./BTC_outlier_detection_eq2_RSCline_27DEC24_2024-05-30.pdf}
\newcommand{\sixmthoutlierKDETwo}{./BTC_outlier_detection_eq2_kde_27DEC24_2024-05-30.pdf}
\newcommand{\sixmthoutlierCDFTwo}{./BTC_outlier_detection_eq2_cdf_27DEC24_2024-05-30.pdf}

\newcommand{\oneyroutlierRSClineTwo}{./BTC_outlier_detection_eq2_RSCline_28MAR25_2024-05-30.pdf}
\newcommand{\oneyroutlierKDETwo}{./BTC_outlier_detection_eq2_kde_28MAR25_2024-05-30.pdf}
\newcommand{\oneyroutlierCDFTwo}{./BTC_outlier_detection_eq2_cdf_28MAR25_2024-05-30.pdf}

\begin{figure}[htbp]
    \centering
    \begin{subfigure}[b]{0.32\textwidth}
        \centering
        \includegraphics[width=\textwidth]{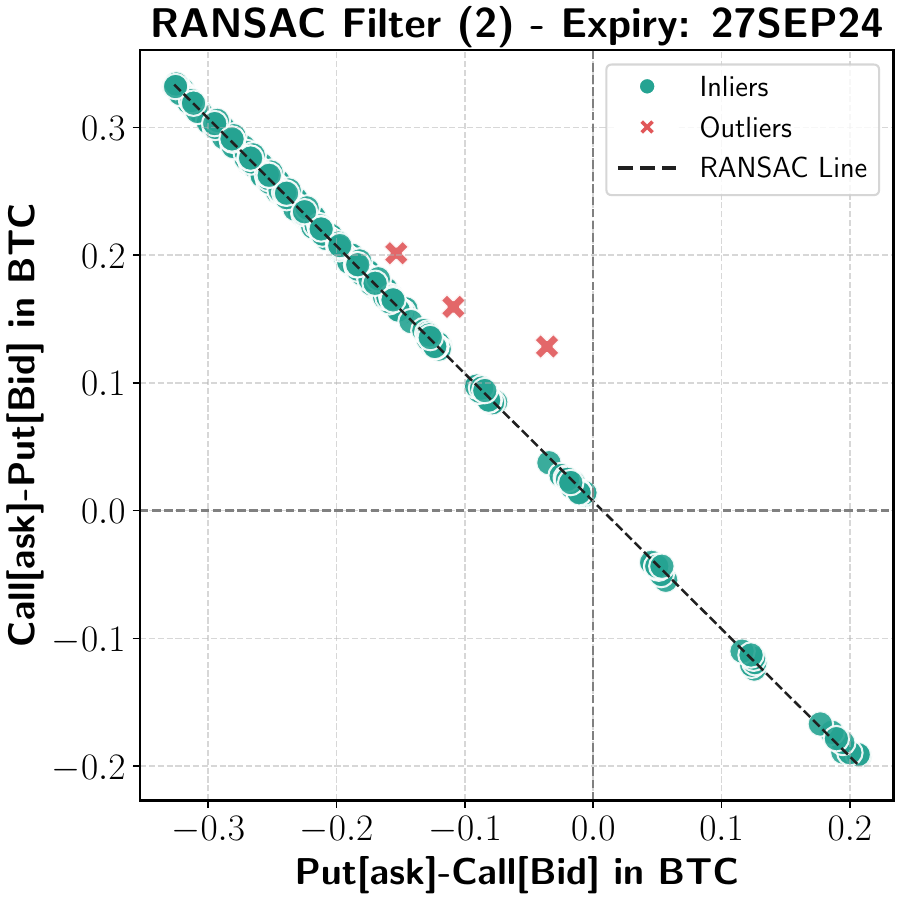}
    \end{subfigure}
    \hspace{0.00\textwidth}
    \begin{subfigure}[b]{0.32\textwidth}
        \centering
        \includegraphics[width=\textwidth]{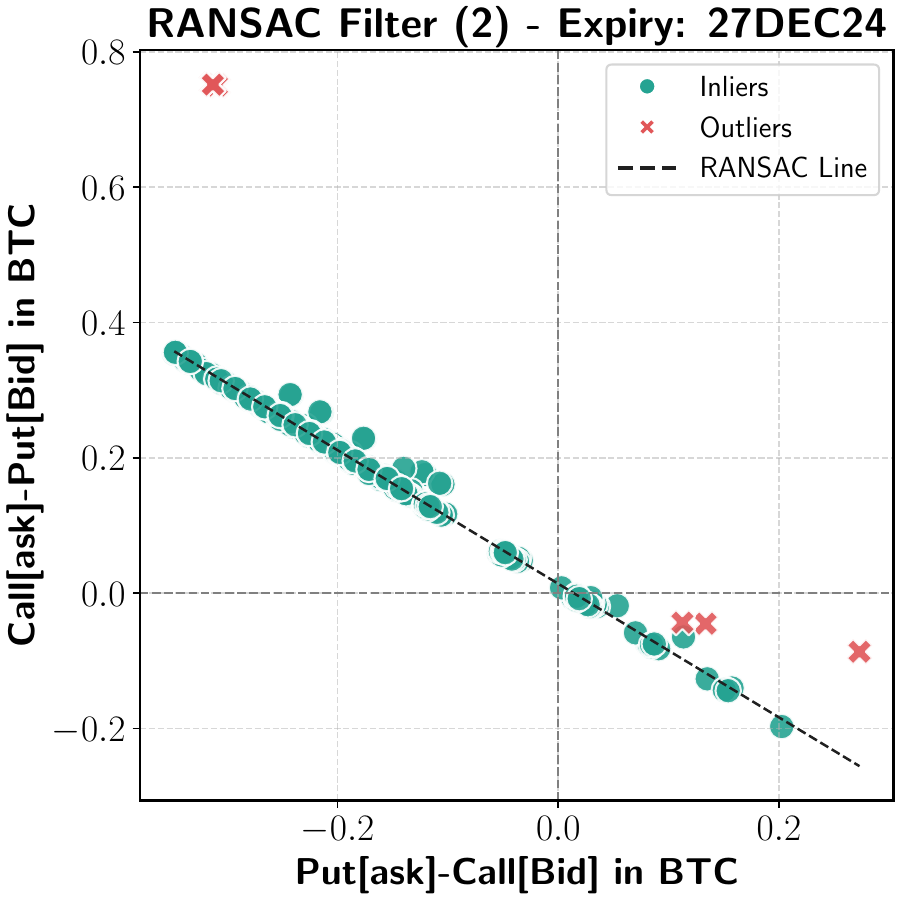}
    \end{subfigure}
    \hspace{0.00\textwidth}
    \begin{subfigure}[b]{0.32\textwidth}
        \centering
        \includegraphics[width=\textwidth]{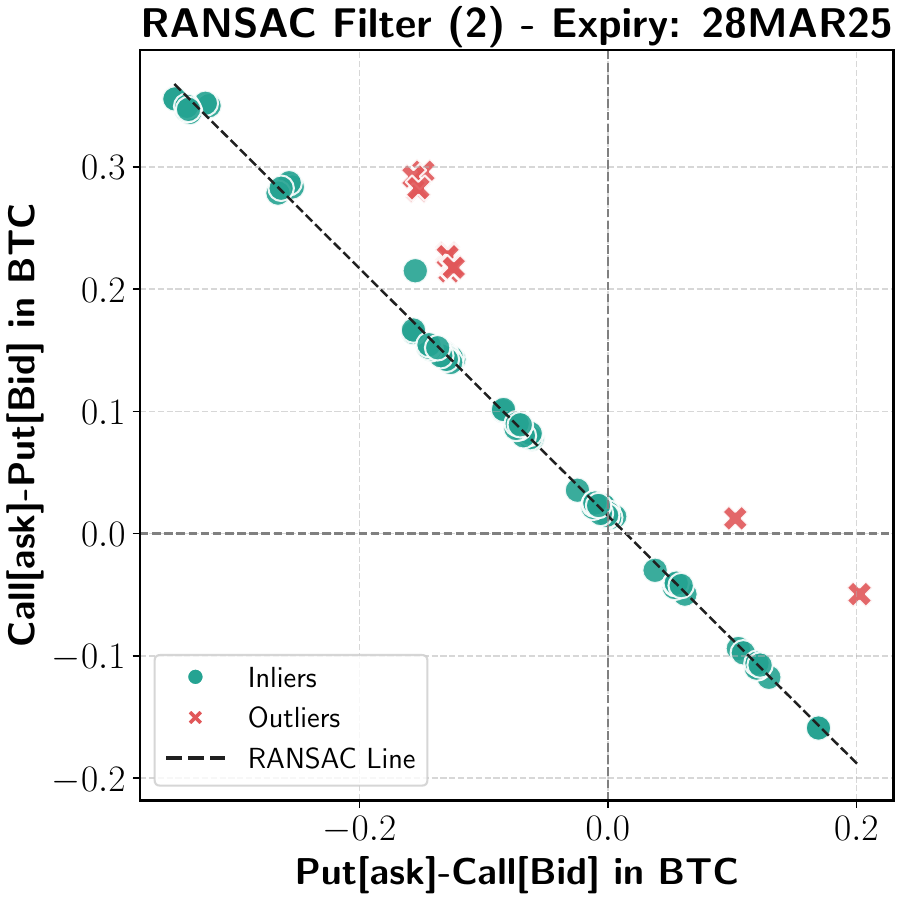}
    \end{subfigure}

    \centering
    \begin{subfigure}[b]{0.32\textwidth}
        \centering
        \includegraphics[width=\textwidth]{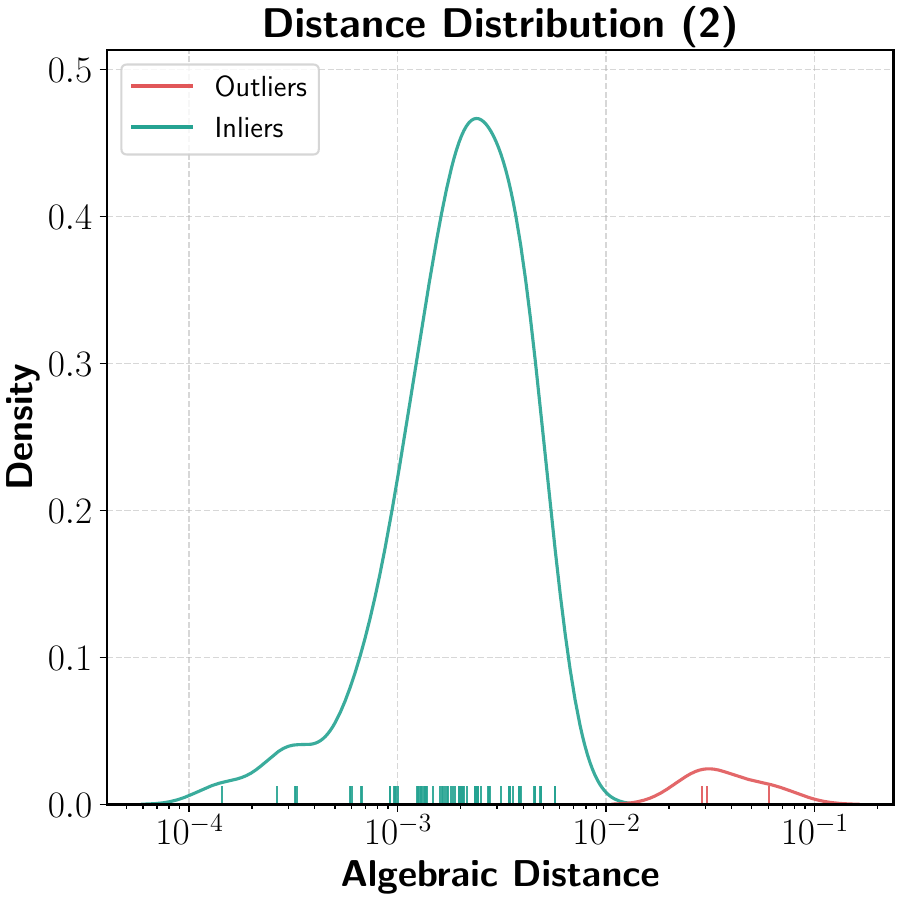}
    \end{subfigure}
    \hspace{0.00\textwidth}
    \begin{subfigure}[b]{0.32\textwidth}
        \centering
        \includegraphics[width=\textwidth]{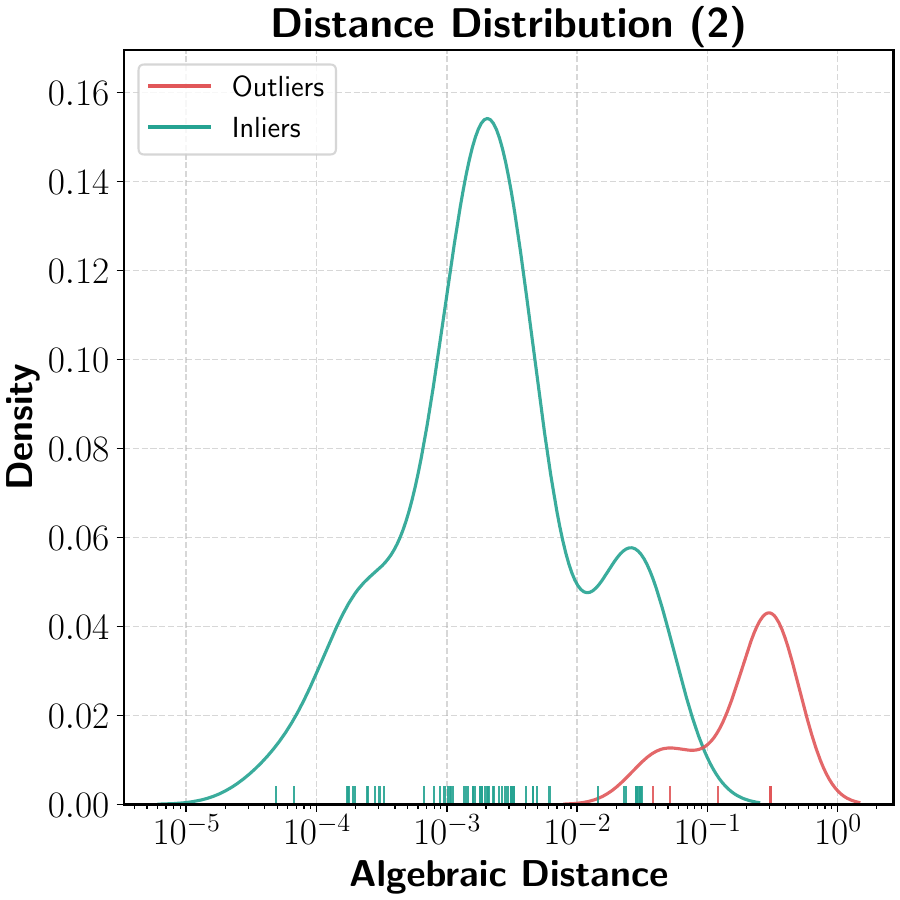}
    \end{subfigure}
    \hspace{0.00\textwidth}
    \begin{subfigure}[b]{0.32\textwidth}
        \centering
        \includegraphics[width=\textwidth]{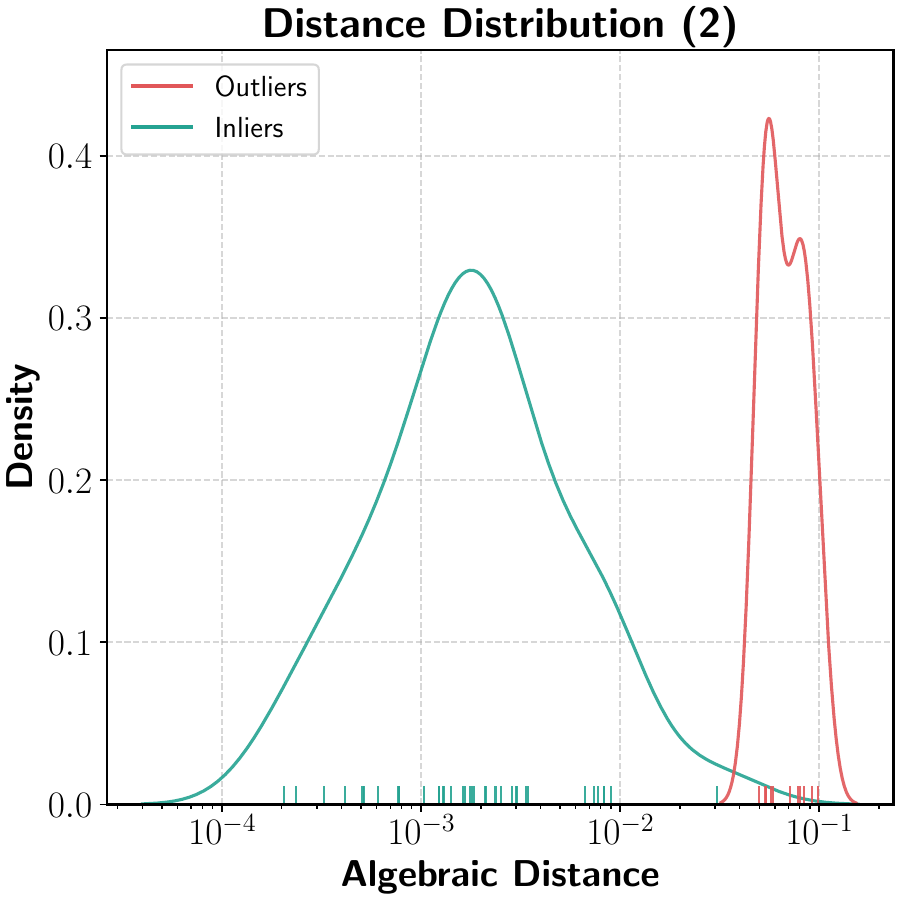}
    \end{subfigure}

    \vspace{1.2em}

    \centering
    \begin{subfigure}[b]{0.32\textwidth}
        \centering
        \includegraphics[width=\textwidth]{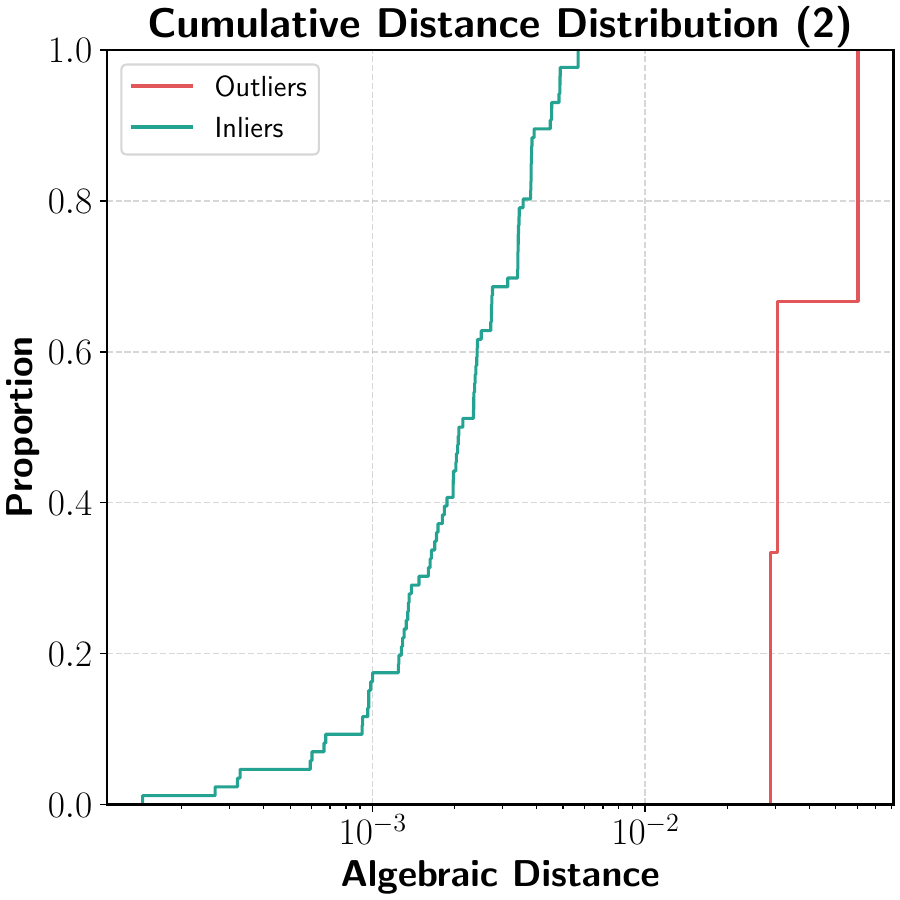}
    \end{subfigure}
    \hspace{0.00\textwidth}
    \begin{subfigure}[b]{0.32\textwidth}
        \centering
        \includegraphics[width=\textwidth]{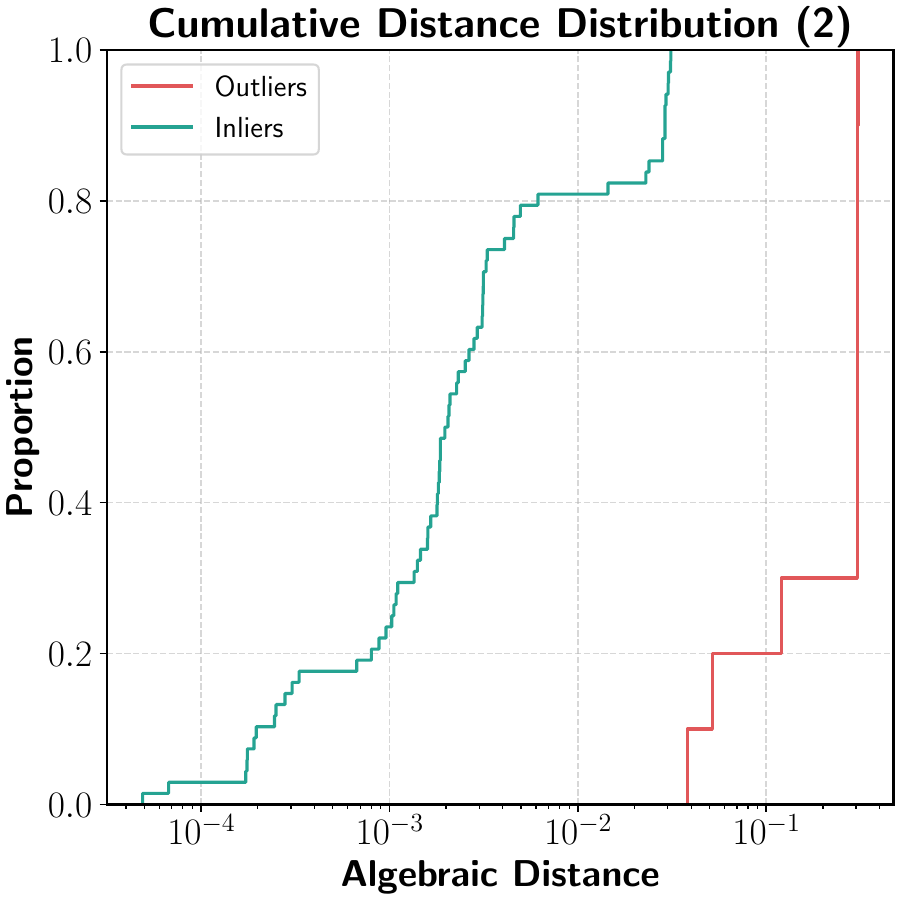}
    \end{subfigure}
    \hspace{0.00\textwidth}
    \begin{subfigure}[b]{0.32\textwidth}
        \centering
        \includegraphics[width=\textwidth]{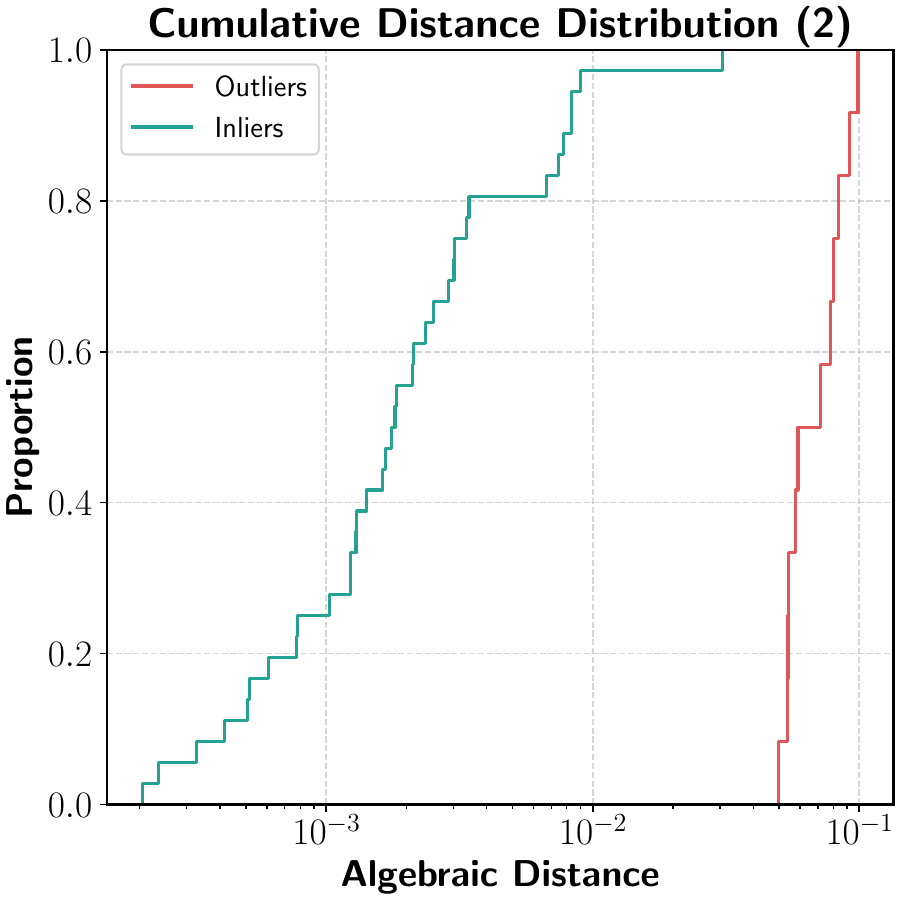}
    \end{subfigure}
    \caption{RANSAC outlier detection for BTC inverse options on May~30, 2024, using orientation~(2) of Eq.~\eqref{eq:spread_relation}. The columns correspond to expiry dates September~27, 2024 (left), December~27, 2024 (middle), and March~28, 2025 (right).}
\end{figure}

\newcommand{\BTCrtenorthreemthD}{./BTC_rate_r_tenor_60D.pdf}
\newcommand{\BTCrtenorsixmthD}{./BTC_rate_r_tenor_90D.pdf}
\newcommand{\BTCrtenoroneyrD}{./BTC_rate_r_tenor_180D.pdf}

\newcommand{\BTCqtenorthreemthD}{./BTC_rate_q_tenor_60D.pdf}
\newcommand{\BTCqtenorsixmthD}{./BTC_rate_q_tenor_90D.pdf}
\newcommand{\BTCqtenoroneyrD}{./BTC_rate_q_tenor_180D.pdf}

\begin{figure}[htbp]
    \centering

    \begin{subfigure}[b]{0.5\textwidth}
        \centering
        \includegraphics[width=\textwidth]{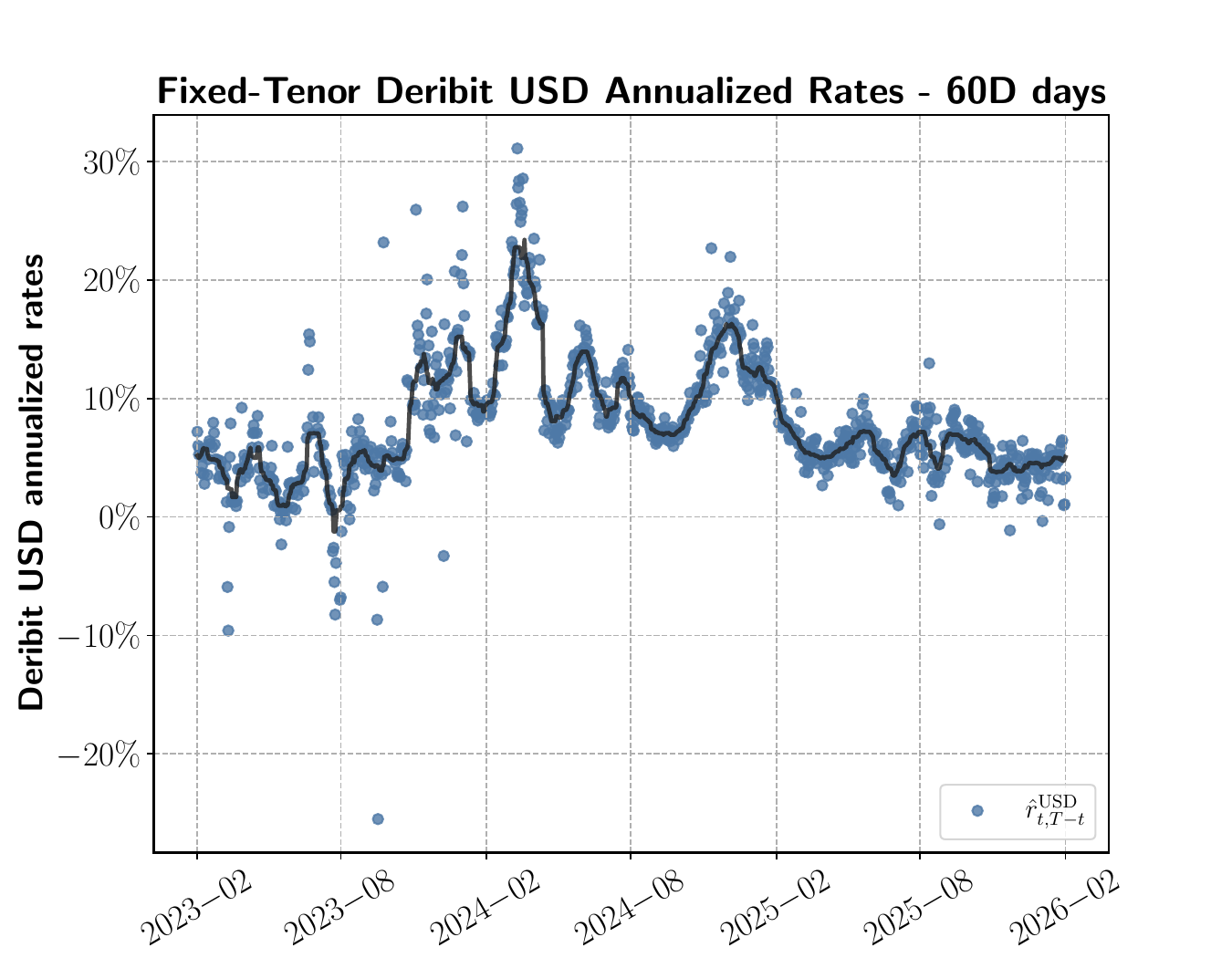}
    \end{subfigure}
    \hspace{-0.02\textwidth}
    \begin{subfigure}[b]{0.5\textwidth}
        \centering
        \includegraphics[width=\textwidth]{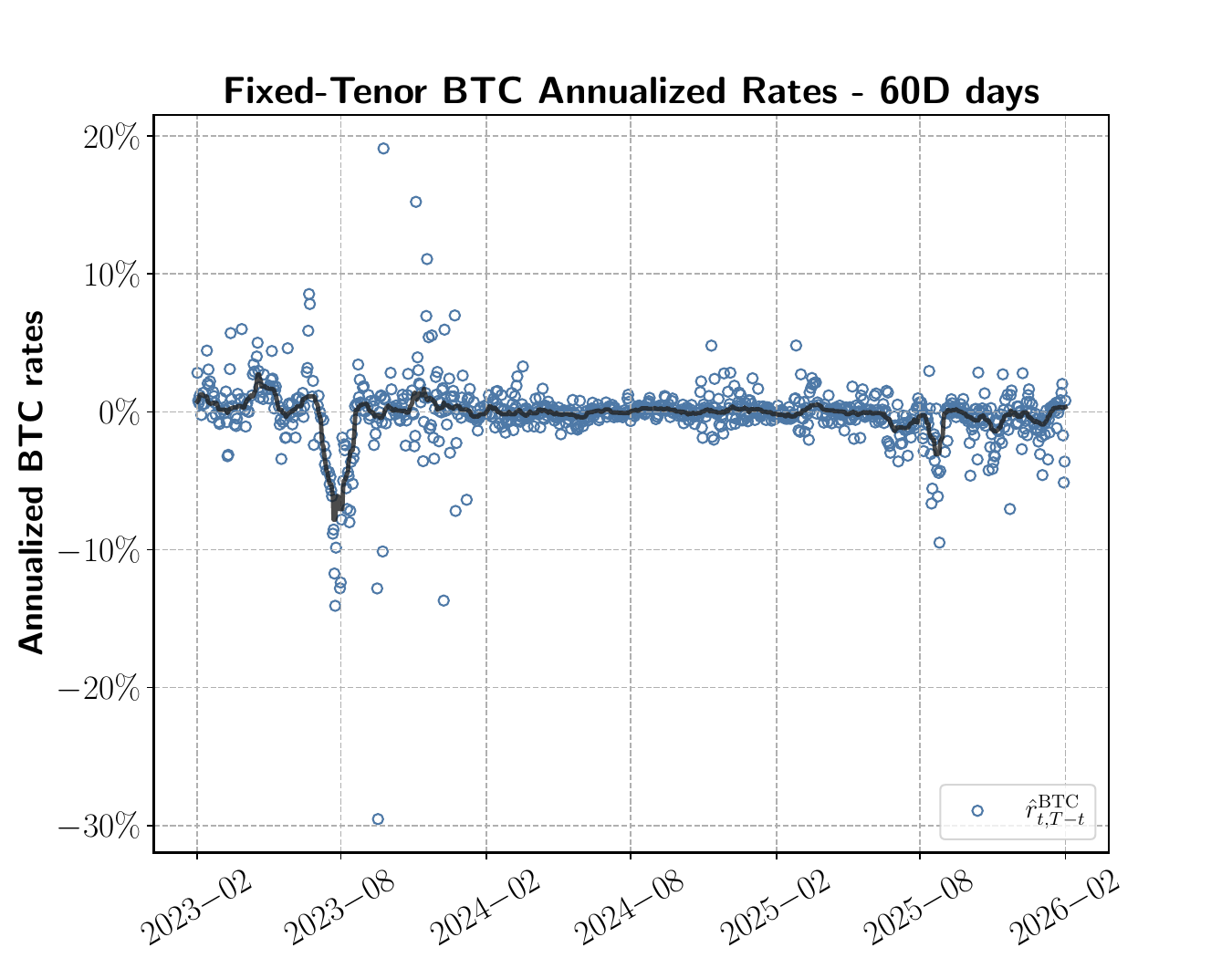}
    \end{subfigure}

    \begin{subfigure}[b]{0.5\textwidth}
        \centering
        \includegraphics[width=\textwidth]{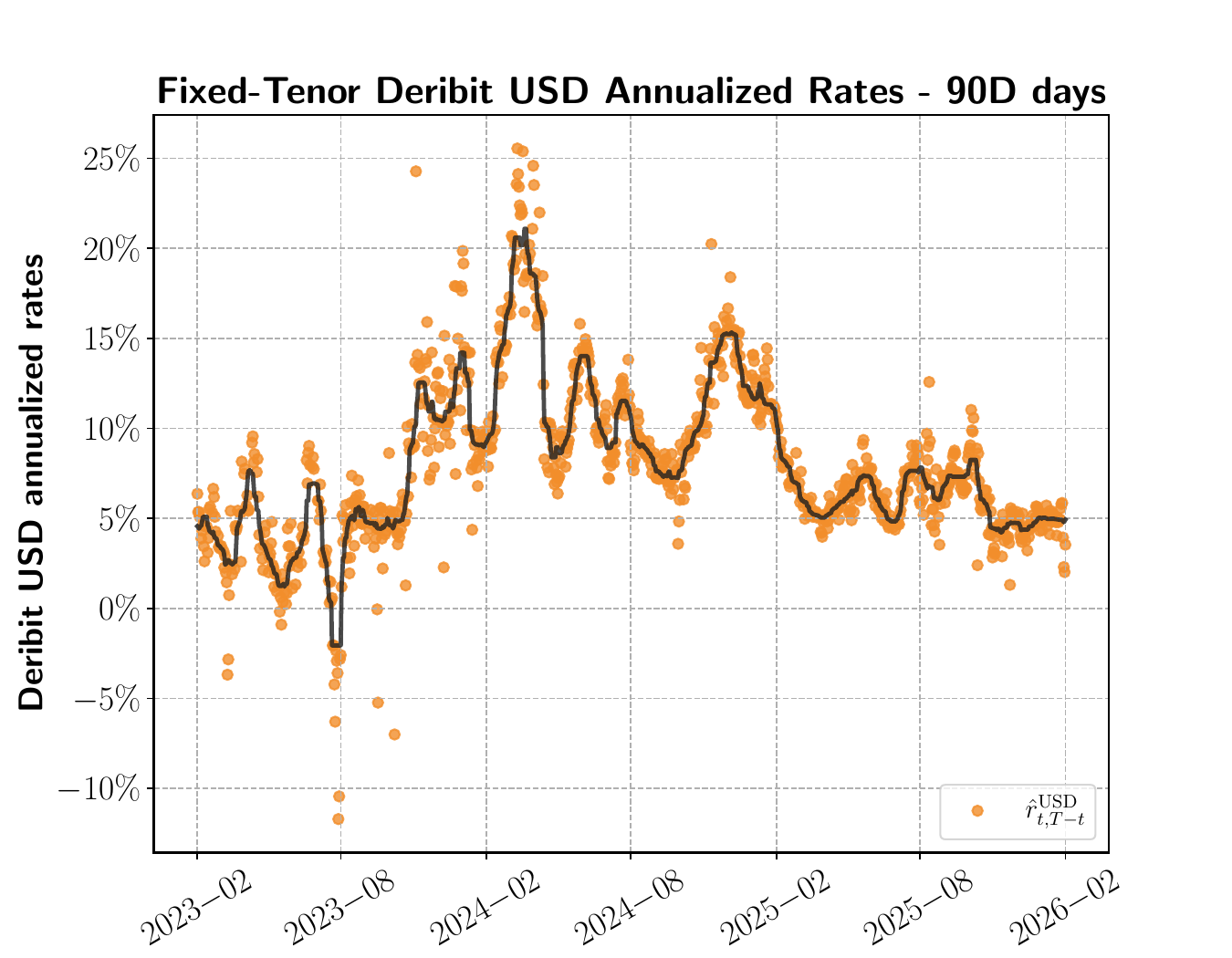}
    \end{subfigure}
    \hspace{-0.02\textwidth}
    \begin{subfigure}[b]{0.5\textwidth}
        \centering
        \includegraphics[width=\textwidth]{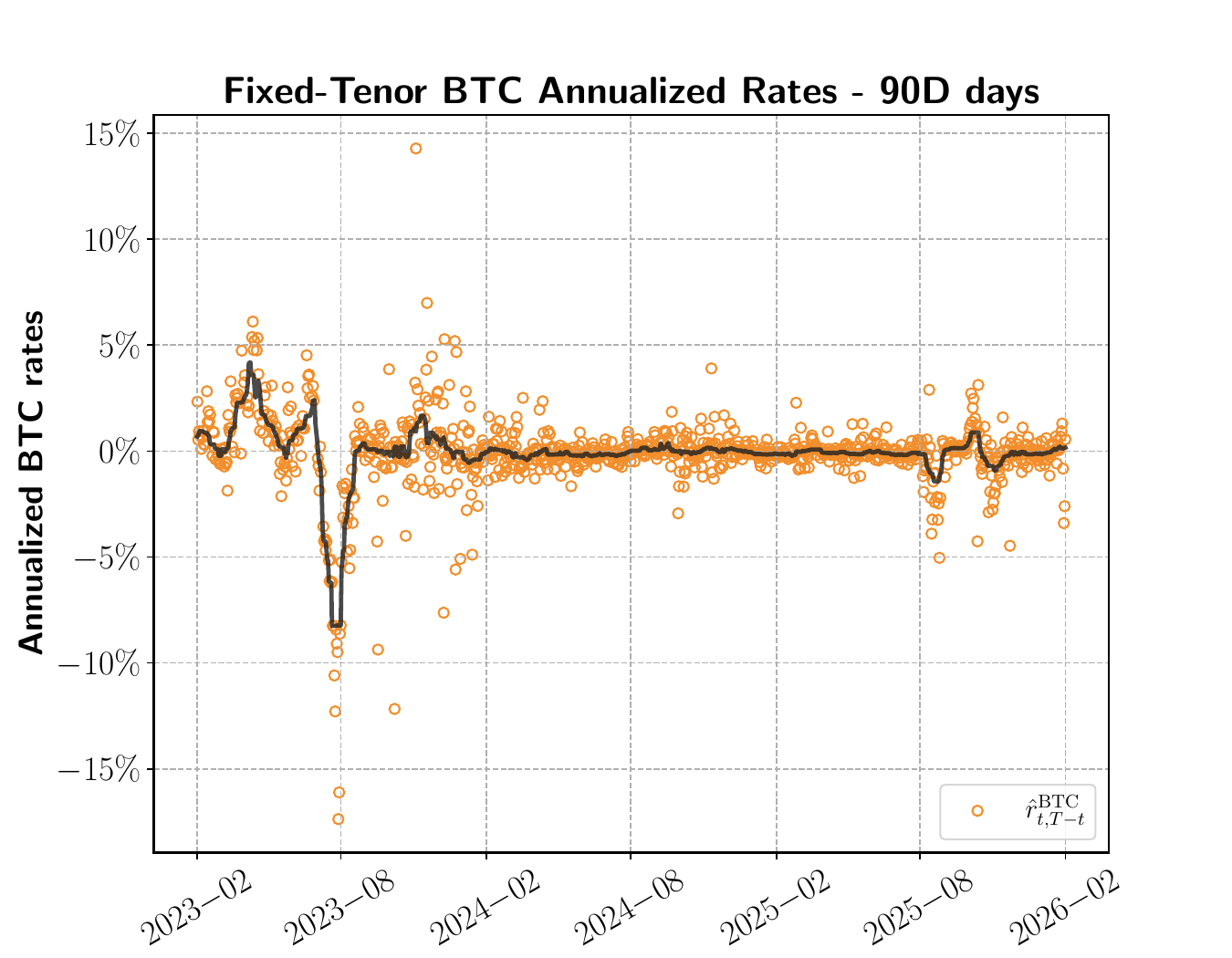}
    \end{subfigure}

    \begin{subfigure}[b]{0.5\textwidth}
        \centering
        \includegraphics[width=\textwidth]{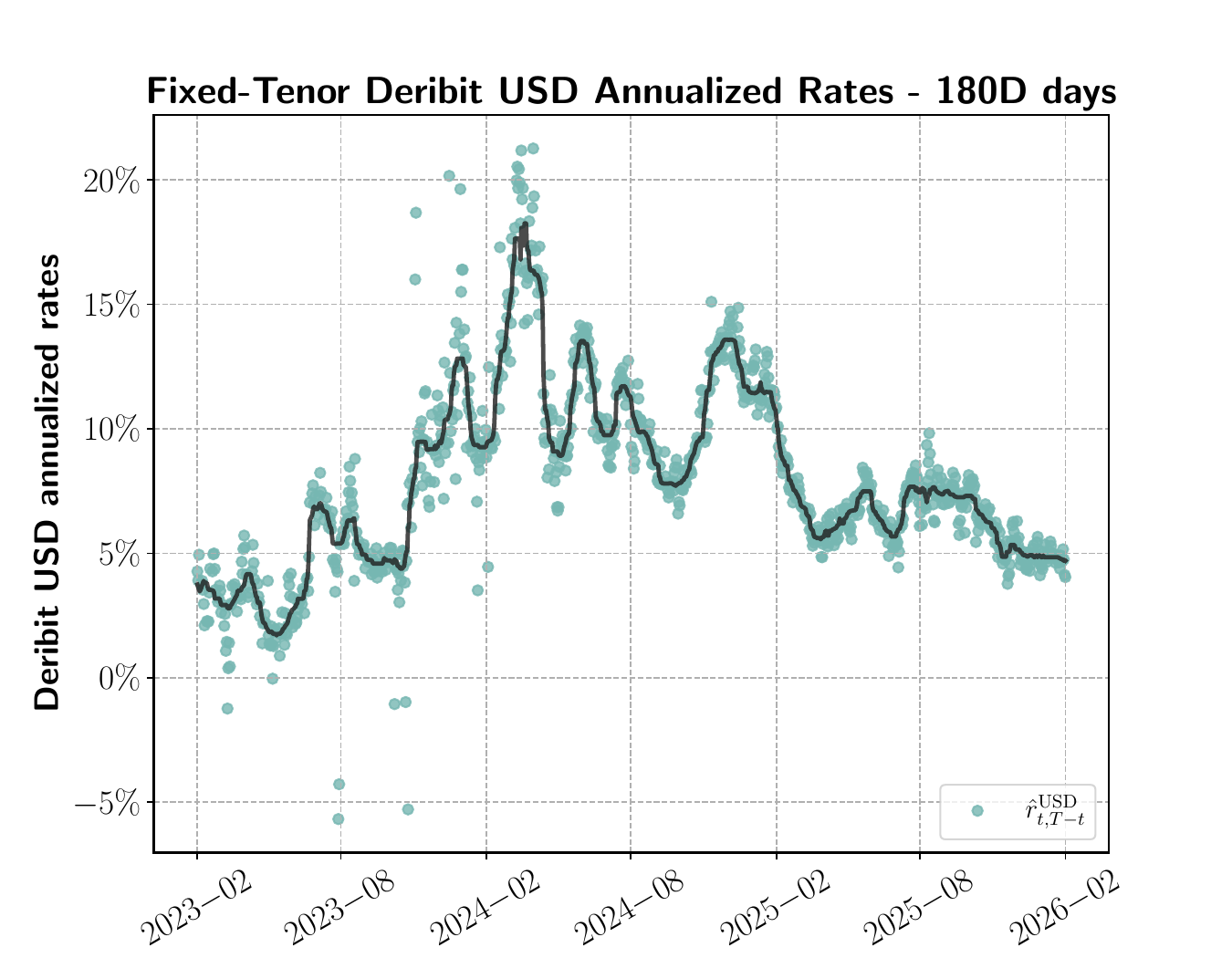}
    \end{subfigure}
    \hspace{-0.02\textwidth}
    \begin{subfigure}[b]{0.5\textwidth}
        \centering
        \includegraphics[width=\textwidth]{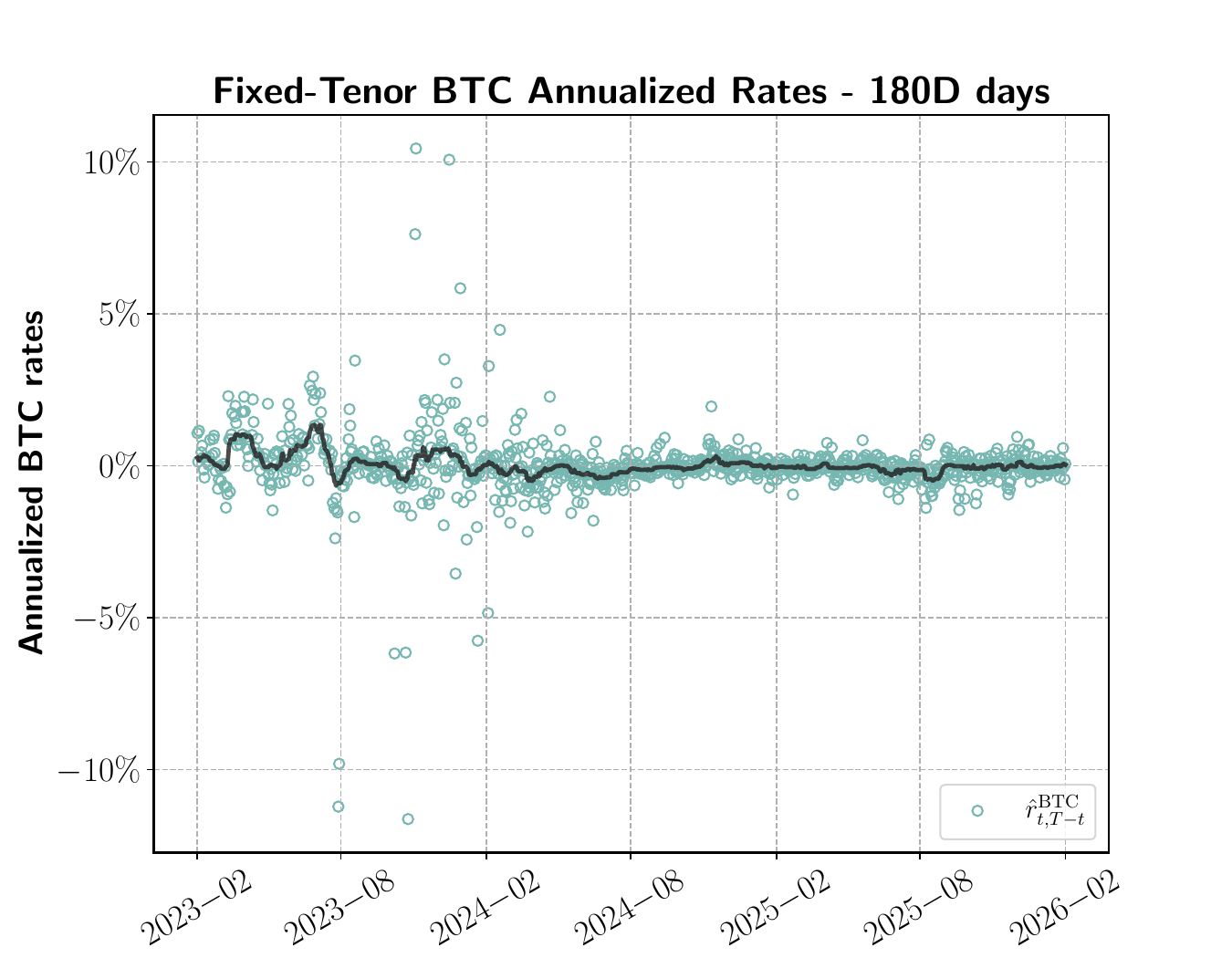}
    \end{subfigure}
    \caption{Raw time series of Deribit USD interest rates (left) and BTC interest rates (right). Maturities: 60~days~(top), 90~days~(middle), and 180~days~(bottom). The black line represents the centered 21-day rolling median.}    \label{fig:BTC_timeseries_rate_curve}
\end{figure}

\newcommand{\ETHrtenorthreemthD}{./ETH_rate_r_tenor_60D.pdf}
\newcommand{\ETHrtenorsixmthD}{./ETH_rate_r_tenor_90D.pdf}
\newcommand{\ETHrtenoroneyrD}{./ETH_rate_r_tenor_180D.pdf}

\newcommand{\ETHqtenorthreemthD}{./ETH_rate_q_tenor_60D.pdf}
\newcommand{\ETHqtenorsixmthD}{./ETH_rate_q_tenor_90D.pdf}
\newcommand{\ETHqtenoroneyrD}{./ETH_rate_q_tenor_180D.pdf}

\begin{figure}[htbp]
    \centering

    \begin{subfigure}[b]{0.5\textwidth}
        \centering
        \includegraphics[width=\textwidth]{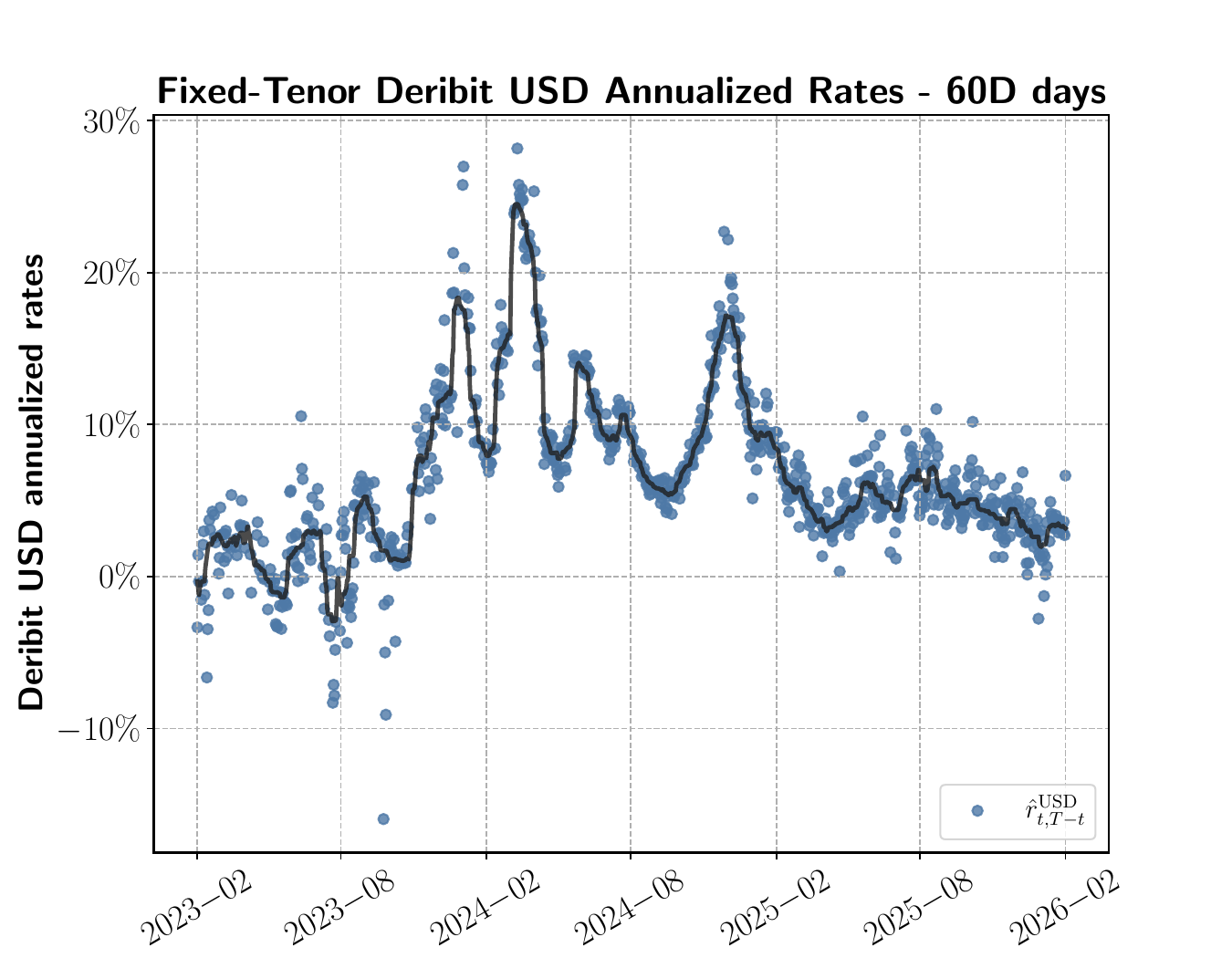}
    \end{subfigure}
    \hspace{-0.02\textwidth}
    \begin{subfigure}[b]{0.5\textwidth}
        \centering
        \includegraphics[width=\textwidth]{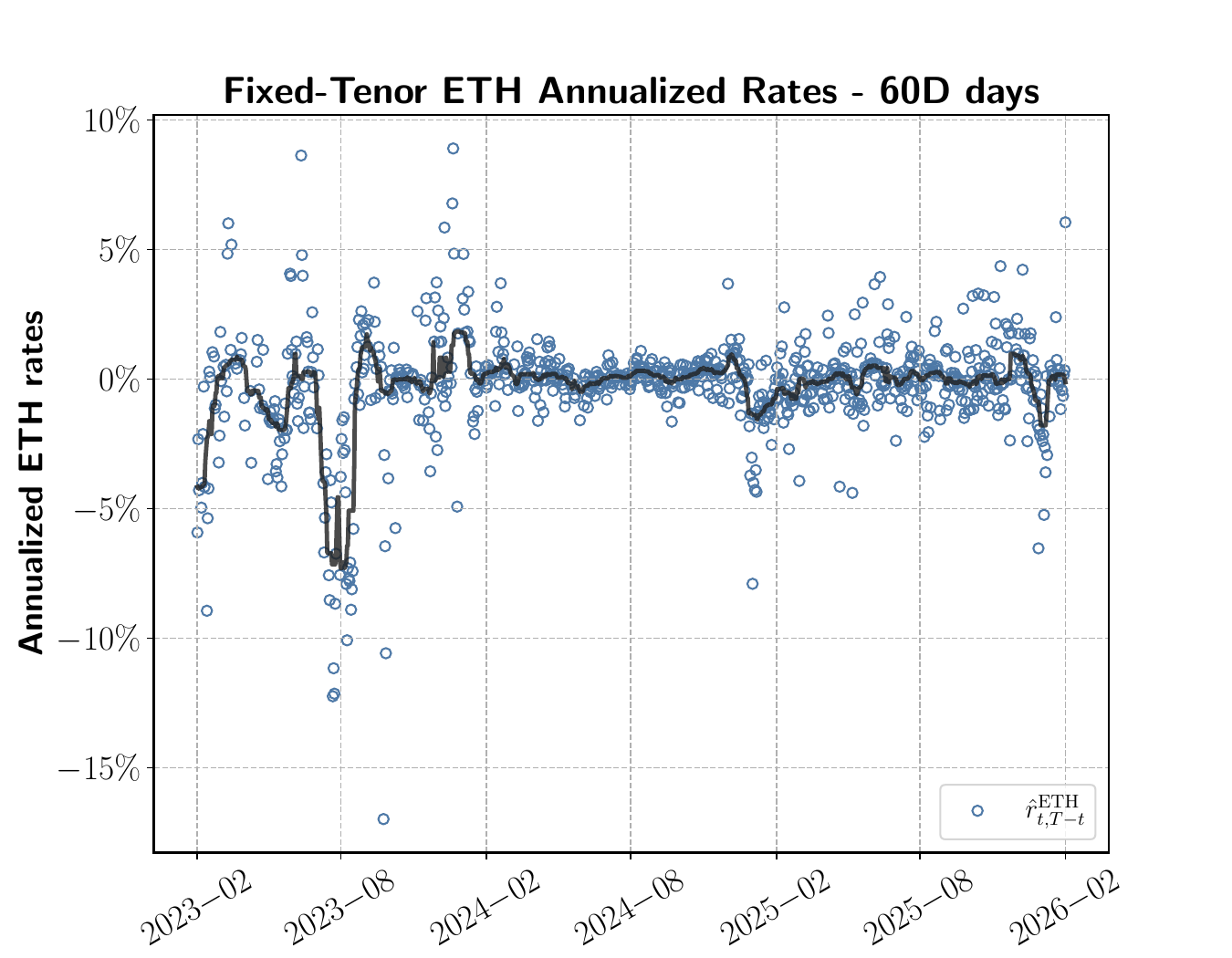}
    \end{subfigure}

    \begin{subfigure}[b]{0.5\textwidth}
        \centering
        \includegraphics[width=\textwidth]{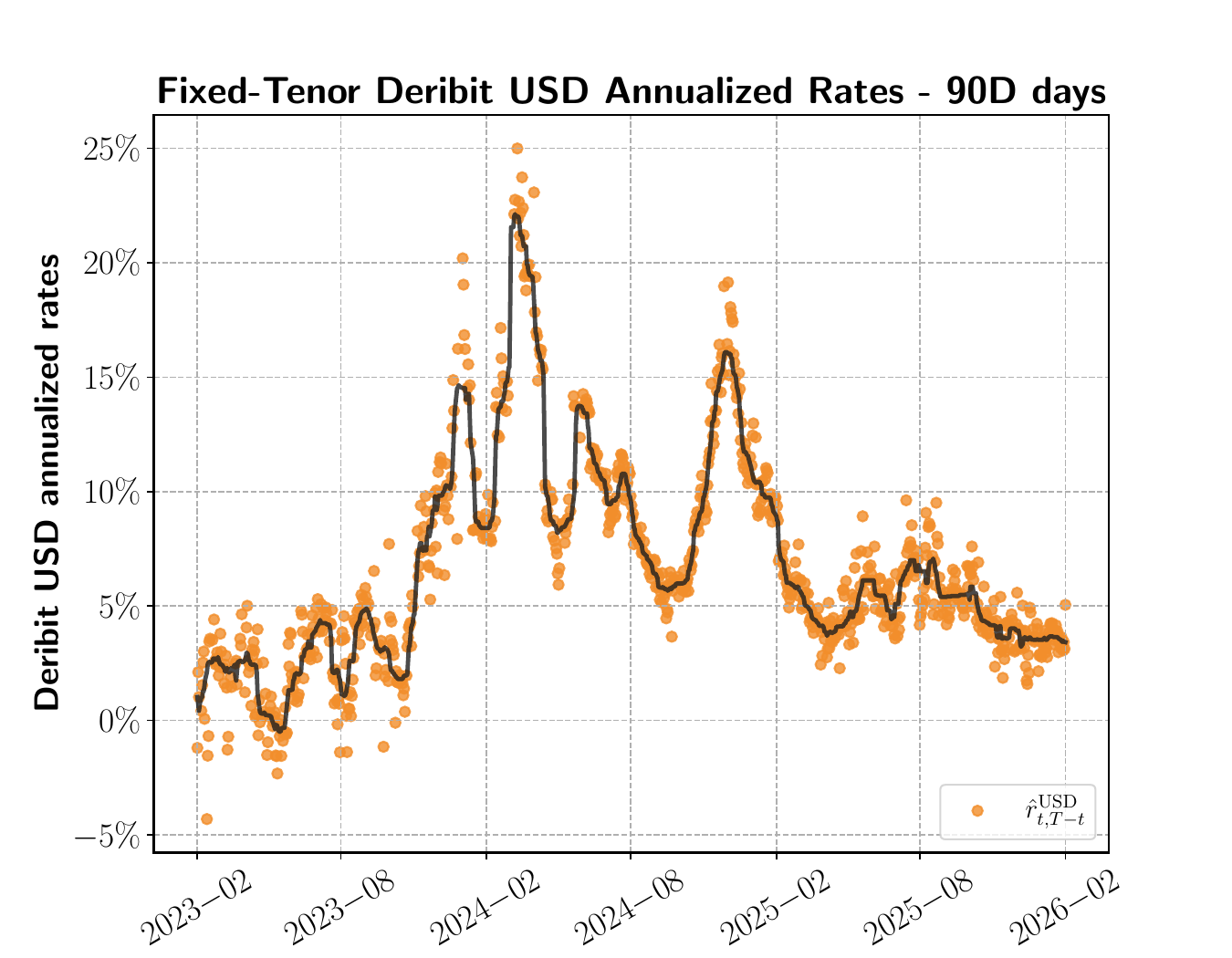}
    \end{subfigure}
    \hspace{-0.02\textwidth}
    \begin{subfigure}[b]{0.5\textwidth}
        \centering
        \includegraphics[width=\textwidth]{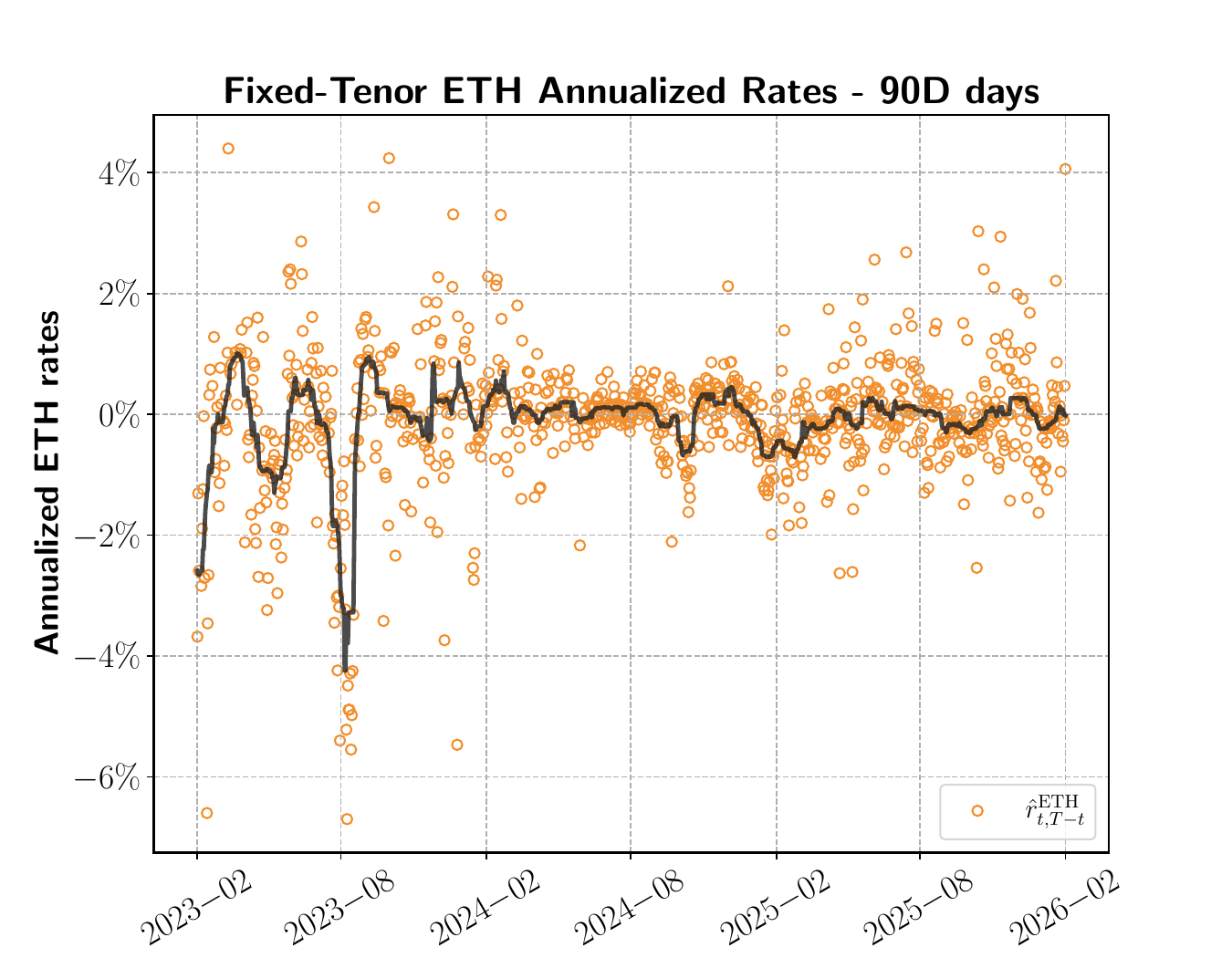}
    \end{subfigure}

    \begin{subfigure}[b]{0.5\textwidth}
        \centering
        \includegraphics[width=\textwidth]{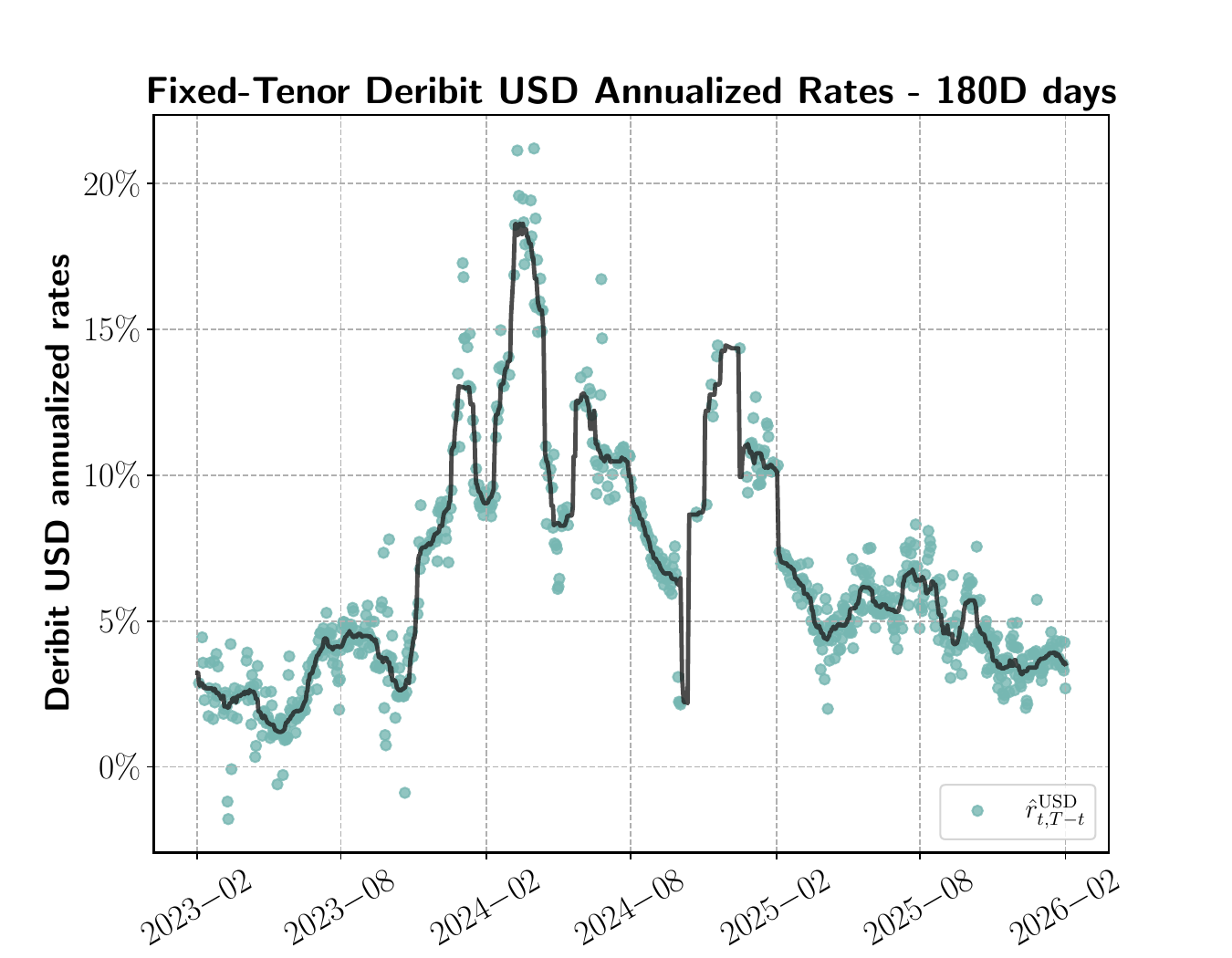}
    \end{subfigure}
    \hspace{-0.02\textwidth}
    \begin{subfigure}[b]{0.5\textwidth}
        \centering
        \includegraphics[width=\textwidth]{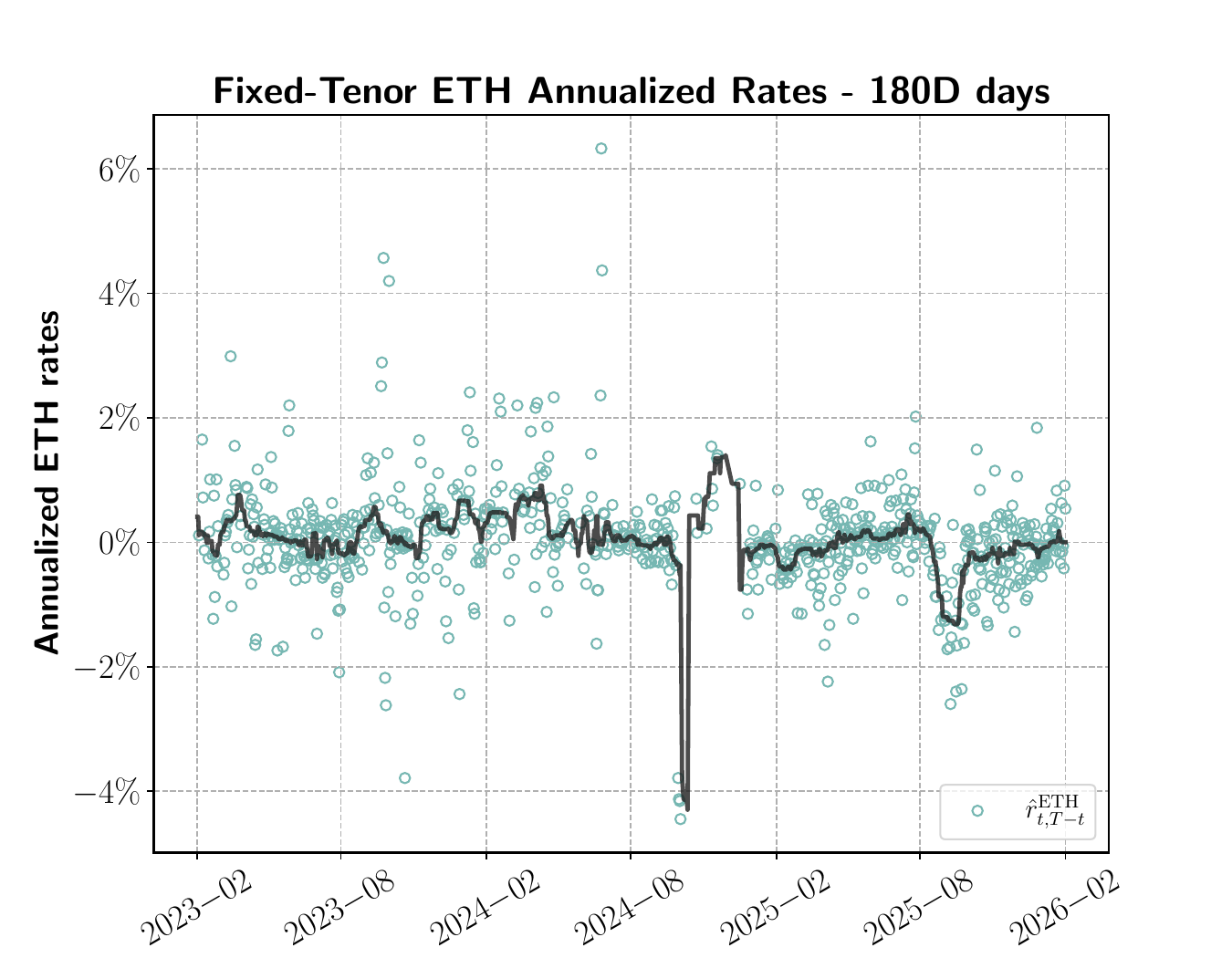}
    \end{subfigure}
    \caption{Raw time series of Deribit USD interest rates (left) and ETH interest rates (right). Maturities: 60~days~(top), 90~days~(middle), and 180~days~(bottom). The black line represents the centered 21-day rolling median.}
    \label{fig:ETH_timeseries_rate_curve}
\end{figure}

\end{document}